\documentclass{aastex7}
\usepackage{graphicx}
\usepackage{mathrsfs}
\begin{document}

\title{Mapping the Cosmic-Ray Ionization Rate in the Local Galaxy with H$_3^+$}
\shorttitle{Mapping the Cosmic-Ray Ionization Rate}

\author[0000-0001-8533-6440]{Nick Indriolo}
\affil{AURA for the European Space Agency (ESA), ESA Office, Space Telescope Science Institute, 3700 San Martin Drive, Baltimore, MD 21218, USA}
\email[show]{nindriolo@stsci.edu}

\author[0000-0002-1590-1018]{Alexei V. Ivlev}
\affil{Max-Planck Institute for Extraterrestrial Physics, Garching, Munich, D-85748, Germany}
\email{ivlev@mpe.mpg.de}

\author{T. Pellegrin}
\affil{Max-Planck Institute for Extraterrestrial Physics, Garching, Munich, D-85748, Germany}
\email{pellegrintheodore@gmail.com}

\author{M. Obolentseva}
\affil{Max-Planck Institute for Extraterrestrial Physics, Garching, Munich, D-85748, Germany}
\email{marta.obolentseva@yandex.ru}

\author[0000-0003-1481-7911]{Paola Caselli}
\affil{Max-Planck Institute for Extraterrestrial Physics, Garching, Munich, D-85748, Germany}
\email{caselli@mpe.mpg.de}

\author[0000-0001-7838-3425]{A. M. Jacob}
\affil{I. Physikalisches Institut der Universit{\"a}t zu K{\"o}ln, Z{\"u}lpicher Str. 77, 50937 K{\"o}ln, Germany}
\affil{Max-Planck-Institut f{\"u}r Radioastronomie, Auf dem H{\"u}gel 69, 53121 Bonn, Germany}
\email{ajacob@ph1.uni-koeln.de}

\author[0000-0001-8341-1646]{D. A. Neufeld}
\affil{William H. Miller III Department of Physics \& Astronomy, Johns Hopkins University, Baltimore, MD 21218, USA}
\email{neufeld@jhu.edu}

\author[0000-0003-1572-0505]{Kedron Silsbee}
\affil{Department of Physics, University of Texas at El Paso, El Paso, TX, 79968, USA}
\email{ksilsbee@gmail.com}

\author[0000-0003-0030-9510]{M. G. Wolfire}
\affil{Department of Astronomy, University of Maryland, College Park, MD 20742-2421, USA}
\email{mwolfire@gmail.com}

\begin{abstract}
Chemistry in diffuse molecular clouds relies primarily on rapid ion-molecule reactions. Formation of the initial ions, H$^+$ and H$_2^+$, is dominated by cosmic-ray ionization of H and H$_2$, making the cosmic-ray ionization rate (denoted $\zeta({\rm X})$ for species X) an important parameter for chemical modeling. We have made observations targeting absorption lines of H$_3^+$, one of the most reliable tracers of $\zeta({\rm H_2})$, toward diffuse molecular cloud sight lines where the H$_2$ column density has been directly measured in the ultraviolet, detecting H$_3^+$ in 12 out of 27 sight lines. The 3D-PDR modeling method introduced by \citet{obolentseva2024} was used to infer cosmic-ray ionization rates in the clouds along these sight lines, and our combined sample has a mean ionization rate of $5.3\times10^{-17}$~s$^{-1}$ with standard deviation $2.5\times10^{-17}$~s$^{-1}$. By associating H$_3^+$ absorption with gas density peaks derived from the differential extinction maps of \citet{edenhofer2024} we have constructed a sparsely sampled 3D map of the cosmic-ray ionization rate in targeted regions within about 1~kpc of the Sun. Specific regions show reasonably uniform ionization rates over length scales of tens of parsecs, with the average ionization rate in each region being different. Large differences (factor of 5) in $\zeta({\rm H_2})$ are found over length scales of about 100 pc. This supports a picture where the cosmic-ray ionization rate varies smoothly over small size scales, but is not uniform everywhere in the Galactic disk, likely being controlled by proximity to particle acceleration sites.
\end{abstract}

\section{Introduction}

The importance of ion-molecule reactions in interstellar chemistry has been established now for over half a century \citep{watson1973,herbst1973}. At the low densities ($n\lesssim100$~cm$^{-3}$) and temperatures ($T\lesssim100$~K) in the diffuse, molecular, interstellar medium (ISM) the most rapid reactions are two-body and exothermic, and the induced-dipole reactions that occur between a neutral and ionic species proceed roughly at the Langevin rate (of order $10^{-9}$~cm$^3$~s$^{-1}$), much faster than typical neutral-neutral reactions  \citep[e.g.,][]{tielens2013}. The limitation for ion-molecule chemistry then, is creation of the initial ions, primarily H$^+$ and H$_2^+$, which require energies above 13.6~eV and 15.4~eV, respectively. Photons at these energies are mainly produced by massive stars, but the prevalence of atomic H in the ISM means that most such photons are absorbed reasonably close to their points of origin. Those that do travel far enough to encounter a molecular cloud will be absorbed in the outermost layers of that cloud, and so will not contribute to ionization in regions where the gas is primarily molecular.
As a result, throughout most of the diffuse molecular ISM, cosmic rays serve as the dominant source of ionization.

Estimates of the cosmic-ray ionization rate (number of ionizations per atom/molecule per unit time; herein denoted $\zeta({\rm X})$ for species X) have been made both from theoretical calculations and observations. Theoretical ionization rates are primarily based on the observed proton spectrum extrapolated to lower energies, and studies have most recently utilized the proton spectrum measured outside of the heliosphere by {\it Voyager 1} \citep{cummings2016} and {\it Voyager 2} \citep{stone2019}, finding $\zeta({\rm H})\approx1.6\times10^{-17}$~s$^{-1}$. These proton spectra, and others, can also be propagated through some column density of material to account for various energy-loss mechanisms and infer the ionization rate inside of molecular clouds \citep[e.g.,][]{padovani2009,padovani2018,padovani2020}. Ionization rates due to cosmic-ray electrons and heavier nuclei can also be computed \citep[e.g.,][]{padovani2018}, but the impact of electron ionization on the total ionization rate is highly uncertain because the process is dominated by electrons with energies below those measured by the {\it Voyager} probes.

Observational constraints on the cosmic-ray ionization rate utilize specific molecular species that have abundances dependent on the H or H$_2$ ionization rate (e.g., OH, OH$^+$, H$_3^+$, ArH$^+$). Of these species, the simple chemistry surrounding the trihydrogen cation (H$_3^+$), makes it the most direct tracer of $\zeta({\rm H_2})$ \citep{dalgarno2006}. Note that while in principle $\zeta({\rm H_2})$ inferred from molecular abundances is the total ionization rate due to all mechanisms (e.g., cosmic rays, X-ray photons, EUV photons, etc.), in general it is a good approximation for the cosmic-ray ionization rate in the environments under consideration. Contributions from X-ray photons are only important in regions with high X-ray fluxes, while EUV photons do not reach the interiors of diffuse molecular clouds. Previous surveys of H$_3^+$ in diffuse molecular clouds determined an average ionization rate of $\zeta({\rm H_2})\approx3.5\times10^{-16}$~s$^{-1}$ \citep{indriolo2007,indriolo2012}, but recent downward revisions to interstellar gas densities \citep{neufeld2024} suggest that that estimate may be about an order of magnitude too high, as revealed by a new analysis of the original H$_3^+$ data \citep{obolentseva2024}. 

Expanding upon our previous work, we have completed a new survey of H$_3^+$ in diffuse molecular clouds, targeting sight lines where the H$_2$ column density is directly measured from prior UV absorption line observations \citep[e.g.,][]{shull2021}. In combination with 3D differential extinction maps derived from {\it Gaia} data \citep[e.g.,][]{edenhofer2024}, this enables application of the 3D-PDR \citep{bisbas2012} modeling techniques introduced by \citet{obolentseva2024} for the purpose of inferring the cosmic-ray ionization rate within a localized cloud along the line of sight. We have thus begun to construct a map of the cosmic-ray ionization rate in the local ISM within the Galaxy.

\section{Observations and Data Reduction}

Observations targeting H$_3^+$ presented herein were made using the iSHELL spectrograph \citep{rayner2022} at the NASA Infrared Telescope Facility (IRTF) and the Cryogenic Infrared Echelle Spectrograph \citep[CRIRES;][]{kaufl2004} at the Very Large Telescope (VLT). IRTF/iSHELL observations utilized the 0\farcs375 slit to provide a resolving power (resolution) of about 80,000 (4~km~s$^{-1}$). Use of the {\it Lp2} mode enabled simultaneous coverage of the $R(1,1)^u$, $R(1,0)$, $R(1,1)^l$, and $Q(1,1)$ transitions at 3.66808~$\mu$m, 3.66852~$\mu$m, 3.71548~$\mu$m, and 3.92863~$\mu$m, respectively. Spectra were obtained in an ABBA nodding pattern to facilitate the removal of atmospheric emission via image pair subtraction. To account for slight variations in the fringe pattern due to instrument flexure, a series of flat field images was taken for every $\sim$30 min of exposure time on target. Observations of each science target were immediately preceded or followed by the observations of a bright, early type star at similar airmass for use as a telluric standard star. VLT/CRIRES observations utilized the 0\farcs2 slit to provide a resolving power (resolution) of about 100,000 (3~km~s$^{-1}$). A reference wavelength of 3715~nm was used to place the $R(1,1)^u$ and $R(1,0)$ transitions on detector 1 and the $R(1,1)^l$ transition on detector 3. To maximize starlight passing through the narrow slit, the adaptive optics system was employed. Spectra were obtained in an ABBA nodding pattern, with 10\arcsec\ separation between nod positions, and 3\arcsec\ jitter about those positions. 
The log of observations from IRTF programs 2021B075, 2022B027, and 2023A033, and VLT program 088.C-0351 is presented in Table \ref{tbl_obslog}. 

Reduction of IRTF data was performed primarily using Spextool \citep{cushing2004} version 5.0.3 for iSHELL. Steps completed with this software package include: combination of flat field images within each series; derivation of the wavelength solution using atmospheric emission features; non-linearity correction; image pair subtraction; flat field correction; identification and tracing of extraction apertures; spectral extraction; background subtraction; combination of spectra using the robust weighted mean method; and division of science target spectra by telluric standard spectra. 

Further data reduction was performed using scripts in a jupyter notebook. Ratioed spectra were normalized by division by a moving boxcar average (15 pixels wide) that was interpolated across absorption features. Because Earth's orbital motion causes astrophysical lines to shift with respect to atmospheric lines throughout the year, wavelength scales for all spectra were converted to the local standard of rest (LSR) frame using radial velocity corrections calculated with astropy. For any targets that were observed on multiple nights, the normalized spectra were combined using a weighted average (weighted by 1/$\sigma^2$, where $\sigma$ is the standard deviation of the line-free continuum in each spectrum). The normalized spectra resulting from these reduction procedures---focused on narrow wavelength windows about the H$_3^+$ transitions---are presented in Figures \ref{fig_h3p_spectra} and \ref{fig_h3p_spectra_no}.

Raw CRIRES images were processed using the CRIRES pipeline version 2.3.3. Standard calibration techniques, including subtraction of dark frames, division by flat fields, interpolation over bad pixels, and correction for detector non-linearity effects, were applied. Consecutive A and B nod position images were subtracted from each other to remove sky emission features, and all images from each nod position were combined to create average A and B images. Spectra were extracted from these images using the \texttt{apall} routine in \textsc{iraf}\footnote{\url{https://iraf-community.github.io/}} and then imported to IGOR Pro.\footnote{\url{https://www.wavemetrics.com}} Wavelength calibration was performed using atmospheric absorption lines, and is accurate to $\pm1$~km~s$^{-1}$. Spectra from the A and B nod positions were then averaged onto a common wavelength scale.

To remove baseline fluctuations and atmospheric features the science target spectra were divided by telluric standard star spectra using custom macros developed in IGOR Pro that allow for stretching and shifting of the telluric standard spectrum in the wavelength axis, and scaling of the telluric standard intensity according to Beer's law \citep{mccall2001}.  The resulting ratioed spectra were then divided by a 30 pixel boxcar average of the continuum level (interpolated across absorption lines) to remove residual fluctuations and produce normalized spectra. The same techniques applied to IRTF/iSHELL data for conversion to the LSR frame and combination of observations from different nights were also applied to VLT/CRIRES data, and the normalized spectra are presented in Figure \ref{fig_h3p_spectra_no}.

\section{Analysis}

\subsection{Spectral Analysis} \label{sect_spectral_analysis}
Clear detections of the $R(1,1)^u$ and $R(1,0)$ transitions are made toward the twelve sight lines presented in Figure \ref{fig_h3p_spectra}. The $R(1,1)^l$ transition is intrinsically weaker than the first two, and is only detected toward eight sight lines. The $Q(1,1)$ transition is intrinsically the weakest of all four transitions, and the continuum level S/N near 3.9~$\mu$m is lower than at 3.7~$\mu$m, such that this transition is only detected toward HD 224151. Due to the higher noise levels and weaker transition, the $Q(1,1)$ transition does not provide any additional meaningful information for our analysis, and is ignored from this point forward.
 
Each absorption line is fit with a gaussian function to determine equivalent width ($W_\lambda$), line center velocity, and line width. The resulting fit parameters are presented in Table \ref{tbl_measurements}, along with the column densities inferred from equivalent widths given the standard equation for optically thin absorption:
\begin{equation}
N(J,K)=\left(\frac{3hc}{8\pi^3}\right)\frac{W_\lambda}{\lambda}\frac{1}{|\mu|^2},
\label{eq_column}
\end{equation}
where $N(J,K)$ is the column density in the state from which the transition arises, $h$ is Planck's constant, $c$ is the speed of light, $\lambda$ is the transition wavelength, and $|\mu|^2$ is the square of the transition dipole moment \citep[see Table 2 in ][for values]{goto2002}. In diffuse molecular clouds only the $(J,K)=(1,0)$ and $(1,1)$ states of H$_3^+$ (ground {\it ortho} and ground {\it para} states, respectively) are expected to be significantly populated, so the total column density of H$_3^+$ is well approximated by the sum of column densities in those two states. 

When H$_3^+$ absorption is not detected, upper limits to the equivalent width are determined using the method described in \citet{obolentseva2024}, where CH absorption is used to define the expected H$_3^+$ absorption profile. Observations covering the 4300~\AA\ line of CH were obtained from multiple facilities and instruments, including UVES at the VLT \citep{eso_uves}\footnote{\url{https://doi.org/10.18727/archive/50}},
HIRES at Keck (D. Welty 2024, private communication), HERMES at the Mercator Telescope \citep[MELCHIORS archive;][]{royer2024}\footnote{\url{https://www.royer.se/melchiors.html}}, and the Coud\'{e} feed telescope at KPNO (D. Welty 2024, private communication). Spectra were normalized in intensity, and wavelength was converted to LSR velocity. If the spectral resolution of the CH spectrum was finer than that of the H$_3^+$ spectrum, then the CH spectrum was degraded to match the H$_3^+$ spectrum. CH absorption profiles were then fit using 1--3 gaussian components based on their complexity, and the line centers and line widths were saved for use in H$_3^+$ fitting. If the spectral resolution of the CH spectrum was poorer than that of the H$_3^+$ spectrum, then the line widths returned by the fit were reduced to that expected for the finer resolution, assuming the absorption line is unresolved. For a given sight line each H$_3^+$ absorption line was then fit using the sum of $N$ gaussian components, where $N$ was the number of components used to fit the CH profile, and the line centers and line widths of these gaussians were fixed to the values returned by the CH fitting procedure; only the line depths were allowed to vary as free parameters. An example of this CH and H$_3^+$ fitting procedure is shown in Figure \ref{fig_example_CHfit}. Taking the line depths returned by the fit and their uncertainties, we computed the equivalent widths and their uncertainties. These ``measured'' equivalent widths do not correspond to detections of H$_3^+$ absorption, but are a useful means to constrain the potential signal from absorption in coherent features that happen to be at the expected location in the spectrum, and are presented in Table \ref{tbl_upperlimits}. These values are used to compute column densities in the respective states via Equation (\ref{eq_column}). The $R(1,1)^u$ and $R(1,1)^l$ transitions provide independent estimates of the column density in the $(J,K)=(1,1)$ state, and we adopt the {\it smaller} value of $N(1,1)+\sigma(N(1,1))$ as the upper limit for this state. The upper limit on the total H$_3^+$ column density is taken to be $N(1,0)+\sigma(N(1,0))+N(1,1)+\sigma(N(1,1))$, and these values are presented in Table \ref{tbl_LoS}, along with values of $N({\rm H}_3^+)$ for sight lines where H$_3^+$ was detected.

\subsection{The Underlying Chemistry}

The simple chemistry by which H$_3^+$ is formed
\begin{equation}
{\rm H}_2 + {\rm CR}\rightarrow {\rm H}_2^+ + e^- + {\rm CR}',
\label{re_CR_H2}
\end{equation}
\begin{equation}
{\rm H}_2^+ + {\rm H}_2\rightarrow {\rm H}_3^+ + {\rm H}
\label{re_H2_H2+}
\end{equation}
and destroyed
\begin{equation}
{\rm H}_3^+ + e^-\rightarrow {\rm H}_2 + {\rm H~or~H + H + H},
\label{re_H3+_e}
\end{equation}
in diffuse clouds makes its abundance useful for inferring cosmic-ray ionization rates. In the simplistic scenario where we assume steady state chemistry, constant gas density, constant electron density, and co-location of H$_2$ and H$_3^+$ along the line of sight, the cosmic-ray ionization rate of molecular hydrogen can be inferred analytically from the expression
\begin{equation}
\zeta({\rm H}_2)=k_{e}x_{e}n_{\rm H}\frac{N({\rm H}_3^+)}{N({\rm H}_2)},
\label{eq_crir}
\end{equation}
where $\zeta({\rm H}_2)$ is the total cosmic-ray ionization rate of H$_2$ (i.e., accounting for ionization by both hadronic and leptonic cosmic rays, as well as by secondary electrons), $k_e$ is the H$_3^+$-electron recombination rate coefficient, $n_{\rm H}\equiv n({\rm H})+2n({\rm H}_2)$ is the density of hydrogen nuclei, $x_{e}\equiv n_e/n_{\rm H}$ is the electron fraction, and $N({\rm H}_2)$ and $N({\rm H}_3^+)$ are column densities. Ionization rates calculated from equation (\ref{eq_crir}) are in reasonable agreement with those inferred from 1D chemical models \citep{neufeld2017}, and are within a factor of 2 of those inferred from the 3D-PDR modeling presented in \citet{obolentseva2024} for most sight lines when the same gas density is adopted. Given the reasonable agreement between methods, in cases where the line-of-sight gas density is complex and the H$_3^+$ absorption cannot be confidently localized to a single cloud, we use a re-arranged version of equation (\ref{eq_crir}) to provide an estimate of $\zeta({\rm H}_2)/n_{\rm H}$. For sight lines where the absorbing gas can be assigned to a single well-defined cloud though, we utilize the 3D-PDR analysis introduced by \citet{obolentseva2024}.

\subsection{Simulations with 3D-PDR} \label{sec_sim3DPDR}

From the full sample of spectra with H$_3^+$ detections presented in Figure \ref{fig_h3p_spectra}, nine sight lines have properties conducive to the full 3D-PDR modeling method: HD~23180, HD~281159, HD~170740, HD~179406, HD~203374, HD~206165, HD~206267, HD~207198, and HD~224151. Additionally, we consider the sight line toward HD~27778 for which H$_3^+$ was reported in \citet{albertsson2014}. A detailed description of the 3D-PDR simulation setup can be found in \citet{obolentseva2024}; below we summarize the essential aspects involved in the modeling.

The simulation of individual clouds along target sight lines relies on the differential extinction maps of \citet{edenhofer2024}, converted into 3D gas density maps. This requires assumptions regarding the extinction curve and gas-to-dust ratio, and we adopt the same conversion as in our previous work: $n_{\rm H}=1710(dE_{GRZ}/ds)$~cm$^{-3}$, where $E_{ZRG}$ is the extinction defined by \citet{zhang2023} and $s$ is the distance along the line of sight in pc \citep[see][and references therein for more details]{neufeld2024,obolentseva2024}. The resulting 1D gas density profiles along our target sight lines and 2D gas density maps containing these sight lines can be found in Figures \ref{fig_gaia_hd23180} through \ref{fig_gaia_hd217312}. In the present paper we use the mean density distribution\footnote{Differences between the simulation results reported in \citet{obolentseva2024} for a selected map realization and for the mean map were negligible.}, which is derived from 12 posterior map samples  \citep{edenhofer2024}.

For individual sight lines we find some deviations between the total hydrogen column densities ($N_{\rm H}\equiv N({\rm H})+2N({\rm H}_2)$) obtained from measurements (Obs.~$N_{\rm H}$) and the values derived from the extinction map (Map~$N_{\rm H}$), as summarized in Table~\ref{tbl_LoS}. Similar to the targets studied in \citet{obolentseva2024}, there is a moderate (around 10--50\%) systematic underestimate of the total gas column derived from the map with respect to observations (associated with possible uncertainties in the extinction curve and gas-to-dust ratio, see discussion therein). Following \citet{obolentseva2024}, we compensate for this systematic effect in our simulations by multiplying the gas density $n_{\rm H}$ deduced from the map by the scaling factor equal to the column density ratio (Obs.~$N_{\rm H}$)/(Map~$N_{\rm H}$). Our modeling is therefore made assuming that the density distribution corresponds to the mean measured value of the total gas column density.

For each selected cloud, we define an ellipsoidal simulation domain centered at the density peak, with one axis oriented along the line of sight. The exact size of the ellipsoid is different for each modeled cloud, and is defined so that the entire cloud (individual coherent structure in the 3D gas density map) is fully contained within the ellipsoid. The two semi-axes perpendicular to the line of sight are constrained to have the same size (varied from 30--45~pc among our sample of modeled clouds), while the semi-axis along the line of sight is independent (varied from 30--60~pc among our sample). The nature of the ellipsoid is not further constrained, and simulation domains include spheres, prolate spheroids, and oblate spheroids.
Within the simulation domain, the mean gas density map is interpolated onto a uniform Cartesian grid with a 1~pc resolution\footnote{Simulations were tested at multiple spatial resolutions and found to converge for elements with size 1$\times$1$\times$1~pc, precluding the need for finer sampling.}, resulting in a total of $(1-5)\times 10^{5}$ elements, and serves as the underlying gas distribution for the model. 
To compute the far ultraviolet (FUV) radiation field, we follow the method described in \citet{obolentseva2024} and divide the field in the domain into two components: point-source contributions of individual ``nearby'' hot stars and a continuous ``far-field'' component. The far-field component in the proximity of all simulated clouds is found to be in the range $0.2\leq\chi\leq 0.5$, where $\chi$ is the FUV field from \citet{draine1978}. Coordinates of individual hot stars contributing noticeably to the local FUV field are taken from Gaia DR3 \citep{gaiaDR32022}, the StarHorse catalog \citep{anders2019}, or from Hipparcos~2 data \citep{vanleeuwen2007}; we found the contribution of such stars to be important only for the sight lines to HD~23180 and HD~281159.

To probe the possible expected range of cosmic-ray ionization rates we perform simulations for four different characteristic values of $\zeta({\rm H_2})$: $1\times$, $5\times$, $10\times$, and $20\times10^{-17}$~s$^{-1}$. The results of simulations for each detection sight line are shown in Figure~\ref{fig_modelresults}, plotting the computed values of $N({\rm H}_2)$ and $N({\rm H}_3^+)$ versus $\zeta({\rm H_2})$ together with the observed values and their uncertainties from Table~\ref{tbl_LoS}. Similar to the results reported in \citet{obolentseva2024}, Figure~\ref{fig_modelresults} shows a moderate deviation of the computed amount of H$_2$ from the measured values, with the overall underestimate of $N({\rm H}_2)$ in our simulations. As discussed therein, this discrepancy might be attributed to the fact that we compute equilibrium H$_2$ abundance and assume the clouds to be quiescent \citep[see also][]{Bialy2019}. To correct for a moderate effect of this discrepancy on the inferred value of $\zeta({\rm H}_2)$, we follow the approach implemented in \citet{obolentseva2024} and require the ratio $N({\rm H}_3^+)/N({\rm H}_2)$ that is derived from the simulation to be equal to the ratio of the measured mean values -- which ensures that cosmic rays generate $N({\rm H}_3^+)$ in measured proportions to $N({\rm H}_2)$. The black symbols in the bottom panels of Figure~\ref{fig_modelresults} show the resulting ``corrected'' values, obtained by multiplying the computed $N({\rm H}_3^+)$ (blue symbols) by the ratio of the measured to computed $N({\rm H}_2)$. The intersection between the black curve and the red horizontal line determines the cosmic-ray ionization rate for each cloud, and these values are reported in Table \ref{tbl_zeta}.

\subsubsection{HD~23180 and HD~281159} 
These two sight lines represent a remarkable case where directions to the target stars practically coincide. By comparing Figures~\ref{fig_gaia_hd23180} and \ref{fig_gaia_hd281159} we conclude that the former sight line primarily probes the near side of the Per-Tau shell \citep{bialy2021} at a distance of $\approx150$~pc, while the latter sight line is essentially a sum of that and of the contribution from the far side at $\approx300$~pc. With an on-sky angular separation of 8.2 arcmin the material being probed by these two sight lines is separated by only 0.35~pc in the near cloud and 0.7~pc in the far cloud, well below the resolution of the 3D dust map. This fact allows us to disentangle the two contributions for the sight line to HD~281159 by assuming that the near cloud has the same H$_2$ and H$_3^+$ column densities as those measured toward HD~23180, and thus that the far side cloud has column densities equal to the difference between the HD~281159 and HD~23180 line of sight values. The panels for HD~281159 in Figure~\ref{fig_modelresults} show the simulation results plotted versus $\zeta({\rm H}_2)$ in the far side cloud, with the $N({\rm H_3^+})$ contribution from the near side cloud set equal to that measured toward HD~23180. Note that the H$_3^+$ absorption lines are only detected at a $2\sigma$ level toward HD~281159 though, so the far cloud analysis results are less robust than for other sight lines.

\subsubsection{HD~167971, HD~216532, and HD~216898}
These three sight lines showing H$_3^+$ absorption were excluded from the 3D-PDR analysis. HD~167971 (see Figure \ref{fig_gaia_hd167971}) is located well beyond the extent of the \citet{edenhofer2024} extinction map, and a significant portion of the total hydrogen column along the sight line may be coming from that unmapped region, as can be seen from the more extensive maps from \citet{lallement2022}.
In particular, the H$_3^+$ absorption observed toward HD~167971 is at $v_{\rm LSR}\approx27$~km~s$^{-1}$, which suggests that the probed gas corresponds to the component in the \citet{lallement2022} dust map beyond 1~kpc\footnote{This assumes that larger velocities are more likely to occur in more distant material.}. As for HD~216532 and HD~216898 (see Figures~\ref{fig_gaia_hd216532} and \ref{fig_gaia_hd216898}), these neighboring sight lines contain multiple gas clumps of comparable column densities. The major contribution to the total gas column is provided by the peaks at a distance around 800 pc, located in the vicinity of the Cep OB3 association, which has three O stars and tens of B stars \citep{blaauw1959} producing an estimated local FUV field of $\chi>30$. Given the degraded accuracy of the extinction map at such distances and the critical importance of knowing gas distribution to compute the formation of H$_2$ in such extreme environments, we concluded that reasonable modeling of these two sight lines cannot be performed.

\subsubsection{Sources of Uncertainty} \label{sec_uncertainty}

Uncertainties on the inferred ionization rates are difficult to quantify, 
as there are contributions from several different sources,
including: in the measured H$_2$ and H$_3^+$ column densities; in the construction of differential extinction maps; in the conversion from extinction to gas density; in the model assumptions that routinely under-predict $N({\rm H_2})$. Accounting for uncertainties in $N({\rm H}_3^+)$ is conceptually the most straightforward, as we can simply consider the overlap between the curves and shaded regions in Figure \ref{fig_modelresults}, rather than just the intersection point. \citet{obolentseva2024} accounted for uncertainties in $N({\rm H_2)}$ and $N({\rm H})$ by running a larger grid of 3D-PDR models, where the total column density through the cloud was set equal to the measured $\pm1\sigma$ limits, and used a reduced-$\chi^2$ analysis to determine the optimum value of the cosmic-ray ionization rate and its uncertainties. Through this analysis they determined that the contribution from $\sigma(N({\rm H}_3^+))$ dominated that from $\sigma(N({\rm H}))$ and $\sigma(N({\rm H}_2))$ in the total uncertainty on the ionization rate. In Table \ref{tbl_zeta} we choose to report uncertainties on $\zeta({\rm H_2})$ based only on the measured values of $N({\rm H}_3^+)\pm\sigma(N({\rm H}_3^+))$, but caution the reader that total uncertainties are likely larger due to the reasons given below.

Quantifying the uncertainties related to the extinction maps and models is more difficult. While the analysis of \citet{zhang2023} assumed a universal extinction curve with $R(V)=3.1$ (where $R(V)\equiv A(V)/E(B-V)$ is the ratio of total-to-selective visual extinction), there are large scale spatial variations in $R(V)$ within 2~kpc of the Sun \citep{zhang2025}. Individual sight lines show deviations from the average relation \citep[$2.3\leq R(V)\leq5.6$,][and references therein]{gordon2023}, and this could impact the gas density profile derived along any particular line of sight. The reliability of the differential extinction maps themselves is related to the density of background stars per unit solid angle in a given region on-sky, and for some sight lines the 1D density profile is more trustworthy than others. Additionally, the assumptions made by different reconstruction methods can produce differences in the final maps (see, e.g., Figure \ref{fig_gaia_hd42087}). Discrepancies between the observed and map-integrated values of $N_{\rm H}$ for the same sight line demonstrate that the gas density profiles used in our 3D-PDR modeling may deviate from ground truth, even with the scaling method described above. This may be especially important for clouds that are farther away from the Sun, as the resolution along the sight line becomes coarser, and narrow, high-density peaks can be smoothed out. Evidence for differential extinction maps underestimating the gas density was reported by \citet{neufeld2024}, where they showed that the ratio between gas densities derived from differential extinction maps and gas densities derived from an analysis of the rotational excitation of C$_2$ decreased as a function of distance from the Sun. Beyond about 400~pc the C$_2$-derived densities are about 10 times larger than the extinction-derived densities, suggesting that the extinction maps are incapable of resolving the density structure within clouds at larger distances, and so become less reliable in serving as the basis for 3D-PDR models. Adopting higher gas densities results in higher inferred ionization rates, but the extent of higher density peaks probed by C$_2$ along any given sight line is unclear, and the observed H$_3^+$ must reside in gas at a range of densities, likely bounded by C$_2$ and extinction map estimates. All of these effects introduce systematic uncertainties to our analysis that can be different for every sight line and are not quantifiable in a straightforward manner.


\subsection{Analytical Calculations} \label{sec_analytical}

While the complications described above make the 3D-PDR modeling approach infeasible for HD~167971, HD~216532, and HD~216898, it is still possible to derive information about the cosmic-ray ionization rate from these H$_3^+$ detections.
The line of sight toward HD~43384---H$_3^+$ detection originally presented in \citet{albertsson2014} but never before analyzed in terms of the cosmic-ray ionization rate---shares similarities with these three sight lines, and here we choose to add it to our analysis. Finally, while a 3D-PDR analysis of HD~41117 was presented in \citet{obolentseva2024}, they noted that the sight line is an outlier both in terms of observed vs. map-integrated total hydrogen column density and 1D gas density profile complexity. Their derived ionization rate is an average line-of-sight value, and cannot be localized to a specific interstellar cloud. Because the HD~41117 sight line has characteristics that make it a poor candidate for the 3D-PDR modeling, we choose to reanalyze it here using the methods applied to other, similarly complex sight lines.

Despite these sight lines having complex CH and K~\textsc{i} absorption profiles, H$_3^+$ absorption in each sight line is limited to a single velocity component, as seen in Figure \ref{fig_h3p_spectra} and Table \ref{tbl_measurements} \citep[and Figure 1 of][]{albertsson2014}. Total H$_2$ column densities in each sight line include contributions from all of the absorbing clouds, so calculations using those values would underestimate the ionization rate in the cloud with H$_3^+$. We take advantage of the linear relationship between H$_2$ and CH column densities \citep[$N({\rm CH})/N({\rm H_2})=3.5^{+2.1}_{-1.4}\times10^{-8}$;][]{sheffer2008} and use observations of the 4300~\AA\ transition of CH to estimate $N({\rm H_2})$ in the cloud component where H$_3^+$ absorption is observed. Observations toward HD~216532 and HD~216898 including CH absorption are presented in \citet{pan2004}, while spectra covering 4300~\AA\ toward HD~43384 and HD~167971 observed with UVES are available in the ESO Science Archive \citep{eso_uves}. Using the same CH profile fitting methods discussed previously, we determine equivalent widths for the various components and compute $N({\rm CH})$ for the component that matches the H$_3^+$ absorption in velocity \citep[CH transition oscillator strength taken from][]{larsson1983a}. Estimated H$_2$ column densities in the specific cloud giving rise to H$_3^+$ absorption are given in Table \ref{tbl_zeta_n}. The K~\textsc{i} profile toward HD~41117 shows two narrow components \citep{welty2001}, but the spectral resolution in published data \citep{crane1995} and archival UVES data covering the CH line is too coarse to separate the components, so we must use the total line of sight H$_2$ column in this case. Although we can estimate $N({\rm H_2})$ in the cloud with H$_3^+$ in most cases, the inability to assign the absorption to a peak in the 1D line-of-sight density profile means that we cannot easily determine the gas density, so instead of computing $\zeta({\rm H_2})$ we use a simple re-arrangement of equation (\ref{eq_crir}) to compute $\zeta({\rm H_2})/n_{\rm H}$. In determining $k_e$ we adopt the expression from \citet{mccall2004}, and assume that the electron temperature is equal to the H$_2$ spin temperature derived from the measured $J=0$ and $J=1$ column densities. The electron fraction, $x_e$, is assumed to be equal to the average abundance of C$^+$ in diffuse molecular clouds \citep[$1.5\times10^{-4}$;][]{sofia2004}, which is a reasonable approximation for environments where the dominant source of electrons is photoionized carbon. The results of these calculations are presented in Table \ref{tbl_zeta_n}. Uncertainties in $\zeta({\rm H_2})/n_{\rm H}$ account for $\sigma(N({\rm H}_3^+))$ and $\sigma(N({\rm H}_2))$, with the latter based directly on H$_2$ observations for HD~41117, and on $\sigma(N({\rm CH}))$ and the scatter in the $N({\rm CH})/N({\rm H_2})$ relationship for the other four sight lines.

\subsection{Upper Limits}

For sight lines where H$_3^+$ is not detected, the analysis described in Section \ref{sect_spectral_analysis} provides an upper limit to the H$_3^+$ column density along the entire line of sight. In cases where there is a single density peak along the line of sight we can constrain the ionization rate within that cloud, but for cases with multiple density peaks we can only constrain the average ionization rate along the sight line. Rather than using the full 3D-PDR modeling technique for these sight lines, we instead employ the simpler analytical method, again computing $\zeta({\rm H_2})/n_{\rm H}$. Results are given in Table \ref{tbl_zeta_n}. Note that this method of computing upper limits is less rigorous than the ``optimum value'' method employed by \citet{obolentseva2024}, and results in lower values.

\subsection{Method Comparison} \label{sec_methodcompare}
In sight lines where the 3D-PDR modeling analysis has been employed, it is also possible to infer cosmic-ray ionization rates using equation (\ref{eq_crir}). The values of $N({\rm H}_3^+)$ and $N({\rm H}_2)$ are taken from Table \ref{tbl_LoS}, while $x_e$ and $k_e$ are treated as in Section \ref{sec_analytical}. Gas densities are determined from the analysis of C$_2$ rotational level populations \citep{neufeld2024} and from the \citet{edenhofer2024} differential extinction maps, and both sets of densities are presented in Table \ref{tbl_zeta}. In the case of the differential extinction maps, we take the mean of the peak densities in the absorbing cloud for all twelve map realizations, which differs from the peak density in the mean realization since the peak can occur at slightly different distances along the line of sight. The $dE$-map derived densities are then rescaled by the ratio (Obs.~$N_{\rm H}$)/(Map~$N_{\rm H}$), as described in Section \ref{sec_sim3DPDR}. Using both sets of densities we infer cosmic-ray ionization rates via the simple analytical expression in the modeled clouds, and present these results in Table \ref{tbl_zeta}.

A comparison of the ionization rates inferred using the different methods in presented in Figure \ref{fig_method_compare}. The left panel compares ionization rates inferred from equation (\ref{eq_crir}) using gas densities derived from C$_2$ excitation to ionization rates inferred from the 3D-PDR analysis. The center panel shows the same, but in this case the analytical expression uses gas densities derived from the differential extinction maps. When C$_2$ derived densities are used the inferred cosmic-ray ionization rates are larger than those found from the 3D-PDR modeling by a factor of 2--3 typically, although some sight lines are closer to a factor of 10. When $dE$ derived gas densities are used both methods are generally in agreement within a factor of 2. Of course, better agreement is expected for the latter case since both methods use the same gas density. This suggests that the different densities used by the 3D-PDR analysis and C$_2$ excitation analysis may drive the different ionization rates inferred. As mentioned in Section \ref{sec_uncertainty}, the $dE$-derived densities may systematically decrease with increasing distance from the Sun. The right panel of Figure \ref{fig_method_compare} shows the ratio of $\zeta({\rm H}_2)$ inferred from the analytical analysis using C$_2$-derived gas densities to $\zeta({\rm H}_2)$ inferred from the 3D-PDR analysis as a function of cloud distance from the Sun. While the two highest ratios come from clouds at the two largest distances, it is difficult to determine if there is an overall trend due to the lack of data points at distances beyond 500~pc. As a result, we do not currently think that estimates of the cosmic-ray ionization rate from the 3D-PDR analysis are strongly biased by the distance to the cloud where H$_3^+$ absorption is detected. For the remainder of this work we adopt the values of $\zeta({\rm H}_2)$ derived from the 3D-PDR models for all analysis and discussion points.

\section{Results and Discussion} \label{sec_discussion}

Our 3D-PDR analysis of ten sight lines provides estimates of the cosmic-ray ionization rate in individual clouds that can be discussed together with the clouds analyzed by \citet{obolentseva2024}, thus providing a larger sample with which to explore relationships between $\zeta({\rm H_2})$ and other line-of-sight parameters. Two relationships generally of interest are those between $\zeta({\rm H_2})$ and $N_{\rm H}$---displayed in Figure \ref{fig_zeta_vs_NH}---and between $\zeta({\rm H_2})$ and location---displayed in Figures \ref{fig_crirmap} and \ref{fig_crirmapzoom} as maps of the cosmic-ray ionization rate in the nearby Galaxy. Before addressing those specific topics, we begin with a general discussion on the distribution of ionization rates in comparison to previous work. 

\subsection{Comparison to Previous Work}

The most extensive surveys of the cosmic-ray ionization rate in diffuse atomic and molecular clouds rely on observations of OH$^+$ \citep{indriolo2015oxy,bacalla2019,jacob2020} and H$_3^+$ \citep{indriolo2007,indriolo2012}. As discussed in our recent work \citep{obolentseva2024,neufeld2024}, downward revisions to the interstellar gas density impact the previous ionization rate surveys, reducing the inferred values of $\zeta$ by a factor of about 9 on average. Scaling the values from those studies by this factor results in an average ionization rate of $\zeta({\rm H_2})\approx4\times10^{-17}$~s$^{-1}$, with lower and upper $1\sigma$ standard deviations on the distribution of ionization rates ranging from about $(1 -9)\times10^{-17}$~s$^{-1}$. This is in good agreement with our newly derived results, where the mean and standard deviation are $(5.3\pm2.5)\times10^{-17}$~s$^{-1}$.

As discussed above, it is difficult to fully quantify the uncertainties on our inferred values of the cosmic-ray ionization rate. Given the regional nature of inferred ionization rates (see Section \ref{sec_regions} below) it seems unlikely that the uncertainties alone are large enough to explain the full range of values. The differences in our inferred ionization rates support previous findings that the cosmic-ray ionization rate is not uniform throughout the Galaxy \citep{indriolo2012,indriolo2015oxy,redaelli2025}, and that individual clouds must be affected by local sources of cosmic-rays. The most conspicuous cases of elevated ionization rates come from observations made in close proximity to supernova remnants \citep[e.g.,][]{indriolo2010ic443,ceccarelli2011,vaupre2014,indriolo2023}.  Studies have shown that the cosmic-ray ionization rate can vary even within a dense molecular cloud complex \citep{ceccarelli2014,fontani2017,favre2018,cabedo2023,pineda2024,socci2024}, possibly tracing local acceleration of cosmic-rays by protostars \citep{gaches2018}, so variations between different diffuse cloud sight lines is quite reasonable. While we have not probed any sight lines that are knowingly adjacent to supernova remnants, it is possible that the distance between the observed cloud and the nearest site of particle acceleration has an impact on the inferred ionization rates. Such a detailed analysis is beyond the scope of this paper, and will be addressed in future work.

\subsection{Ionization rate vs gas column}

Models of cosmic-ray propagation into interstellar gas clouds that account for energy losses due to particle interactions with the ambient medium predict that the cosmic-ray ionization rate should decrease with increasing hydrogen column density \citep[e.g.,][]{padovani2009,padovani2018,silsbee2019}. This is because the low-energy cosmic rays that are most efficient at ionizing hydrogen are quickly lost in the outer layers of the cloud, either due to ionization or to exclusion by magnetic effects \citep[e.g.,][]{padovani2011,silsbee2020,bustard2021}. The exact, theoretical relationship between $\zeta({\rm H_2})$ and $N_{\rm H}$ depends on the mode of particle propagation (diffusion vs. free streaming) and the input cosmic-ray spectrum, but in general there should be a measurable difference in $\zeta({\rm H_2})$ across the range of $N_{\rm H}$ probed by our observations \citep{padovani2009,padovani2018,silsbee2019}. Note, however, that the observed hydrogen column density along a sight line may not represent the amount of material that cosmic rays must traverse. If the cloud being probed is elongated in one or more dimensions, then the measured value of $N_{\rm H}$ could over- or under-estimate the amount of material through which cosmic rays must pass, depending on the orientation of the cloud with respect to the line of sight. Keeping this uncertainty in mind, as well as the fact that our inferred ionization rates are averaged over the cloud along the line of sight and not representative of a specific depth into the cloud, the relationship between $\zeta({\rm H_2})$ and $N_{\rm H}$ determined for observations presented herein and in \citet{obolentseva2024} is presented in Figure \ref{fig_zeta_vs_NH}. The data points seem to suggest a marginal decrease in ionization rate with increasing hydrogen column, although there is significant scatter in the relationship. This scatter may be attributed to the cloud geometry, magnetic field orientation, and/or to differences in the cosmic-ray spectra impinging on the clouds. Different particle fluxes incident on two different clouds will produce different ionization rates at the same depth into the clouds, as is demonstrated by the two theoretical curves that also appear in Figure \ref{fig_zeta_vs_NH}. These curves correspond to the model $\mathscr{L}$ (low ionization rate) and model $\mathscr{H}$ (high ionization rate) cosmic-ray spectra, assuming the model of cosmic-ray propagation focused on energy losses \citep{padovani2018}. The low spectrum corresponds to the proton spectrum directly measured in the local ISM by {\it Voyager} 1 \citep{cummings2016} and {\it Voyager} 2 \citep{stone2019}, extrapolated to lower energies, while the high spectrum has an increasing particle flux toward lower energies, and was developed primarily as a potential explanation for the high cosmic-ray ionization rates being inferred from H$_3^+$ over the past two decades \citep{mccall2003,indriolo2007,indriolo2012}. It is clear that the data points favor the model $\mathscr{L}$ curve over the model $\mathscr{H}$ curve, suggesting that a {\it Voyager}-like spectrum is preferred. This is not surprising given the findings of \citet{neufeld2024} and \citet{obolentseva2024} that overestimates of the gas density led to overestimates of the cosmic-ray ionization rate by a factor of about 10 in many studies. An order of magnitude decrease to the average cosmic-ray ionization rate removes the underlying motivation for creating model $\mathscr{H}$ in the first place, and our results suggest that this model can likely be ignored henceforth, at least when considering gas that is far away from any site of particle acceleration.

\subsection{Galactic Trends}

The new cosmic-ray ionization rates derived herein allow us to expand our map of $\zeta({\rm H_2})$ in the local Galaxy. A map of the cosmic-ray ionization rate within 1.25~kpc of the Sun is presented in Figure \ref{fig_crirmap}. The distribution of target sight lines on-sky is shown in the top panel in Galactic coordinates, and the distribution of clouds within the Galactic plane as viewed from above is shown in the bottom panel, with ionization rates indicated by color. The bottom panel of our map can be overlaid on Figure 5 from \citet{edenhofer2024} to visualize the dust structures in which we are inferring ionization rates. Broadly speaking, there is no large-scale gradient or trend in $\zeta({\rm H_2})$ throughout the nearby Galaxy, although our sampling is currently very sparse.

One trend that has previously been suggested is a gradient in the cosmic-ray ionization rate with Galactocentric radius \citep{indriolo2015oxy,jacob2020}. This finding resulted from observations of OH$^+$ with {\it Herschel}/HIFI and APEX toward H~\textsc{ii} regions distributed throughout the Galaxy, at distances of up to about 13~kpc, and including targets within the central molecular zone. Regions closer to the Galactic center show higher ionization rates, while beyond a radius of about 5~kpc the values become uniform, albeit with large scatter. This is the expectation for a higher density of star formation (and thus particle acceleration sites) in the inner Galaxy, with cosmic rays slowly diffusing outward. Our current survey, limited to the region within about 1~kpc of the Sun, does not probe a wide enough range of radii within the Galaxy to place any new constraints on this relationship.

\subsection{Regional Properties} \label{sec_regions}

While our survey of cosmic-ray ionization rates in the nearby Galaxy does not reveal any large-scale trends, there are a few regions where we have multiple measurements of $\zeta({\rm H_2})$ in close proximity. We can use these closely-spaced results to better understand the cosmic-ray ionization rate on relatively small spatial scales in a few different regions.

\subsubsection{Per-Tau Shell} \label{sec_pertau}

The Per-Tau Shell is a roughly spherical structure in gas and dust bounded by the Taurus Molecular Cloud ($d\sim125$~pc) on the near side, and the Perseus Molecular Cloud ($d\sim320$~pc) on the far side \citep{bialy2021}. Sight lines toward HD~23180, HD~281159, HD~27778, {\it HD~21856}, {\it HD~22951}, {\it HD~24398}, and {\it HD~24534} probe this structure\footnote{Italicized targets were analyzed and presented in \citet{obolentseva2024}.}. The outline of the shell can be seen in the 2D gas density map presented in Figure \ref{fig_gaia_hd23180}, centered at about $(X,Z)=(200,-80)$. HD~24398 and HD~27778 are located within the shell, probing only material on the near side. HD~23180 is located within the cloud on the far side of the shell, but the sight line predominantly probes gas on the near side of the shell. HD~21856, HD~22951, HD~24534, and HD~281159 lie fully behind the shell, but HD~21856 and HD~22951 probe regions of lower density ($n_{\rm H}\lesssim20$~cm$^{-3}$) on both sides. HD~24534 predominantly probes the near side of the shell, while HD~281159 probes both sides. Zoomed-in views of this region in Galactic latitude vs. longitude and Galactic longitude vs. distance are shown in the top and bottom-left panels of Figure \ref{fig_crirmapzoom}, respectively.

Detections of H$_3^+$ toward HD~23180, HD~24398, HD~24534, and HD~27778 provide estimates of the cosmic-ray ionization rate on the near side of the shell. The latter three sight lines give consistent results of 7.6, 7.0, and 7.5 ($\times10^{-17}$~s$^{-1}$), while HD~23180 is the outlier at $3.4\times10^{-17}$~s$^{-1}$. One possible reason for this deviation may be related to the fact that the target star in this case is thought to be located {\it within} the gas density peak on the far side of the Per-Tau shell, making this a unique case where a small change to the adopted distance to the star has a large impact on the amount of observed H$_2$ attributed to the clouds on the near and far sides of the shell. Furthermore, the density reconstruction for HD~23180 and HD~281159 is less reliable than for other sight lines due to a limited number of background stars used to reconstruct the extinction map in this specific direction (G. Edenhofer 2024, private communication). If we assume that HD~23180 is only 3~pc farther away than our initially adopted distance of 301~pc---well within the $1\sigma$ limits reported by Gaia \citep{bailer-jones2021}---then our model returns a cosmic-ray ionization rate in the near side cloud that is consistent with the three nearby sight lines. Given this possibility, we conclude that the near side of the Per-Tau shell at $d\approx150$~pc \citep[average location of the density peak from][]{edenhofer2024} has an ionization rate of about $7\times10^{-17}$~s$^{-1}$. At this distance the angular separations between HD~24398, HD~24534, and HD~23180 correspond to physical separations of $\lesssim7$~pc in the plane of the sky, with HD~27778 being about 30~pc away. HD~21856 and HD~22951, the sight lines with H$_3^+$ non-detections \citep{indriolo2012}, are also about 10~pc away, and \citet{obolentseva2024} determined an upper limit of $\zeta({\rm H_2})<11.4\times10^{-17}$~s$^{-1}$ toward the latter\footnote{This result was based on Keck/NIRSPEC observations, not the IRTF/iSHELL observations presented herein.}. These results indicate that the cosmic-ray ionization rate is relatively uniform across a region tens of parsecs wide on the near side of the Per-Tau shell.

As previously discussed, using the H$_2$ and H$_3^+$ column densities toward both HD~23180 and HD~281159 allows us to derive the ionization rate in the far side cloud toward HD~281159. 
This is the only sight line where we can attribute H$_3^+$ absorption to the far cloud, for which we derive an ionization rate of $5.0\times10^{-17}$~s$^{-1}$. While the HD~22951 and HD~21856 sight lines probe both sides of the Per-Tau shell, they happen to pass through regions of lower gas density and have significantly less H$_2$ than the other sight lines, such that the H$_3^+$ absorption lines were below the detection limits of \citet{indriolo2012}. We again failed to detect H$_3^+$ toward HD~22951 in this work, but can still analyze the sight line in more detail for additional insight. If we assume that H$_2$ is equally distributed between the near and far clouds along this sight line and use the extinction map derived densities (rescaled as before using the ratio of the observed to map-integrated total hydrogen column densities) in equation (\ref{eq_crir}), we find upper limits of $\leq9.1\times10^{-17}$~s$^{-1}$ and $\leq6.1\times10^{-17}$~s$^{-1}$ in the near and far clouds, respectively. These are consistent with the mean value inferred in the near side cloud, and the result in the far side cloud. 


\subsubsection{Cep OB2}

HD~203374, HD~206165, HD~206267, HD~207198, and {\it HD~210839} are all members of the Cep OB2 association, located about 870~pc away \citep{contreras2002}. Line of sight gas density profiles for these targets are shown in Figures \ref{fig_gaia_hd203374}--\ref{fig_gaia_hd207198}, and in all cases the cloud with the most material is about 420--460~pc away from the Sun. HD~203374, HD~206165, and HD~207198 are clustered about 5$^\circ$ above the other two sight lines in Galactic longitude, and all probe the density peak at $d\sim425$~pc (see Figure \ref{fig_crirmapzoom}). For a cloud that is 425~pc away from the Sun, these three sight lines span a distance of about 23~pc on-sky. HD~206267 and HD~210839 are about 35~pc away from the other three sight lines on-sky, and probe gas that is 30--40~pc farther away from the Sun, for a total separation of about 50~pc. H$_3^+$ is detected in all five sight lines, and we infer ionization rates of 2.5--4.7$\times10^{-17}$~s$^{-1}$ across the group. There is no evidence of a difference in $\zeta({\rm H_2})$ between the three sight lines probing the slightly nearer cloud and the two sight lines probing the farther cloud. All of the values are reasonably consistent with each other, indicating a fairly uniform cosmic-ray ionization rate across this 50~pc wide region.

While HD~224151 is not part of the Cep OB2 association, the gas probed by this sight line is the closest in proximity to that region for which the 3D-PDR analysis was performed. The H$_3^+$ absorption toward HD~224151 occurs in a cloud at $d\approx366$~pc, and the on-sky separation between this sight line and the nearest Cep OB2 sight line amounts to about 90~pc. In total then, this cloud is about 120~pc away from the Cep OB2 region. The ionization rate toward HD~224151 is $11.5\times10^{-17}$~s$^{-1}$, substantially higher than toward Cep OB2, so this provides a constraint on the distances over which $\zeta({\rm H_2})$ remains uniform.

\subsubsection{Cep OB3}

The Cep OB3 association lies at a distance of about 900~pc \citep{moreno-corral1993}, and includes the stars HD~216532, HD~216898, HD~217035, and HD~217312. The gas density profiles along these sight lines are highly complex, with multiple strong and weak peaks; e.g., Figure \ref{fig_gaia_hd216532}. Similar complexity is present in the absorption profiles of CH and K~\textsc{i} \citep{pan2004}, which serve as tracers of H$_2$ and total hydrogen content, respectively. Assigning observed absorption components in velocity space to density peaks along the sight lines is non-trivial. While HD~217035 and HD~217312 have very similar inferred line of sight density profiles, the former has its strongest CH absorption at $v_{\rm LSR}\approx-14$~km~s$^{-1}$, while CH absorption in the latter is strongest at $v_{\rm LSR}\approx7$~km~s$^{-1}$. We can speculate that the H$_3^+$ absorption observed toward HD~216532 and HD~216898 arises in one of the two clouds between about 800--900~pc, but as previously described, we could not reliably perform the detailed 3D-PDR analysis for these sight lines.
As a result, we only compute $\zeta/n_{\rm H}$ for these sight lines, and are not able to constrain the cosmic-ray ionization rate in this region. Still, the values of $\zeta/n_{\rm H}=5.4\times10^{-18}$~cm$^3$~s$^{-1}$ and 
$5.0\times10^{-18}$~cm$^3$~s$^{-1}$ inferred for HD~216532 and HD~216898, respectively, suggest consistent ionization rates in the region if the gas densities in the two clouds are comparable.

\subsubsection{Gem OB1}

{\it HD~41117}, HD~42087, and HD~43384 are all in the direction of the Gem OB1 association, which is at a distance of about 1.5~kpc \citep[][and references therein]{carpenter1995}. Gas density profiles towards this region contain several peaks of similar magnitude, none of which exceed 5~cm$^{-3}$ in the \citet{edenhofer2024} map. The \citet{lallement2022} map shows similar structure at greater distances (see Figures \ref{fig_gaia_hd43384} and \ref{fig_gaia_hd42087}), and it is not possible to conclusively determine which density peaks give rise to the observed molecular absorption. A full 3D-PDR analysis of HD~41117 was presented in \citet{obolentseva2024}, but due to the complicated nature of the sight line the ionization rate was not well constrained in the three intervening clouds. Our more conservative approach of leaving the gas density unconstrained results in $\zeta({\rm H_2})/n_{\rm H}=2.7\times10^{-18}$~cm$^3$~s$^{-1}$. The same analysis for HD~43384 \citep[H$_3^+$ absorption presented in][]{albertsson2014} gives $1.7\times10^{-18}$~cm$^3$~s$^{-1}$, and again these two results are reasonably consistent. No H$_3^+$ is detected toward HD~42087. Overall, we cannot reliably infer the cosmic-ray ionization rate in this direction due to the complexities in the gas density and absorption profiles.

\subsubsection{Sco OB2}
The agglomeration of massive stars closest to the Sun is the Sco OB2 (or Sco-Cen) association, with mean distances to its three sub-groups between 118~pc and 145~pc away \citep{preibisch2008}. HD~145502, HD~147933, {\it HD~148184}, and {\it HD~149757} are all members of this association, while {\it HD~149404} is more distant but in the same region on-sky.  All of these sight lines probe gas that is about 100--150~pc away in the direction toward the inner Galaxy, and H$_3^+$ has never been detected in this region \citep{indriolo2007,indriolo2012}. If we take the upper limits on $\zeta({\rm H_2})/n_{\rm H}$ reported in Table \ref{tbl_zeta_n} and adopt the peak gas densities from the 1D profiles, then the minimum upper limit on the cosmic-ray ionization rate in this region is about $<8\times10^{-17}$~s$^{-1}$. Despite the lack of H$_3^+$ detections in this direction then, we cannot conclude that the ionization rate is substantially different from elsewhere.





\subsection{Comparison to Other Cosmic-Ray Tracers}

\subsubsection{Gamma Rays}

The ionization of H and H$_2$ by cosmic rays is thought to be dominated by protons with $E\lesssim 1$~GeV, as determined by the dominant contribution to the integral of the product between the proton spectrum and hydrogen ionization cross section \citep{padovani2009}.
Protons above $E\sim240$~MeV can produce neutral pions ($\pi^0$) via inelastic collisions with ambient gas, and these $\pi^0$ particles rapidly decay into pairs of $\gamma$-ray photons that 
can be detected with high-energy observatories (e.g., Fermi-LAT, MAGIC, HESS, VERITAS). As Fermi-LAT has continued to map the $\gamma$-ray sky building up better statistics, various groups have analyzed the $\gamma$-ray spectra of both the diffuse ISM \citep[e.g.,][]{ackermann2012} and concentrated molecular clouds to determine the flux of protons streaming through these regions. This provides complementary information about the cosmic-ray flux at higher energies compared to the lower energy flux probed by our ionization rate measurements.

Several studies have investigated the gamma-ray spectra observed from nearby molecular clouds in the Gould belt \citep{yang2014,neronov2017,aharonian2020,baghmanyan2020}, with more recent studies benefiting from improved photon statistics. Some of these molecular clouds, e.g., Perseus, Taurus, Cepheus, $\rho$ Ophiuchus, are in close proximity to the regions that we have probed with H$_3^+$ observations, and we can compare the inferred high-energy cosmic-ray flux in a cloud to the average ionization rate in the nearby region. The two regions where we have the best constraints on the cosmic-ray ionization rate are Taurus and Cepheus, where average values are $6.1\times10^{-17}$~s$^{-1}$ and $3.5\times10^{-17}$~s$^{-1}$, respectively. It is interesting then, that in a relative sense \citet{baghmanyan2020} find the opposite result, with the Taurus cloud having a proton spectrum consistent with that measured in Earth-orbit, while the Cepheus molecular cloud requires an excess of high-energy cosmic rays to match the observed $\gamma$-ray spectrum. Their findings apply to protons in the 30--1000~GeV range though, so it is possible that the excess flux does not persist to the lower energies important for ionization.

\subsubsection{Lithium Isotope Ratios}

The two stable isotopes of lithium, $^6$Li and $^7$Li, are both produced by cosmic-ray spallation of heavier nuclei (e.g., C, N, and O) and by $\alpha+\alpha$ fusion reactions, but only $^7$Li was formed during big bang nucleosynthesis, and only $^7$Li is formed today by additional processes in stars \citep[][and references therein]{knauth2017}. Measurements of the $^7$Li/$^6$Li abundance ratio can be used to place constraints on the formation mechanism for lithium, with a value of about 2 predicted for purely cosmic-ray driven formation \citep{reeves1970,meneguzzi1971}, and additional $^7$Li production mechanisms required to explain the value of 12.2 observed in the solar system \citep{lodders2003}. Observations of $^6$Li and $^7$Li have been made toward some of the same stars in the Per OB2 region for which we present H$_3^+$ observations, revealing $2\lesssim$~$^7$Li/$^6$Li$~\lesssim 5$ toward HD~23180 and HD~281159, and $^7$Li/$^6$Li~$\approx10$ toward HD~24398 and HD~24534 \citep{knauth2000,knauth2003,knauth2017}. The reduced $^7$Li/$^6$Li ratios toward HD~23180 and HD~281159 have been interpreted as evidence for enhanced cosmic-ray fluxes toward the clouds probed by these sight lines, potentially due to protons accelerated and focused by a supernova remnant \citep{hartquist1983}. Our analysis of H$_3^+$ observations suggests the opposite: cosmic-ray ionization rates toward HD~23180 and HD~281159 are lower than (or consistent with, if the distance to HD~23180 is fine tuned; see Section \ref{sec_pertau}) those toward HD~24398 and HD~24534. As discussed by \citet{knauth2017}, these conflicting results can be reconciled by considering that H$_3^+$ probes the instantaneous cosmic-ray flux experienced by a cloud at present times, while the $^7$Li/$^6$Li ratio traces the integrated history of the cosmic-ray flux experienced by the ISM; perhaps the clouds toward HD~23180 and HD~281159 experienced a much higher cosmic-ray flux in the recent past compared to the present day.


\section{Summary}

We have performed a survey searching for H$_3^+$ absorption in sight lines where the H$_2$ column density has been directly measured from observations. Twelve sight lines show new detections of H$_3^+$, while observations toward fifteen sight lines result in non-detections. This more than doubles (from 9 to 21) the sample of sight lines with direct measurements of both $N({\rm H_2})$ and $N({\rm H}_3^+)$, enabling an expanded analysis using our 3D-PDR cloud modeling method. Cosmic-ray ionization rates derived from our analysis follow the recent findings of \citet{obolentseva2024} and \citet{neufeld2024} that previous values were too high by a factor of about 9, due to overestimates of the gas density in diffuse molecular clouds. Combining our sample with that from \citet{obolentseva2024}, we find a mean ionization rate of $5.3\times10^{-17}$~s$^{-1}$ with standard deviation $2.5\times10^{-17}$~s$^{-1}$, which agrees well with previous results from H$_3^+$ and OH$^+$ if they are scaled down using the factor above.

Three dimensional gas density maps---derived from {\it Gaia} differential extinction maps---allow for the localization of H$_3^+$ absorption to specific clouds in the local Galaxy (within $\sim1$~kpc of the Sun). This, in turn, enables the construction of a map of cosmic-ray ionization rates. We do not find any large scale gradients in the local Galaxy, but this may be in part due to our sparse sampling. We do, however, find that ionization rates within specific regions tend to be somewhat uniform, and that different regions have different average ionization rates. Diffuse gas at $d\sim150$~pc near the Taurus cloud shows an average cosmic-ray ionization rate of $7.4\times10^{-17}$~s$^{-1}$, while diffuse gas at $d\sim430$~pc near the Cepheus cloud has an average value of $\zeta({\rm H_2})=3.5\times10^{-17}$~s$^{-1}$.  The maximum ionization rate in our sample, $11.5\times10^{-17}$~s$^{-1}$ toward HD~224151, is found in gas that is about 120~pc away from the Cepheus cloud region, indicating that the uniformity of the ionization rate across the $\sim50$~pc wide Cepheus region does not extend to this larger distance scale. These finding suggest that the cosmic-ray ionization rate remains uniform across local (tens of parsec) regions, but varies across the Galactic disk on $\sim100$~pc length scales. More extensive surveys of molecules that trace the cosmic-ray ionization rate are needed to increase the density of sampling throughout our map, and thus better understand how this parameter varies within the Galaxy.

The authors thank D. Welty for providing CH and K~\textsc{i} spectra toward multiple targets for use in our upper limit calculations, and are grateful to G. Edenhofer for his help with analyzing 3D dust extinction maps. A.~M.~J. would like to acknowledge the support of the Max Planck Society and SFB 1601. Visiting Astronomer at the Infrared Telescope Facility, which is operated by the University of Hawaii under contract 80HQTR19D0030 with the National Aeronautics and Space Administration. Partially based on observations collected at the European Southern Observatory under ESO programmes 088.C-0351(AB) and 194.C-0833. This research has used data, tools or materials developed as part of the EXPLORE project that has received funding from the European Union’s Horizon 2020 research and innovation programme under grant agreement No 101004214.
\software{3D-PDR \citep{bisbas2012}, Astropy \citep{astropy2013,astropy2018,astropy2022}, IRAF \citep{iraf1986,iraf1993}, Matplotlib \citep{matplotlib2007}, Scipy \citep{scipy2019arxiv}, Spextool \citep{cushing2004}, Spextool - Xtellcorr \citep{vacca2003}}

\bibliographystyle{aasjournalv7}
\bibliography{indy_master}

@article{astropy2022,
	adsurl = {https://ui.adsabs.harvard.edu/abs/2022ApJ...935..167A},
	archiveprefix = {arXiv},
	author = {{Astropy Collaboration} and {Price-Whelan}, Adrian M. and {Lim}, Pey Lian and {Earl}, Nicholas and {Starkman}, Nathaniel and {Bradley}, Larry and {Shupe}, David L. and {Patil}, Aarya A. and {Corrales}, Lia and {Brasseur}, C.~E. and {N{\"o}the}, Maximilian and {Donath}, Axel and {Tollerud}, Erik and {Morris}, Brett M. and {Ginsburg}, Adam and {Vaher}, Eero and {Weaver}, Benjamin A. and {Tocknell}, James and {Jamieson}, William and {van Kerkwijk}, Marten H. and {Robitaille}, Thomas P. and {Merry}, Bruce and {Bachetti}, Matteo and {G{\"u}nther}, H. Moritz and {Aldcroft}, Thomas L. and {Alvarado-Montes}, Jaime A. and {Archibald}, Anne M. and {B{\'o}di}, Attila and {Bapat}, Shreyas and {Barentsen}, Geert and {Baz{\'a}n}, Juanjo and {Biswas}, Manish and {Boquien}, M{\'e}d{\'e}ric and {Burke}, D.~J. and {Cara}, Daria and {Cara}, Mihai and {Conroy}, Kyle E. and {Conseil}, Simon and {Craig}, Matthew W. and {Cross}, Robert M. and {Cruz}, Kelle L. and {D'Eugenio}, Francesco and {Dencheva}, Nadia and {Devillepoix}, Hadrien A.~R. and {Dietrich}, J{\"o}rg P. and {Eigenbrot}, Arthur Davis and {Erben}, Thomas and {Ferreira}, Leonardo and {Foreman-Mackey}, Daniel and {Fox}, Ryan and {Freij}, Nabil and {Garg}, Suyog and {Geda}, Robel and {Glattly}, Lauren and {Gondhalekar}, Yash and {Gordon}, Karl D. and {Grant}, David and {Greenfield}, Perry and {Groener}, Austen M. and {Guest}, Steve and {Gurovich}, Sebastian and {Handberg}, Rasmus and {Hart}, Akeem and {Hatfield-Dodds}, Zac and {Homeier}, Derek and {Hosseinzadeh}, Griffin and {Jenness}, Tim and {Jones}, Craig K. and {Joseph}, Prajwel and {Kalmbach}, J. Bryce and {Karamehmetoglu}, Emir and {Ka{\l}uszy{\'n}ski}, Miko{\l}aj and {Kelley}, Michael S.~P. and {Kern}, Nicholas and {Kerzendorf}, Wolfgang E. and {Koch}, Eric W. and {Kulumani}, Shankar and {Lee}, Antony and {Ly}, Chun and {Ma}, Zhiyuan and {MacBride}, Conor and {Maljaars}, Jakob M. and {Muna}, Demitri and {Murphy}, N.~A. and {Norman}, Henrik and {O'Steen}, Richard and {Oman}, Kyle A. and {Pacifici}, Camilla and {Pascual}, Sergio and {Pascual-Granado}, J. and {Patil}, Rohit R. and {Perren}, Gabriel I. and {Pickering}, Timothy E. and {Rastogi}, Tanuj and {Roulston}, Benjamin R. and {Ryan}, Daniel F. and {Rykoff}, Eli S. and {Sabater}, Jose and {Sakurikar}, Parikshit and {Salgado}, Jes{\'u}s and {Sanghi}, Aniket and {Saunders}, Nicholas and {Savchenko}, Volodymyr and {Schwardt}, Ludwig and {Seifert-Eckert}, Michael and {Shih}, Albert Y. and {Jain}, Anany Shrey and {Shukla}, Gyanendra and {Sick}, Jonathan and {Simpson}, Chris and {Singanamalla}, Sudheesh and {Singer}, Leo P. and {Singhal}, Jaladh and {Sinha}, Manodeep and {Sip{\H{o}}cz}, Brigitta M. and {Spitler}, Lee R. and {Stansby}, David and {Streicher}, Ole and {{\v{S}}umak}, Jani and {Swinbank}, John D. and {Taranu}, Dan S. and {Tewary}, Nikita and {Tremblay}, Grant R. and {de Val-Borro}, Miguel and {Van Kooten}, Samuel J. and {Vasovi{\'c}}, Zlatan and {Verma}, Shresth and {de Miranda Cardoso}, Jos{\'e} Vin{\'\i}cius and {Williams}, Peter K.~G. and {Wilson}, Tom J. and {Winkel}, Benjamin and {Wood-Vasey}, W.~M. and {Xue}, Rui and {Yoachim}, Peter and {Zhang}, Chen and {Zonca}, Andrea and {Astropy Project Contributors}},
	doi = {10.3847/1538-4357/ac7c74},
	eid = {167},
	eprint = {2206.14220},
	journal = {\apj},
	month = aug,
	number = {2},
	pages = {167},
	primaryclass = {astro-ph.IM},
	title = {{The Astropy Project: Sustaining and Growing a Community-oriented Open-source Project and the Latest Major Release (v5.0) of the Core Package}},
	volume = {935},
	year = 2022}

@article{astropy2018,
	adsurl = {https://ui.adsabs.harvard.edu/abs/2018AJ....156..123A},
	archiveprefix = {arXiv},
	author = {{Astropy Collaboration} and {Price-Whelan}, A.~M. and {Sip{\H{o}}cz}, B.~M. and {G{\"u}nther}, H.~M. and {Lim}, P.~L. and {Crawford}, S.~M. and {Conseil}, S. and {Shupe}, D.~L. and {Craig}, M.~W. and {Dencheva}, N. and {Ginsburg}, A. and {VanderPlas}, J.~T. and {Bradley}, L.~D. and {P{\'e}rez-Su{\'a}rez}, D. and {de Val-Borro}, M. and {Aldcroft}, T.~L. and {Cruz}, K.~L. and {Robitaille}, T.~P. and {Tollerud}, E.~J. and {Ardelean}, C. and {Babej}, T. and {Bach}, Y.~P. and {Bachetti}, M. and {Bakanov}, A.~V. and {Bamford}, S.~P. and {Barentsen}, G. and {Barmby}, P. and {Baumbach}, A. and {Berry}, K.~L. and {Biscani}, F. and {Boquien}, M. and {Bostroem}, K.~A. and {Bouma}, L.~G. and {Brammer}, G.~B. and {Bray}, E.~M. and {Breytenbach}, H. and {Buddelmeijer}, H. and {Burke}, D.~J. and {Calderone}, G. and {Cano Rodr{\'\i}guez}, J.~L. and {Cara}, M. and {Cardoso}, J.~V.~M. and {Cheedella}, S. and {Copin}, Y. and {Corrales}, L. and {Crichton}, D. and {D'Avella}, D. and {Deil}, C. and {Depagne}, {\'E}. and {Dietrich}, J.~P. and {Donath}, A. and {Droettboom}, M. and {Earl}, N. and {Erben}, T. and {Fabbro}, S. and {Ferreira}, L.~A. and {Finethy}, T. and {Fox}, R.~T. and {Garrison}, L.~H. and {Gibbons}, S.~L.~J. and {Goldstein}, D.~A. and {Gommers}, R. and {Greco}, J.~P. and {Greenfield}, P. and {Groener}, A.~M. and {Grollier}, F. and {Hagen}, A. and {Hirst}, P. and {Homeier}, D. and {Horton}, A.~J. and {Hosseinzadeh}, G. and {Hu}, L. and {Hunkeler}, J.~S. and {Ivezi{\'c}}, {\v{Z}}. and {Jain}, A. and {Jenness}, T. and {Kanarek}, G. and {Kendrew}, S. and {Kern}, N.~S. and {Kerzendorf}, W.~E. and {Khvalko}, A. and {King}, J. and {Kirkby}, D. and {Kulkarni}, A.~M. and {Kumar}, A. and {Lee}, A. and {Lenz}, D. and {Littlefair}, S.~P. and {Ma}, Z. and {Macleod}, D.~M. and {Mastropietro}, M. and {McCully}, C. and {Montagnac}, S. and {Morris}, B.~M. and {Mueller}, M. and {Mumford}, S.~J. and {Muna}, D. and {Murphy}, N.~A. and {Nelson}, S. and {Nguyen}, G.~H. and {Ninan}, J.~P. and {N{\"o}the}, M. and {Ogaz}, S. and {Oh}, S. and {Parejko}, J.~K. and {Parley}, N. and {Pascual}, S. and {Patil}, R. and {Patil}, A.~A. and {Plunkett}, A.~L. and {Prochaska}, J.~X. and {Rastogi}, T. and {Reddy Janga}, V. and {Sabater}, J. and {Sakurikar}, P. and {Seifert}, M. and {Sherbert}, L.~E. and {Sherwood-Taylor}, H. and {Shih}, A.~Y. and {Sick}, J. and {Silbiger}, M.~T. and {Singanamalla}, S. and {Singer}, L.~P. and {Sladen}, P.~H. and {Sooley}, K.~A. and {Sornarajah}, S. and {Streicher}, O. and {Teuben}, P. and {Thomas}, S.~W. and {Tremblay}, G.~R. and {Turner}, J.~E.~H. and {Terr{\'o}n}, V. and {van Kerkwijk}, M.~H. and {de la Vega}, A. and {Watkins}, L.~L. and {Weaver}, B.~A. and {Whitmore}, J.~B. and {Woillez}, J. and {Zabalza}, V. and {Astropy Contributors}},
	doi = {10.3847/1538-3881/aabc4f},
	eid = {123},
	eprint = {1801.02634},
	journal = {\aj},
	month = sep,
	number = {3},
	pages = {123},
	primaryclass = {astro-ph.IM},
	title = {{The Astropy Project: Building an Open-science Project and Status of the v2.0 Core Package}},
	volume = {156},
	year = 2018}

@misc{eso_uves,
	author = {{European Southern Observatory (ESO)}},
	copyright = {Data Access Policy for ESO Data held in the ESO Science Archive Facility},
	doi = {10.18727/ARCHIVE/50},
	language = {en},
	publisher = {European Southern Observatory (ESO)},
	title = {UVES reduced data obtained by standard ESO pipeline processing},
	url = {https://doi.eso.org/10.18727/archive/50},
	year = {2020}}

@article{zhang2025,
	adsurl = {https://ui.adsabs.harvard.edu/abs/2025Sci...387.1209Z},
	archiveprefix = {arXiv},
	author = {{Zhang}, Xiangyu and {Green}, Gregory M.},
	doi = {10.1126/science.ado9787},
	eprint = {2407.14594},
	journal = {Science},
	month = mar,
	number = {6739},
	pages = {1209-1214},
	primaryclass = {astro-ph.GA},
	title = {{Three-dimensional maps of the interstellar dust extinction curve within the Milky Way galaxy}},
	volume = {387},
	year = 2025}

@article{knauth2000,
	adsurl = {https://ui.adsabs.harvard.edu/abs/2000Natur.405..656K},
	author = {{Knauth}, D.~C. and {Federman}, S.~R. and {Lambert}, David L. and {Crane}, P.},
	doi = {10.1038/35015028},
	journal = {\nat},
	month = jun,
	number = {6787},
	pages = {656-658},
	title = {{Newly synthesized lithium in the interstellar medium}},
	volume = {405},
	year = 2000}

@article{bisbas2012,
	adsurl = {https://ui.adsabs.harvard.edu/abs/2012MNRAS.427.2100B},
	archiveprefix = {arXiv},
	author = {{Bisbas}, T.~G. and {Bell}, T.~A. and {Viti}, S. and {Yates}, J. and {Barlow}, M.~J.},
	doi = {10.1111/j.1365-2966.2012.22077.x},
	eprint = {1209.1091},
	journal = {\mnras},
	month = dec,
	number = {3},
	pages = {2100-2118},
	primaryclass = {astro-ph.SR},
	title = {{3D-PDR: a new three-dimensional astrochemistry code for treating photodissociation regions}},
	volume = {427},
	year = 2012}

@article{shull2019,
	adsurl = {https://ui.adsabs.harvard.edu/abs/2019ApJ...882..180S},
	archiveprefix = {arXiv},
	author = {{Shull}, J. Michael and {Danforth}, Charles W.},
	doi = {10.3847/1538-4357/ab357d},
	eid = {180},
	eprint = {1907.13148},
	journal = {\apj},
	month = sep,
	number = {2},
	pages = {180},
	primaryclass = {astro-ph.SR},
	title = {{Distances to Galactic OB Stars: Photometry versus Parallax}},
	volume = {882},
	year = 2019}

@article{cabedo2023,
	adsurl = {https://ui.adsabs.harvard.edu/abs/2023A&A...669A..90C},
	archiveprefix = {arXiv},
	author = {{Cabedo}, Victoria and {Maury}, Ana{\"e}lle and {Girart}, Josep Miquel and {Padovani}, Marco and {Hennebelle}, Patrick and {Houde}, Martin and {Zhang}, Qizhou},
	doi = {10.1051/0004-6361/202243813},
	eid = {A90},
	eprint = {2204.10043},
	journal = {\aap},
	month = jan,
	pages = {A90},
	primaryclass = {astro-ph.GA},
	title = {{Magnetically regulated collapse in the B335 protostar?. II. Observational constraints on gas ionization and magnetic field coupling}},
	volume = {669},
	year = 2023}

@article{favre2018,
	adsurl = {https://ui.adsabs.harvard.edu/abs/2018ApJ...859..136F},
	archiveprefix = {arXiv},
	author = {{Favre}, C. and {Ceccarelli}, C. and {L{\'o}pez-Sepulcre}, A. and {Fontani}, F. and {Neri}, R. and {Manigand}, S. and {Kama}, M. and {Caselli}, P. and {Jaber Al-Edhari}, A. and {Kahane}, C. and {Alves}, F. and {Balucani}, N. and {Bianchi}, E. and {Caux}, E. and {Codella}, C. and {Dulieu}, F. and {Pineda}, J.~E. and {Sims}, I.~R. and {Theul{\'e}}, P.},
	doi = {10.3847/1538-4357/aabfd4},
	eid = {136},
	eprint = {1804.07825},
	journal = {\apj},
	month = jun,
	number = {2},
	pages = {136},
	primaryclass = {astro-ph.GA},
	title = {{SOLIS IV. Hydrocarbons in the OMC-2 FIR4 Region, a Probe of Energetic Particle Irradiation of the Region}},
	volume = {859},
	year = 2018}

@article{fontani2017,
	adsurl = {https://ui.adsabs.harvard.edu/abs/2017A&A...605A..57F},
	archiveprefix = {arXiv},
	author = {{Fontani}, F. and {Ceccarelli}, C. and {Favre}, C. and {Caselli}, P. and {Neri}, R. and {Sims}, I.~R. and {Kahane}, C. and {Alves}, F.~O. and {Balucani}, N. and {Bianchi}, E. and {Caux}, E. and {Jaber Al-Edhari}, A. and {Lopez-Sepulcre}, A. and {Pineda}, J.~E. and {Bachiller}, R. and {Bizzocchi}, L. and {Bottinelli}, S. and {Chacon-Tanarro}, A. and {Choudhury}, R. and {Codella}, C. and {Coutens}, A. and {Dulieu}, F. and {Feng}, S. and {Rimola}, A. and {Hily-Blant}, P. and {Holdship}, J. and {Jimenez-Serra}, I. and {Laas}, J. and {Lefloch}, B. and {Oya}, Y. and {Podio}, L. and {Pon}, A. and {Punanova}, A. and {Quenard}, D. and {Sakai}, N. and {Spezzano}, S. and {Taquet}, V. and {Testi}, L. and {Theul{\'e}}, P. and {Ugliengo}, P. and {Vastel}, C. and {Vasyunin}, A.~I. and {Viti}, S. and {Yamamoto}, S. and {Wiesenfeld}, L.},
	doi = {10.1051/0004-6361/201730527},
	eid = {A57},
	eprint = {1707.01384},
	journal = {\aap},
	month = sep,
	pages = {A57},
	primaryclass = {astro-ph.GA},
	title = {{Seeds of Life in Space (SOLIS). I. Carbon-chain growth in the Solar-type protocluster OMC2-FIR4}},
	volume = {605},
	year = 2017}

@article{ceccarelli2014,
	adsurl = {https://ui.adsabs.harvard.edu/abs/2014ApJ...790L...1C},
	author = {{Ceccarelli}, C. and {Dominik}, C. and {L{\'o}pez-Sepulcre}, A. and {Kama}, M. and {Padovani}, M. and {Caux}, E. and {Caselli}, P.},
	doi = {10.1088/2041-8205/790/1/L1},
	eid = {L1},
	journal = {\apjl},
	month = jul,
	number = {1},
	pages = {L1},
	title = {{Herschel Finds Evidence for Stellar Wind Particles in a Protostellar Envelope: Is This What Happened to the Young Sun?}},
	volume = {790},
	year = 2014}

@article{gordon2023,
	adsurl = {https://ui.adsabs.harvard.edu/abs/2023ApJ...950...86G},
	archiveprefix = {arXiv},
	author = {{Gordon}, Karl D. and {Clayton}, Geoffrey C. and {Decleir}, Marjorie and {Fitzpatrick}, E.~L. and {Massa}, Derck and {Misselt}, Karl A. and {Tollerud}, Erik J.},
	doi = {10.3847/1538-4357/accb59},
	eid = {86},
	eprint = {2304.01991},
	journal = {\apj},
	month = jun,
	number = {2},
	pages = {86},
	primaryclass = {astro-ph.GA},
	title = {{One Relation for All Wavelengths: The Far-ultraviolet to Mid-infrared Milky Way Spectroscopic R(V)-dependent Dust Extinction Relationship}},
	volume = {950},
	year = 2023}

@article{jacob2020,
	adsurl = {https://ui.adsabs.harvard.edu/abs/2020A&A...643A..91J},
	archiveprefix = {arXiv},
	author = {{Jacob}, Arshia M. and {Menten}, Karl M. and {Wyrowski}, Friedrich and {Winkel}, Benjamin and {Neufeld}, David A.},
	doi = {10.1051/0004-6361/202039197},
	eid = {A91},
	eprint = {2010.02258},
	journal = {\aap},
	month = nov,
	pages = {A91},
	primaryclass = {astro-ph.GA},
	title = {{Extending the view of ArH$^{+}$ chemistry in diffuse clouds}},
	volume = {643},
	year = 2020}

@article{indriolo2023,
	adsurl = {https://ui.adsabs.harvard.edu/abs/2023ApJ...950...64I},
	archiveprefix = {arXiv},
	author = {{Indriolo}, Nick},
	doi = {10.3847/1538-4357/acc6c4},
	eid = {64},
	eprint = {2303.13689},
	journal = {\apj},
	month = jun,
	number = {1},
	pages = {64},
	primaryclass = {astro-ph.HE},
	title = {{Absorption-line Observations of \{\{H\}\}\_\{3\}(+) and CO in Sight Lines Toward the Vela and W28 Supernova Remnants}},
	volume = {950},
	year = 2023}

@article{cummings2016,
	adsurl = {https://ui.adsabs.harvard.edu/abs/2016ApJ...831...18C},
	author = {{Cummings}, A.~C. and {Stone}, E.~C. and {Heikkila}, B.~C. and {Lal}, N. and {Webber}, W.~R. and {J{\'o}hannesson}, G. and {Moskalenko}, I.~V. and {Orlando}, E. and {Porter}, T.~A.},
	doi = {10.3847/0004-637X/831/1/18},
	eid = {18},
	journal = {\apj},
	month = nov,
	number = {1},
	pages = {18},
	title = {{Galactic Cosmic Rays in the Local Interstellar Medium: Voyager 1 Observations and Model Results}},
	volume = {831},
	year = 2016}

@article{bustard2021,
	adsurl = {https://ui.adsabs.harvard.edu/abs/2021ApJ...913..106B},
	archiveprefix = {arXiv},
	author = {{Bustard}, Chad and {Zweibel}, Ellen G.},
	doi = {10.3847/1538-4357/abf64c},
	eid = {106},
	eprint = {2012.06585},
	journal = {\apj},
	month = jun,
	number = {2},
	pages = {106},
	primaryclass = {astro-ph.HE},
	title = {{Cosmic-Ray Transport, Energy Loss, and Influence in the Multiphase Interstellar Medium}},
	volume = {913},
	year = 2021}

@article{blaauw1959,
	adsurl = {https://ui.adsabs.harvard.edu/abs/1959ApJ...130...69B},
	author = {{Blaauw}, A. and {Hiltner}, W.~A. and {Johnson}, H.~L.},
	doi = {10.1086/146697},
	journal = {\apj},
	month = jul,
	pages = {69},
	title = {{Photoelectric Photometry of the Association III Cephei.}},
	volume = {130},
	year = 1959}

@dataset{gaiaDR32022,
	adsurl = {https://ui.adsabs.harvard.edu/abs/2022yCat.1355....0G},
	author = {{Gaia Collaboration}},
	doi = {10.26093/cds/vizier.1355},
	eid = {I/355},
	howpublished = {VizieR On-line Data Catalog: I/355. Originally published in: doi:10.1051/0004-63},
	month = may,
	title = {{VizieR Online Data Catalog: Gaia DR3 Part 1. Main source (Gaia Collaboration, 2022)}},
	year = 2022}

@article{anders2019,
	adsurl = {https://ui.adsabs.harvard.edu/abs/2019A&A...628A..94A},
	archiveprefix = {arXiv},
	author = {{Anders}, F. and {Khalatyan}, A. and {Chiappini}, C. and {Queiroz}, A.~B. and {Santiago}, B.~X. and {Jordi}, C. and {Girardi}, L. and {Brown}, A.~G.~A. and {Matijevi{\v{c}}}, G. and {Monari}, G. and {Cantat-Gaudin}, T. and {Weiler}, M. and {Khan}, S. and {Miglio}, A. and {Carrillo}, I. and {Romero-G{\'o}mez}, M. and {Minchev}, I. and {de Jong}, R.~S. and {Antoja}, T. and {Ramos}, P. and {Steinmetz}, M. and {Enke}, H.},
	doi = {10.1051/0004-6361/201935765},
	eid = {A94},
	eprint = {1904.11302},
	journal = {\aap},
	month = aug,
	pages = {A94},
	primaryclass = {astro-ph.GA},
	title = {{Photo-astrometric distances, extinctions, and astrophysical parameters for Gaia DR2 stars brighter than G = 18}},
	volume = {628},
	year = 2019}

@article{bialy2019,
	adsurl = {https://ui.adsabs.harvard.edu/abs/2019ApJ...885..109B},
	archiveprefix = {arXiv},
	author = {{Bialy}, Shmuel and {Neufeld}, David and {Wolfire}, Mark and {Sternberg}, Amiel and {Burkhart}, Blakesley},
	doi = {10.3847/1538-4357/ab487b},
	eid = {109},
	eprint = {1909.12305},
	journal = {\apj},
	month = nov,
	number = {2},
	pages = {109},
	primaryclass = {astro-ph.GA},
	title = {{Chemical Abundances in a Turbulent Medium-H$_{2}$, OH$^{+}$, H$_{2}$O$^{+}$, ArH$^{+}$}},
	volume = {885},
	year = 2019}

@article{gaches2018,
	adsurl = {https://ui.adsabs.harvard.edu/abs/2018ApJ...861...87G},
	archiveprefix = {arXiv},
	author = {{Gaches}, Brandt A.~L. and {Offner}, Stella S.~R.},
	doi = {10.3847/1538-4357/aac94d},
	eid = {87},
	eprint = {1805.03215},
	journal = {\apj},
	month = jul,
	number = {2},
	pages = {87},
	primaryclass = {astro-ph.GA},
	title = {{Exploration of Cosmic-ray Acceleration in Protostellar Accretion Shocks and a Model for Ionization Rates in Embedded Protoclusters}},
	volume = {861},
	year = 2018}

@article{baghmanyan2020,
	adsurl = {https://ui.adsabs.harvard.edu/abs/2020ApJ...901L...4B},
	archiveprefix = {arXiv},
	author = {{Baghmanyan}, Vardan and {Peron}, Giada and {Casanova}, Sabrina and {Aharonian}, Felix and {Zanin}, Roberta},
	doi = {10.3847/2041-8213/abb5f8},
	eid = {L4},
	eprint = {2009.08893},
	journal = {\apjl},
	month = sep,
	number = {1},
	pages = {L4},
	primaryclass = {astro-ph.HE},
	title = {{Evidence of Cosmic-Ray Excess from Local Giant Molecular Clouds}},
	volume = {901},
	year = 2020}

@article{zhang2023,
	adsurl = {https://ui.adsabs.harvard.edu/abs/2023MNRAS.524.1855Z},
	archiveprefix = {arXiv},
	author = {{Zhang}, Xiangyu and {Green}, Gregory M. and {Rix}, Hans-Walter},
	doi = {10.1093/mnras/stad1941},
	eprint = {2303.03420},
	journal = {\mnras},
	month = sep,
	number = {2},
	pages = {1855-1884},
	primaryclass = {astro-ph.SR},
	title = {{Parameters of 220 million stars from Gaia BP/RP spectra}},
	volume = {524},
	year = 2023}

@incollection{preibisch2008,
	adsurl = {https://ui.adsabs.harvard.edu/abs/2008hsf2.book..235P},
	author = {{Preibisch}, T. and {Mamajek}, E.},
	booktitle = {Handbook of Star Forming Regions, Volume II},
	doi = {10.48550/arXiv.0809.0407},
	editor = {{Reipurth}, B.},
	pages = {235},
	title = {{The Nearest OB Association: Scorpius-Centaurus (Sco OB2)}},
	volume = {5},
	year = 2008}

@article{neufeld2024,
	adsurl = {https://ui.adsabs.harvard.edu/abs/2024ApJ...973..143N},
	archiveprefix = {arXiv},
	author = {{Neufeld}, David A. and {Welty}, Daniel E. and {Ivlev}, Alexei V. and {Caselli}, Paola and {Edenhofer}, Gordian and {Indriolo}, Nick and {Obolentseva}, Marta and {Silsbee}, Kedron and {Sonnentrucker}, Paule and {Wolfire}, Mark G.},
	doi = {10.3847/1538-4357/ad7264},
	eid = {143},
	eprint = {2408.11108},
	journal = {\apj},
	month = oct,
	number = {2},
	pages = {143},
	primaryclass = {astro-ph.GA},
	title = {{The Densities in Diffuse and Translucent Molecular Clouds: Estimates from Observations of C$_{2}$ and from Three-dimensional Extinction Maps}},
	volume = {973},
	year = 2024}

@article{fan2024,
	adsurl = {https://ui.adsabs.harvard.edu/abs/2024A&A...681A...6F},
	archiveprefix = {arXiv},
	author = {{Fan}, Haoyu and {Rocha}, Carlos M.~R. and {Cordiner}, Martin and {Linnartz}, Harold and {Cox}, Nick L.~J. and {Farhang}, Amin and {Smoker}, Jonathan and {Roueff}, Evelyne and {Ehrenfreund}, Pascale and {Salama}, Farid and {Foing}, Bernard H. and {Lallement}, Rosine and {MacIsaac}, Heather and {Kulik}, Klay and {Sarre}, Peter and {van Loon}, Jacco Th. and {Cami}, Jan},
	doi = {10.1051/0004-6361/202243910},
	eid = {A6},
	eprint = {2310.03259},
	journal = {\aap},
	month = jan,
	pages = {A6},
	primaryclass = {astro-ph.GA},
	title = {{The EDIBLES survey. VII. A survey of C$_{2}$ and C$_{3}$ in interstellar clouds}},
	volume = {681},
	year = 2024}

@article{moreno-corral1993,
	adsurl = {https://ui.adsabs.harvard.edu/abs/1993A&A...273..619M},
	author = {{Moreno-Corral}, M.~A. and {C.}, Chavarria-K. and {de La Ra}, E. and {Wagner}, S.},
	journal = {\aap},
	month = jun,
	pages = {619-632},
	title = {{H-alpha interferometric, optical and near IR photometric studies of star forming regions. I. The Cepheus B/Sh2-155/Cepheus OB3 association complex.}},
	volume = {273},
	year = 1993}

@article{contreras2002,
	adsurl = {https://ui.adsabs.harvard.edu/abs/2002AJ....124.1585C},
	author = {{Contreras}, Maria E. and {Sicilia-Aguilar}, Aurora and {Muzerolle}, James and {Calvet}, Nuria and {Berlind}, Perry and {Hartmann}, Lee},
	doi = {10.1086/341825},
	journal = {\aj},
	month = sep,
	number = {3},
	pages = {1585-1592},
	title = {{A Study of Intermediate-Mass Stars in Trumpler 37}},
	volume = {124},
	year = 2002}

@article{socci2024,
	adsurl = {https://ui.adsabs.harvard.edu/abs/2024A&A...687A..70S},
	archiveprefix = {arXiv},
	author = {{Socci}, A. and {Sabatini}, G. and {Padovani}, M. and {Bovino}, S. and {Hacar}, A.},
	doi = {10.1051/0004-6361/202449960},
	eid = {A70},
	eprint = {2404.15754},
	journal = {\aap},
	month = jul,
	pages = {A70},
	primaryclass = {astro-ph.GA},
	title = {{Parsec-scale cosmic-ray ionisation rate in Orion}},
	volume = {687},
	year = 2024}

@article{pineda2024,
	adsurl = {https://ui.adsabs.harvard.edu/abs/2024A&A...686A.162P},
	archiveprefix = {arXiv},
	author = {{Pineda}, Jaime E. and {Sipil{\"a}}, Olli and {Segura-Cox}, Dominique M. and {Valdivia-Mena}, Maria Teresa and {Neri}, Roberto and {Kuffmeier}, Michael and {Ivlev}, Alexei V. and {Offner}, Stella S.~R. and {Maureira}, Maria Jose and {Caselli}, Paola and {Spezzano}, Silvia and {Cunningham}, Nichol and {Schmiedeke}, Anika and {Chen}, Mike},
	doi = {10.1051/0004-6361/202347997},
	eid = {A162},
	eprint = {2402.16202},
	journal = {\aap},
	month = jun,
	pages = {A162},
	primaryclass = {astro-ph.GA},
	title = {{Probing the physics of star formation (ProPStar). I. First resolved maps of the electron fraction and cosmic-ray ionization rate in NGC 1333}},
	volume = {686},
	year = 2024}

@article{knauth2017,
	adsurl = {https://ui.adsabs.harvard.edu/abs/2017ApJ...835L..16K},
	archiveprefix = {arXiv},
	author = {{Knauth}, D.~C. and {Taylor}, C.~J. and {Ritchey}, A.~M. and {Federman}, S.~R. and {Lambert}, D.~L.},
	doi = {10.3847/2041-8213/aa527a},
	eid = {L16},
	eprint = {1704.00357},
	journal = {\apjl},
	month = jan,
	number = {1},
	pages = {L16},
	primaryclass = {astro-ph.GA},
	title = {{Parsec-scale Variations in the $^{7}$Li I/$^{6}$Li I Isotope Ratio Toward IC 348 and the Perseus OB 2 Association}},
	volume = {835},
	year = 2017}

@article{lallement2022,
	adsurl = {https://ui.adsabs.harvard.edu/abs/2022A&A...661A.147L},
	archiveprefix = {arXiv},
	author = {{Lallement}, R. and {Vergely}, J.~L. and {Babusiaux}, C. and {Cox}, N.~L.~J.},
	doi = {10.1051/0004-6361/202142846},
	eid = {A147},
	eprint = {2203.01627},
	journal = {\aap},
	month = may,
	pages = {A147},
	primaryclass = {astro-ph.GA},
	title = {{Updated Gaia-2MASS 3D maps of Galactic interstellar dust}},
	volume = {661},
	year = 2022}

@article{edenhofer2024,
	adsurl = {https://ui.adsabs.harvard.edu/abs/2024A&A...685A..82E},
	archiveprefix = {arXiv},
	author = {{Edenhofer}, Gordian and {Zucker}, Catherine and {Frank}, Philipp and {Saydjari}, Andrew K. and {Speagle}, Joshua S. and {Finkbeiner}, Douglas and {En{\ss}lin}, Torsten A.},
	doi = {10.1051/0004-6361/202347628},
	eid = {A82},
	eprint = {2308.01295},
	journal = {\aap},
	month = may,
	pages = {A82},
	primaryclass = {astro-ph.GA},
	title = {{A parsec-scale Galactic 3D dust map out to 1.25 kpc from the Sun}},
	volume = {685},
	year = 2024}

@article{bialy2021,
	adsurl = {https://ui.adsabs.harvard.edu/abs/2021ApJ...919L...5B},
	archiveprefix = {arXiv},
	author = {{Bialy}, Shmuel and {Zucker}, Catherine and {Goodman}, Alyssa and {Foley}, Michael M. and {Alves}, Jo{\~a}o and {Semenov}, Vadim A. and {Benjamin}, Robert and {Leike}, Reimar and {En{\ss}lin}, Torsten},
	doi = {10.3847/2041-8213/ac1f95},
	eid = {L5},
	eprint = {2109.09763},
	journal = {\apjl},
	month = sep,
	number = {1},
	pages = {L5},
	primaryclass = {astro-ph.GA},
	title = {{The Per-Tau Shell: A Giant Star-forming Spherical Shell Revealed by 3D Dust Observations}},
	volume = {919},
	year = 2021}

@article{vandeputte2023,
	adsurl = {https://ui.adsabs.harvard.edu/abs/2023ApJ...944...33V},
	archiveprefix = {arXiv},
	author = {{Van De Putte}, Dries and {Cartledge}, Stefan I.~B. and {Gordon}, Karl D. and {Clayton}, Geoffrey C. and {Roman-Duval}, Julia},
	doi = {10.3847/1538-4357/ac9902},
	eid = {33},
	eprint = {2210.04972},
	journal = {\apj},
	month = feb,
	number = {1},
	pages = {33},
	primaryclass = {astro-ph.GA},
	title = {{Far-ultraviolet Dust Extinction and Molecular Hydrogen in the Diffuse Milky Way Interstellar Medium}},
	volume = {944},
	year = 2023}

@article{jenkins2009,
	adsurl = {https://ui.adsabs.harvard.edu/abs/2009ApJ...700.1299J},
	archiveprefix = {arXiv},
	author = {{Jenkins}, Edward B.},
	doi = {10.1088/0004-637X/700/2/1299},
	eprint = {0905.3173},
	journal = {\apj},
	month = aug,
	number = {2},
	pages = {1299-1348},
	primaryclass = {astro-ph.GA},
	title = {{A Unified Representation of Gas-Phase Element Depletions in the Interstellar Medium}},
	volume = {700},
	year = 2009}

@article{royer2024,
	adsurl = {https://ui.adsabs.harvard.edu/abs/2024A&A...681A.107R},
	archiveprefix = {arXiv},
	author = {{Royer}, P. and {Merle}, T. and {Dsilva}, K. and {Sekaran}, S. and {Van Winckel}, H. and {Fr{\'e}mat}, Y. and {Van der Swaelmen}, M. and {Gebruers}, S. and {Tkachenko}, A. and {Laverick}, M. and {Dirickx}, M. and {Raskin}, G. and {Hensberge}, H. and {Abdul-Masih}, M. and {Acke}, B. and {Alonso}, M.~L. and {Bandhu Mahato}, S. and {Beck}, P.~G. and {Behara}, N. and {Bloemen}, S. and {Buysschaert}, B. and {Cox}, N. and {Debosscher}, J. and {De Cat}, P. and {Degroote}, P. and {De Nutte}, R. and {De Smedt}, K. and {de Vries}, B. and {Dumortier}, L. and {Escorza}, A. and {Exter}, K. and {Goriely}, S. and {Gorlova}, N. and {Hillen}, M. and {Homan}, W. and {Jorissen}, A. and {Kamath}, D. and {Karjalainen}, M. and {Karjalainen}, R. and {Lampens}, P. and {Lobel}, A. and {Lombaert}, R. and {Marcos-Arenal}, P. and {Menu}, J. and {Merges}, F. and {Moravveji}, E. and {Nemeth}, P. and {Neyskens}, P. and {Ostensen}, R. and {P{\'a}pics}, P.~I. and {Perez}, J. and {Prins}, S. and {Royer}, S. and {Samadi-Ghadim}, A. and {Sana}, H. and {Sans Fuentes}, A. and {Scaringi}, S. and {Schmid}, V. and {Siess}, L. and {Siopis}, C. and {Smolders}, K. and {S{\'o}dor}, {\'A}. and {Thoul}, A. and {Triana}, S. and {Vandenbussche}, B. and {Van de Sande}, M. and {Van De Steene}, G. and {Van Eck}, S. and {van Hoof}, P.~A.~M. and {Van Marle}, A.~J. and {Van Reeth}, T. and {Vermeylen}, L. and {Volpi}, D. and {Vos}, J. and {Waelkens}, C.},
	doi = {10.1051/0004-6361/202346847},
	eid = {A107},
	eprint = {2311.02705},
	journal = {\aap},
	month = jan,
	pages = {A107},
	primaryclass = {astro-ph.SR},
	title = {{MELCHIORS. The Mercator Library of High Resolution Stellar Spectroscopy}},
	volume = {681},
	year = 2024}

@article{obolentseva2024,
	adsurl = {https://ui.adsabs.harvard.edu/abs/2024ApJ...973..142O},
	archiveprefix = {arXiv},
	author = {{Obolentseva}, M. and {Ivlev}, A.~V. and {Silsbee}, K. and {Neufeld}, D.~A. and {Caselli}, P. and {Edenhofer}, G. and {Indriolo}, N. and {Bisbas}, T.~G. and {Lomeli}, D.},
	doi = {10.3847/1538-4357/ad71ce},
	eid = {142},
	eprint = {2408.11511},
	journal = {\apj},
	month = oct,
	number = {2},
	pages = {142},
	primaryclass = {astro-ph.GA},
	title = {{Reevaluation of the Cosmic-Ray Ionization Rate in Diffuse Clouds}},
	volume = {973},
	year = 2024}

@article{vacca2003,
	adsurl = {https://ui.adsabs.harvard.edu/abs/2003PASP..115..389V},
	archiveprefix = {arXiv},
	author = {{Vacca}, William D. and {Cushing}, Michael C. and {Rayner}, John T.},
	doi = {10.1086/346193},
	eprint = {astro-ph/0211255},
	journal = {\pasp},
	month = mar,
	number = {805},
	pages = {389-409},
	primaryclass = {astro-ph},
	title = {{A Method of Correcting Near-Infrared Spectra for Telluric Absorption}},
	volume = {115},
	year = 2003}

@article{padovani2011,
	adsurl = {https://ui.adsabs.harvard.edu/abs/2011A&A...530A.109P},
	archiveprefix = {arXiv},
	author = {{Padovani}, M. and {Galli}, D.},
	doi = {10.1051/0004-6361/201116853},
	eid = {A109},
	eprint = {1104.5445},
	journal = {\aap},
	month = jun,
	pages = {A109},
	primaryclass = {astro-ph.SR},
	title = {{Effects of magnetic fields on the cosmic-ray ionization of molecular cloud cores}},
	volume = {530},
	year = 2011}

@article{silsbee2020,
	adsurl = {https://ui.adsabs.harvard.edu/abs/2020ApJ...902L..25S},
	archiveprefix = {arXiv},
	author = {{Silsbee}, Kedron and {Ivlev}, Alexei V.},
	doi = {10.3847/2041-8213/abbc20},
	eid = {L25},
	eprint = {2009.14208},
	journal = {\apjl},
	month = oct,
	number = {1},
	pages = {L25},
	primaryclass = {astro-ph.HE},
	title = {{Exclusion of Cosmic Rays from Molecular Clouds by Self-generated Electric Fields}},
	volume = {902},
	year = 2020}

@article{silsbee2019,
	adsurl = {https://ui.adsabs.harvard.edu/abs/2019ApJ...879...14S},
	archiveprefix = {arXiv},
	author = {{Silsbee}, Kedron and {Ivlev}, Alexei V.},
	doi = {10.3847/1538-4357/ab22b4},
	eid = {14},
	eprint = {1904.01588},
	journal = {\apj},
	month = jul,
	number = {1},
	pages = {14},
	primaryclass = {astro-ph.HE},
	title = {{Diffusive versus Free-streaming Cosmic-Ray Transport in Molecular Clouds}},
	volume = {879},
	year = 2019}

@article{padovani2018,
	adsurl = {https://ui.adsabs.harvard.edu/abs/2018A&A...614A.111P},
	archiveprefix = {arXiv},
	author = {{Padovani}, Marco and {Ivlev}, Alexei V. and {Galli}, Daniele and {Caselli}, Paola},
	doi = {10.1051/0004-6361/201732202},
	eid = {A111},
	eprint = {1803.09348},
	journal = {\aap},
	month = jun,
	pages = {A111},
	primaryclass = {astro-ph.HE},
	title = {{Cosmic-ray ionisation in circumstellar discs}},
	volume = {614},
	year = 2018}

@article{bailer-jones2021,
	adsurl = {https://ui.adsabs.harvard.edu/abs/2021AJ....161..147B},
	archiveprefix = {arXiv},
	author = {{Bailer-Jones}, C.~A.~L. and {Rybizki}, J. and {Fouesneau}, M. and {Demleitner}, M. and {Andrae}, R.},
	doi = {10.3847/1538-3881/abd806},
	eid = {147},
	eprint = {2012.05220},
	journal = {\aj},
	month = mar,
	number = {3},
	pages = {147},
	primaryclass = {astro-ph.SR},
	title = {{Estimating Distances from Parallaxes. V. Geometric and Photogeometric Distances to 1.47 Billion Stars in Gaia Early Data Release 3}},
	volume = {161},
	year = 2021}

@article{vaupre2014,
	adsurl = {https://ui.adsabs.harvard.edu/abs/2014A&A...568A..50V},
	archiveprefix = {arXiv},
	author = {{Vaupr{\'e}}, S. and {Hily-Blant}, P. and {Ceccarelli}, C. and {Dubus}, G. and {Gabici}, S. and {Montmerle}, T.},
	doi = {10.1051/0004-6361/201424036},
	eid = {A50},
	eprint = {1407.0205},
	journal = {\aap},
	month = aug,
	pages = {A50},
	primaryclass = {astro-ph.GA},
	title = {{Cosmic ray induced ionisation of a molecular cloud shocked by the W28 supernova remnant}},
	volume = {568},
	year = 2014}

@article{shull2021,
	adsurl = {https://ui.adsabs.harvard.edu/abs/2021ApJ...911...55S},
	archiveprefix = {arXiv},
	author = {{Shull}, J. Michael and {Danforth}, Charles W. and {Anderson}, Katherine L.},
	doi = {10.3847/1538-4357/abe707},
	eid = {55},
	eprint = {2102.11301},
	journal = {\apj},
	month = apr,
	number = {1},
	pages = {55},
	primaryclass = {astro-ph.GA},
	title = {{A Far Ultraviolet Spectroscopic Explorer Survey of Interstellar Molecular Hydrogen in the Galactic Disk}},
	volume = {911},
	year = 2021}

@article{rayner2022,
	adsurl = {https://ui.adsabs.harvard.edu/abs/2022PASP..134a5002R},
	author = {{Rayner}, John and {Tokunaga}, Alan and {Jaffe}, Daniel and {Bond}, Timothy and {Bonnet}, Morgan and {Ching}, Gregory and {Connelley}, Michael and {Cushing}, Michael and {Kokubun}, Daniel and {Lockhart}, Charles and {Vacca}, William and {Warmbier}, Eric},
	doi = {10.1088/1538-3873/ac3cb4},
	eid = {015002},
	journal = {\pasp},
	month = jan,
	number = {1031},
	pages = {015002},
	title = {{iSHELL: a 1-5 micron R = 80,000 Immersion Grating Spectrograph for the NASA Infrared Telescope Facility}},
	volume = {134},
	year = 2022}

@article{ceccarelli2011,
	adsurl = {https://ui.adsabs.harvard.edu/abs/2011ApJ...740L...4C},
	archiveprefix = {arXiv},
	author = {{Ceccarelli}, C. and {Hily-Blant}, P. and {Montmerle}, T. and {Dubus}, G. and {Gallant}, Y. and {Fiasson}, A.},
	doi = {10.1088/2041-8205/740/1/L4},
	eid = {L4},
	eprint = {1108.3600},
	journal = {\apjl},
	month = oct,
	number = {1},
	pages = {L4},
	primaryclass = {astro-ph.GA},
	title = {{Supernova-enhanced Cosmic-Ray Ionization and Induced Chemistry in a Molecular Cloud of W51C}},
	volume = {740},
	year = 2011}

@article{stone2019,
	adsurl = {https://ui.adsabs.harvard.edu/abs/2019NatAs...3.1013S},
	author = {{Stone}, Edward C. and {Cummings}, Alan C. and {Heikkila}, Bryant C. and {Lal}, Nand},
	doi = {10.1038/s41550-019-0928-3},
	journal = {Nature Astronomy},
	month = nov,
	pages = {1013-1018},
	title = {{Cosmic ray measurements from Voyager 2 as it crossed into interstellar space}},
	volume = {3},
	year = 2019}

@article{padovani2020,
	adsurl = {https://ui.adsabs.harvard.edu/abs/2020SSRv..216...29P},
	archiveprefix = {arXiv},
	author = {{Padovani}, Marco and {Ivlev}, Alexei V. and {Galli}, Daniele and {Offner}, Stella S.~R. and {Indriolo}, Nick and {Rodgers-Lee}, Donna and {Marcowith}, Alexandre and {Girichidis}, Philipp and {Bykov}, Andrei M. and {Kruijssen}, J.~M. Diederik},
	doi = {10.1007/s11214-020-00654-1},
	eid = {29},
	eprint = {2002.10282},
	journal = {\ssr},
	month = mar,
	number = {2},
	pages = {29},
	primaryclass = {astro-ph.GA},
	title = {{Impact of Low-Energy Cosmic Rays on Star Formation}},
	volume = {216},
	year = 2020}

@article{aharonian2020,
	adsurl = {https://ui.adsabs.harvard.edu/abs/2020PhRvD.101h3018A},
	author = {{Aharonian}, Felix and {Peron}, Giada and {Yang}, Ruizhi and {Casanova}, Sabrina and {Zanin}, Roberta},
	doi = {10.1103/PhysRevD.101.083018},
	eid = {083018},
	journal = {\prd},
	month = apr,
	number = {8},
	pages = {083018},
	title = {{Probing the sea of galactic cosmic rays with Fermi-LAT}},
	volume = {101},
	year = 2020}

@article{scipy2019arxiv,
	adsurl = {https://ui.adsabs.harvard.edu/abs/2019arXiv190710121V},
	archiveprefix = {arXiv},
	author = {{Virtanen}, Pauli and {Gommers}, Ralf and {Oliphant}, Travis E. and {Haberland}, Matt and {Reddy}, Tyler and {Cournapeau}, David and {Burovski}, Evgeni and {Peterson}, Pearu and {Weckesser}, Warren and {Bright}, Jonathan and {van der Walt}, St{\'e}fan J. and {Brett}, Matthew and {Wilson}, Joshua and {Jarrod Millman}, K. and {Mayorov}, Nikolay and {Nelson}, Andrew R.~J. and {Jones}, Eric and {Kern}, Robert and {Larson}, Eric and {Carey}, CJ and {Polat}, {\.I}lhan and {Feng}, Yu and {Moore}, Eric W. and {Vand erPlas}, Jake and {Laxalde}, Denis and {Perktold}, Josef and {Cimrman}, Robert and {Henriksen}, Ian and {Quintero}, E.~A. and {Harris}, Charles R and {Archibald}, Anne M. and {Ribeiro}, Ant{\^o}nio H. and {Pedregosa}, Fabian and {van Mulbregt}, Paul and {Contributors}, SciPy 1. 0},
	eid = {arXiv:1907.10121},
	eprint = {1907.10121},
	journal = {arXiv e-prints},
	month = {Jul},
	pages = {arXiv:1907.10121},
	primaryclass = {cs.MS},
	title = {{SciPy 1.0--Fundamental Algorithms for Scientific Computing in Python}},
	year = {2019}}

@article{matplotlib2007,
	adsurl = {https://ui.adsabs.harvard.edu/abs/2007CSE.....9...90H},
	author = {{Hunter}, John D.},
	doi = {10.1109/MCSE.2007.55},
	journal = {Computing in Science and Engineering},
	month = {May},
	number = {3},
	pages = {90-95},
	title = {{Matplotlib: A 2D Graphics Environment}},
	volume = {9},
	year = {2007}}

@article{astropy2013,
	adsurl = {https://ui.adsabs.harvard.edu/abs/2013A&A...558A..33A},
	archiveprefix = {arXiv},
	author = {{Astropy Collaboration} and {Robitaille}, Thomas P. and {Tollerud}, Erik J. and {Greenfield}, Perry and {Droettboom}, Michael and {Bray}, Erik and {Aldcroft}, Tom and {Davis}, Matt and {Ginsburg}, Adam and {Price-Whelan}, Adrian M. and {Kerzendorf}, Wolfgang E. and {Conley}, Alexander and {Crighton}, Neil and {Barbary}, Kyle and {Muna}, Demitri and {Ferguson}, Henry and {Grollier}, Fr{\'e}d{\'e}ric and {Parikh}, Madhura M. and {Nair}, Prasanth H. and {Unther}, Hans M. and {Deil}, Christoph and {Woillez}, Julien and {Conseil}, Simon and {Kramer}, Roban and {Turner}, James E.~H. and {Singer}, Leo and {Fox}, Ryan and {Weaver}, Benjamin A. and {Zabalza}, Victor and {Edwards}, Zachary I. and {Azalee Bostroem}, K. and {Burke}, D.~J. and {Casey}, Andrew R. and {Crawford}, Steven M. and {Dencheva}, Nadia and {Ely}, Justin and {Jenness}, Tim and {Labrie}, Kathleen and {Lim}, Pey Lian and {Pierfederici}, Francesco and {Pontzen}, Andrew and {Ptak}, Andy and {Refsdal}, Brian and {Servillat}, Mathieu and {Streicher}, Ole},
	doi = {10.1051/0004-6361/201322068},
	eid = {A33},
	eprint = {1307.6212},
	journal = {\aap},
	month = {Oct},
	pages = {A33},
	primaryclass = {astro-ph.IM},
	title = {{Astropy: A community Python package for astronomy}},
	volume = {558},
	year = {2013}}

@inbook{iraf1993,
	adsurl = {https://ui.adsabs.harvard.edu/abs/1993ASPC...52..173T},
	author = {{Tody}, Doug},
	booktitle = {Astronomical Data Analysis Software and Systems II},
	editor = {{Hanisch}, R.~J. and {Brissenden}, R.~J.~V. and {Barnes}, J.},
	pages = {173},
	series = {Astronomical Society of the Pacific Conference Series},
	title = {{IRAF in the Nineties}},
	volume = {52},
	year = {1993}}

@inbook{iraf1986,
	adsurl = {https://ui.adsabs.harvard.edu/abs/1986SPIE..627..733T},
	author = {{Tody}, Doug},
	booktitle = {\procspie},
	doi = {10.1117/12.968154},
	editor = {{Crawford}, David L.},
	pages = {733},
	series = {Society of Photo-Optical Instrumentation Engineers (SPIE) Conference Series},
	title = {{The IRAF Data Reduction and Analysis System}},
	volume = {627},
	year = {1986}}

@article{neronov2017,
	adsurl = {https://ui.adsabs.harvard.edu/abs/2017A&A...606A..22N},
	archiveprefix = {arXiv},
	author = {{Neronov}, Andrii and {Malyshev}, Denys and {Semikoz}, Dmitri V.},
	doi = {10.1051/0004-6361/201731149},
	eid = {A22},
	eprint = {1705.02200},
	journal = {\aap},
	month = {Sep},
	pages = {A22},
	primaryclass = {astro-ph.HE},
	title = {{Cosmic-ray spectrum in the local Galaxy}},
	volume = {606},
	year = {2017}}

@article{ackermann2012,
	adsurl = {https://ui.adsabs.harvard.edu/abs/2012ApJ...750....3A},
	archiveprefix = {arXiv},
	author = {{Ackermann}, M. and {Ajello}, M. and {Atwood}, W.~B. and {Baldini}, L. and {Ballet}, J. and {Barbiellini}, G. and {Bastieri}, D. and {Bechtol}, K. and {Bellazzini}, R. and {Berenji}, B. and {Bland ford}, R.~D. and {Bloom}, E.~D. and {Bonamente}, E. and {Borgland }, A.~W. and {Brandt}, T.~J. and {Bregeon}, J. and {Brigida}, M. and {Bruel}, P. and {Buehler}, R. and {Buson}, S. and {Caliandro}, G.~A. and {Cameron}, R.~A. and {Caraveo}, P.~A. and {Cavazzuti}, E. and {Cecchi}, C. and {Charles}, E. and {Chekhtman}, A. and {Chiang}, J. and {Ciprini}, S. and {Claus}, R. and {Cohen-Tanugi}, J. and {Conrad}, J. and {Cutini}, S. and {de Angelis}, A. and {de Palma}, F. and {Dermer}, C.~D. and {Digel}, S.~W. and {Silva}, E. do Couto e. and {Drell}, P.~S. and {Drlica-Wagner}, A. and {Falletti}, L. and {Favuzzi}, C. and {Fegan}, S.~J. and {Ferrara}, E.~C. and {Focke}, W.~B. and {Fortin}, P. and {Fukazawa}, Y. and {Funk}, S. and {Fusco}, P. and {Gaggero}, D. and {Gargano}, F. and {Germani}, S. and {Giglietto}, N. and {Giordano}, F. and {Giroletti}, M. and {Glanzman}, T. and {Godfrey}, G. and {Grove}, J.~E. and {Guiriec}, S. and {Gustafsson}, M. and {Hadasch}, D. and {Hanabata}, Y. and {Harding}, A.~K. and {Hayashida}, M. and {Hays}, E. and {Horan}, D. and {Hou}, X. and {Hughes}, R.~E. and {J{\'o}hannesson}, G. and {Johnson}, A.~S. and {Johnson}, R.~P. and {Kamae}, T. and {Katagiri}, H. and {Kataoka}, J. and {Kn{\"o}dlseder}, J. and {Kuss}, M. and {Lande}, J. and {Latronico}, L. and {Lee}, S. -H. and {Lemoine-Goumard}, M. and {Longo}, F. and {Loparco}, F. and {Lott}, B. and {Lovellette}, M.~N. and {Lubrano}, P. and {Mazziotta}, M.~N. and {McEnery}, J.~E. and {Michelson}, P.~F. and {Mitthumsiri}, W. and {Mizuno}, T. and {Monte}, C. and {Monzani}, M.~E. and {Morselli}, A. and {Moskalenko}, I.~V. and {Murgia}, S. and {Naumann-Godo}, M. and {Norris}, J.~P. and {Nuss}, E. and {Ohsugi}, T. and {Okumura}, A. and {Omodei}, N. and {Orlando}, E. and {Ormes}, J.~F. and {Paneque}, D. and {Panetta}, J.~H. and {Parent}, D. and {Pesce-Rollins}, M. and {Pierbattista}, M. and {Piron}, F. and {Pivato}, G. and {Porter}, T.~A. and {Rain{\`o}}, S. and {Rando}, R. and {Razzano}, M. and {Razzaque}, S. and {Reimer}, A. and {Reimer}, O. and {Sadrozinski}, H.~F. -W. and {Sgr{\`o}}, C. and {Siskind}, E.~J. and {Spandre}, G. and {Spinelli}, P. and {Strong}, A.~W. and {Suson}, D.~J. and {Takahashi}, H. and {Tanaka}, T. and {Thayer}, J.~G. and {Thayer}, J.~B. and {Thompson}, D.~J. and {Tibaldo}, L. and {Tinivella}, M. and {Torres}, D.~F. and {Tosti}, G. and {Troja}, E. and {Usher}, T.~L. and {Vandenbroucke}, J. and {Vasileiou}, V. and {Vianello}, G. and {Vitale}, V. and {Waite}, A.~P. and {Wang}, P. and {Winer}, B.~L. and {Wood}, K.~S. and {Wood}, M. and {Yang}, Z. and {Ziegler}, M. and {Zimmer}, S.},
	doi = {10.1088/0004-637X/750/1/3},
	eid = {3},
	eprint = {1202.4039},
	journal = {\apj},
	month = {May},
	number = {1},
	pages = {3},
	primaryclass = {astro-ph.HE},
	title = {{Fermi-LAT Observations of the Diffuse {\ensuremath{\gamma}}-Ray Emission: Implications for Cosmic Rays and the Interstellar Medium}},
	volume = {750},
	year = {2012}}

@article{bacalla2019,
	adsurl = {http://adsabs.harvard.edu/abs/2019A%26A...622A..31B},
	archiveprefix = {arXiv},
	author = {{Bacalla}, X.~L. and {Linnartz}, H. and {Cox}, N.~L.~J. and {Cami}, J. and {Roueff}, E. and {Smoker}, J.~V. and {Farhang}, A. and {Bouwman}, J. and {Zhao}, D.},
	doi = {10.1051/0004-6361/201833039},
	eid = {A31},
	eprint = {1811.08662},
	journal = {\aap},
	month = feb,
	pages = {A31},
	title = {{The EDIBLES survey. IV. Cosmic ray ionization rates in diffuse clouds from near-ultraviolet observations of interstellar OH$^{+}$}},
	volume = 622,
	year = 2019}

@article{neufeld2017,
	adsurl = {http://adsabs.harvard.edu/abs/2017ApJ...845..163N},
	archiveprefix = {arXiv},
	author = {{Neufeld}, D.~A. and {Wolfire}, M.~G.},
	doi = {10.3847/1538-4357/aa6d68},
	eid = {163},
	eprint = {1704.03877},
	journal = {\apj},
	month = aug,
	pages = {163},
	title = {{The Cosmic-Ray Ionization Rate in the Galactic Disk, as Determined from Observations of Molecular Ions}},
	volume = 845,
	year = 2017}

@article{indriolo2015oxy,
	adsurl = {http://adsabs.harvard.edu/abs/2015ApJ...800...40I},
	author = {{Indriolo}, N. and {Neufeld}, D.~A. and {Gerin}, M. and {Schilke}, P. and {Benz}, A.~O. and {Winkel}, B. and {Menten}, K.~M. and {Chambers}, E.~T. and {Black}, J.~H. and {Bruderer}, S. and {Falgarone}, E. and {Godard}, B. and {Goicoechea}, J.~R. and {Gupta}, H. and {Lis}, D.~C. and {Ossenkopf}, V. and {Persson}, C.~M. and {Sonnentrucker}, P. and {van der Tak}, F.~F.~S. and {van Dishoeck}, E.~F. and {Wolfire}, M.~G. and {Wyrowski}, F.},
	doi = {10.1088/0004-637X/800/1/40},
	eid = {40},
	journal = {\apj},
	month = feb,
	pages = {40},
	title = {{Herschel Survey of Galactic OH$^{+}$, H$_{2}$O$^{+}$, and H$_{3}$O$^{+}$: Probing the Molecular Hydrogen Fraction and Cosmic-Ray Ionization Rate}},
	volume = 800,
	year = 2015}

@article{cushing2004,
	adsurl = {http://adsabs.harvard.edu/abs/2004PASP..116..362C},
	author = {{Cushing}, M.~C. and {Vacca}, W.~D. and {Rayner}, J.~T.},
	doi = {10.1086/382907},
	journal = {\pasp},
	month = apr,
	pages = {362-376},
	title = {{Spextool: A Spectral Extraction Package for SpeX, a 0.8-5.5 Micron Cross-Dispersed Spectrograph}},
	volume = 116,
	year = 2004}

@article{yang2014,
	adsurl = {http://adsabs.harvard.edu/abs/2014A%26A...566A.142Y},
	archiveprefix = {arXiv},
	author = {{Yang}, R.-z. and {de O{\~n}a Wilhelmi}, E. and {Aharonian}, F.},
	doi = {10.1051/0004-6361/201321044},
	eid = {A142},
	eprint = {1303.7323},
	journal = {\aap},
	month = jun,
	pages = {A142},
	primaryclass = {astro-ph.HE},
	title = {{Probing cosmic rays in nearby giant molecular clouds with the Fermi Large Area Telescope}},
	volume = 566,
	year = 2014}

@article{albertsson2014,
	adsurl = {http://adsabs.harvard.edu/abs/2014ApJ...787...44A},
	archiveprefix = {arXiv},
	author = {{Albertsson}, T. and {Indriolo}, N. and {Kreckel}, H. and {Semenov}, D. and {Crabtree}, K.~N. and {Henning}, T.},
	doi = {10.1088/0004-637X/787/1/44},
	eid = {44},
	eprint = {1403.6533},
	journal = {\apj},
	month = may,
	pages = {44},
	title = {{First Time-dependent Study of H$_{2}$ and H\_3\^{}+ Ortho-Para Chemistry in the Diffuse Interstellar Medium: Observations Meet Theoretical Predictions}},
	volume = 787,
	year = 2014}

@article{vanleeuwen2007,
	adsurl = {http://cdsads.u-strasbg.fr/abs/2007A%26A...474..653V},
	archiveprefix = {arXiv},
	author = {{van Leeuwen}, F.},
	doi = {10.1051/0004-6361:20078357},
	eprint = {0708.1752},
	journal = {\aap},
	month = nov,
	pages = {653-664},
	title = {{Validation of the new Hipparcos reduction}},
	volume = {474},
	year = {2007}}

@article{carpenter1995,
	adsurl = {http://adsabs.harvard.edu/abs/1995ApJ...445..246C},
	author = {{Carpenter}, J.~M. and {Snell}, R.~L. and {Schloerb}, F.~P.},
	doi = {10.1086/175692},
	journal = {\apj},
	month = may,
	pages = {246-268},
	title = {{Anatomy of the Gemini OB1 molecular cloud complex}},
	volume = {445},
	year = {1995}}

@article{crane1995,
	adsurl = {http://adsabs.harvard.edu/abs/1995ApJS...99..107C},
	author = {{Crane}, P. and {Lambert}, D.~L. and {Sheffer}, Y.},
	doi = {10.1086/192180},
	journal = {\apjs},
	month = jul,
	pages = {107-+},
	title = {{A Very High Resolution Survey of Interstellar CH and CH +}},
	volume = {99},
	year = {1995}}

@article{dalgarno2006,
	adsurl = {http://adsabs.harvard.edu/abs/2006PNAS..10312269D},
	author = {{Dalgarno}, A.},
	doi = {10.1073/pnas.0602117103},
	journal = {Proc. Nat. Acad. Sci.},
	month = aug,
	pages = {12269-12273},
	title = {{Interstellar Chemistry Special Feature: The galactic cosmic ray ionization rate}},
	volume = {103},
	year = {2006}}

@article{draine1978,
	adsurl = {http://adsabs.harvard.edu/abs/1978ApJS...36..595D},
	author = {{Draine}, B.~T.},
	doi = {10.1086/190513},
	journal = {\apjs},
	month = apr,
	pages = {595-619},
	title = {{Photoelectric heating of interstellar gas}},
	volume = {36},
	year = {1978}}

@article{goto2002,
	adsurl = {http://adsabs.harvard.edu/abs/2002PASJ...54..951G},
	author = {{Goto}, M. and {McCall}, B.~J. and {Geballe}, T.~R. and {Usuda}, T. and {Kobayashi}, N. and {Terada}, H. and {Oka}, T.},
	eprint = {arXiv:astro-ph/0212159},
	journal = {\pasj},
	month = dec,
	pages = {951-961},
	title = {{Absorption Line Survey of H3+ toward the Galactic Center Sources I. GCS 3-2 and GC IRS3}},
	volume = {54},
	year = {2002}}

@article{hartquist1983,
	adsurl = {http://adsabs.harvard.edu/abs/1983ApJ...266..271H},
	author = {{Hartquist}, T.~W. and {Morfill}, G.~E.},
	doi = {10.1086/160776},
	journal = {\apj},
	month = mar,
	pages = {271-275},
	title = {{Evidence for the stochastic acceleration of cosmic rays in supernova remnants}},
	volume = {266},
	year = {1983}}

@article{herbst1973,
	adsurl = {http://adsabs.harvard.edu/abs/1973ApJ...185..505H},
	author = {{Herbst}, E. and {Klemperer}, W.},
	doi = {10.1086/152436},
	journal = {\apj},
	month = oct,
	pages = {505-534},
	title = {{The Formation and Depletion of Molecules in Dense Interstellar Clouds}},
	volume = {185},
	year = {1973}}

@article{indriolo2010ic443,
	adsurl = {http://adsabs.harvard.edu/abs/2010ApJ...724.1357I},
	archiveprefix = {arXiv},
	author = {{Indriolo}, N. and {Blake}, G.~A. and {Goto}, M. and {Usuda}, T. and {Oka}, T. and {Geballe}, T.~R. and {Fields}, B.~D. and {McCall}, B.~J.},
	doi = {10.1088/0004-637X/724/2/1357},
	eprint = {1010.3252},
	journal = {\apj},
	month = dec,
	pages = {1357-1365},
	primaryclass = {astro-ph.HE},
	title = {{Investigating the Cosmic-ray Ionization Rate Near the Supernova Remnant IC 443 through H$^{+}_{3}$ Observations}},
	volume = {724},
	year = {2010}}

@article{indriolo2007,
	adsurl = {http://adsabs.harvard.edu/abs/2007ApJ...671.1736I},
	archiveprefix = {arXiv},
	author = {{Indriolo}, N. and {Geballe}, T.~R. and {Oka}, T. and {McCall}, B.~J.},
	doi = {10.1086/523036},
	eprint = {0709.1114},
	journal = {\apj},
	month = dec,
	pages = {1736-1747},
	title = {{H$^{+}$$_{3}$ in Diffuse Interstellar Clouds: A Tracer for the Cosmic-Ray Ionization Rate}},
	volume = {671},
	year = {2007}}

@article{indriolo2012,
	adsurl = {http://adsabs.harvard.edu/abs/2012ApJ...745...91I},
	archiveprefix = {arXiv},
	author = {{Indriolo}, N. and {McCall}, B.~J.},
	doi = {10.1088/0004-637X/745/1/91},
	eid = {91},
	eprint = {1111.6936},
	journal = {\apj},
	month = jan,
	pages = {91},
	primaryclass = {astro-ph.GA},
	title = {{Investigating the Cosmic-Ray Ionization Rate in the Galactic Diffuse Interstellar Medium through Observations of H$^{+}_{3}$}},
	volume = {745},
	year = {2012}}

@article{kaufl2004,
	adsurl = {http://adsabs.harvard.edu/abs/2004SPIE.5492.1218K},
	author = {{K\"{a}ufl}, {H.-U.} and {Ballester}, P. and {Biereichel}, P. and {Delabre}, B. and {Donaldson}, R. and {Dorn}, R. and {Fedrigo}, E. and {Finger}, G. and {Fischer}, G. and {Franza}, F. and {Gojak}, D. and {Huster}, G. and {Jung}, Y. and {Lizon}, {J.-L.} and {Mehrgan}, L. and {Meyer}, M. and {Moorwood}, A. and {Pirard}, {J.-F.} and {Paufique}, J. and {Pozna}, E. and {Siebenmorgen}, R. and {Silber}, A. and {Stegmeier}, J. and {Wegerer}, S.},
	booktitle = {Society of Photo-Optical Instrumentation Engineers (SPIE) Conference Series},
	doi = {10.1117/12.551480},
	editor = {{A.~F.~M.~Moorwood \& M.~Iye}},
	journal = {\procspie},
	month = sep,
	pages = {1218-1227},
	series = {Society of Photo-Optical Instrumentation Engineers (SPIE) Conference Series},
	title = {{CRIRES: a high-resolution infrared spectrograph for ESO's VLT}},
	volume = {5492},
	year = {2004}}

@article{knauth2003,
	adsurl = {http://adsabs.harvard.edu/abs/2003ApJ...586..268K},
	author = {{Knauth}, D.~C. and {Federman}, S.~R. and {Lambert}, D.~L.},
	doi = {10.1086/346264},
	eprint = {arXiv:astro-ph/0212233},
	journal = {\apj},
	month = mar,
	pages = {268-285},
	title = {{An Ultra-high-Resolution Survey of the Interstellar ^{7}Li/^{6}Li Isotope Ratio in the Solar Neighborhood}},
	volume = {586},
	year = {2003}}

@article{larsson1983a,
	adsurl = {http://adsabs.harvard.edu/abs/1983JChPh..79.2270L},
	author = {{Larsson}, M. and {Siegbahn}, P.~E.~M.},
	doi = {10.1063/1.446077},
	journal = {J. Chem. Phys.},
	month = sep,
	pages = {2270-2277},
	title = {{A theoretical study of the radiative lifetime of the CH A ^{2}{$\Delta$} state}},
	volume = {79},
	year = {1983}}

@article{lodders2003,
	adsurl = {http://adsabs.harvard.edu/abs/2003ApJ...591.1220L},
	author = {{Lodders}, K.},
	doi = {10.1086/375492},
	journal = {\apj},
	month = jul,
	pages = {1220-1247},
	title = {{Solar System Abundances and Condensation Temperatures of the Elements}},
	volume = {591},
	year = {2003}}

@phdthesis{mccall2001,
	adsurl = {http://adsabs.harvard.edu/abs/2001PhDT.........2M},
	author = {{McCall}, B.~J.},
	school = {The University of Chicago},
	title = {{Spectroscopy of trihydrogen(+) in laboratory and astrophysical plasmas}},
	year = {2001}}

@article{mccall2004,
	adsurl = {http://adsabs.harvard.edu/abs/2004PhRvA..70e2716M},
	author = {{McCall}, B.~J. and {Huneycutt}, A.~J. and {Saykally}, R.~J. and {Djuric}, N. and {Dunn}, G.~H. and {Semaniak}, J. and {Novotny}, O. and {Al-Khalili}, A. and {Ehlerding}, A. and {Hellberg}, F. and {Kalhori}, S. and {Neau}, A. and {Thomas}, R.~D. and {Paal}, A. and {{\"O}sterdahl}, F. and {Larsson}, M.},
	doi = {10.1103/PhysRevA.70.052716},
	journal = {Phys. Rev. A},
	month = nov,
	number = {5},
	pages = {052716-+},
	title = {{Dissociative recombination of rotationally cold H^{+}_{3}}},
	volume = {70},
	year = {2004}}

@article{mccall2003,
	adsurl = {http://adsabs.harvard.edu/abs/2003Natur.422..500M},
	author = {{McCall}, B.~J. and {Huneycutt}, A.~J. and {Saykally}, R.~J. and {Geballe}, T.~R. and {Djuric}, N. and {Dunn}, G.~H. and {Semaniak}, J. and {Novotny}, O. and {Al-Khalili}, A. and {Ehlerding}, A. and {Hellberg}, F. and {Kalhori}, S. and {Neau}, A. and {Thomas}, R. and {{\"O}sterdahl}, F. and {Larsson}, M.},
	doi = {10.1038/nature01498},
	eprint = {arXiv:astro-ph/0302106},
	journal = {\nat},
	month = apr,
	pages = {500-502},
	title = {{An enhanced cosmic-ray flux towards {$\zeta$} Persei inferred from a laboratory study of the H_{3}^{+}-e^{-} recombination rate}},
	volume = {422},
	year = {2003}}

@article{meneguzzi1971,
	adsurl = {http://adsabs.harvard.edu/abs/1971A%26A....15..337M},
	author = {{Meneguzzi}, M. and {Audouze}, J. and {Reeves}, H.},
	journal = {\aap},
	pages = {337-359},
	title = {{The production of the elements Li, Be, B by galactic cosmic rays in space and its relation with stellar observations.}},
	volume = {15},
	year = {1971}}

@article{padovani2009,
	adsurl = {http://adsabs.harvard.edu/abs/2009A%26A...501..619P},
	archiveprefix = {arXiv},
	author = {{Padovani}, M. and {Galli}, D. and {Glassgold}, A.~E.},
	doi = {10.1051/0004-6361/200911794},
	eprint = {0904.4149},
	journal = {\aap},
	month = jul,
	pages = {619-631},
	title = {{Cosmic-ray ionization of molecular clouds}},
	volume = {501},
	year = {2009}}

@article{pan2004,
	adsurl = {http://adsabs.harvard.edu/abs/2004ApJS..151..313P},
	author = {{Pan}, K. and {Federman}, S.~R. and {Cunha}, K. and {Smith}, V.~V. and {Welty}, D.~E.},
	doi = {10.1086/381805},
	eprint = {arXiv:astro-ph/0312095},
	journal = {\apjs},
	month = apr,
	pages = {313-343},
	title = {{Cloud Structure and Physical Conditions in Star-forming Regions from Optical Observations. I. Data and Component Structure}},
	volume = {151},
	year = {2004}}

@article{pan2005,
	adsurl = {http://adsabs.harvard.edu/abs/2005ApJ...633..986P},
	author = {{Pan}, K. and {Federman}, S.~R. and {Sheffer}, Y. and {Andersson}, {B.-G.}},
	doi = {10.1086/491466},
	eprint = {arXiv:astro-ph/0507491},
	journal = {\apj},
	month = nov,
	pages = {986-1004},
	title = {{Cloud Structure and Physical Conditions in Star-forming Regions from Optical Observations. II. Analysis}},
	volume = {633},
	year = {2005}}

@article{rachford2009,
	adsurl = {http://adsabs.harvard.edu/abs/2009ApJS..180..125R},
	author = {{Rachford}, B.~L. and {Snow}, T.~P. and {Destree}, J.~D. and {Ross}, T.~L. and {Ferlet}, R. and {Friedman}, S.~D. and {Gry}, C. and {Jenkins}, E.~B. and {Morton}, D.~C. and {Savage}, B.~D. and {Shull}, J.~M. and {Sonnentrucker}, P. and {Tumlinson}, J. and {Vidal-Madjar}, A. and {Welty}, D.~E. and {York}, D.~G.},
	doi = {10.1088/0067-0049/180/1/125},
	journal = {\apjs},
	month = jan,
	pages = {125-137},
	title = {{Molecular Hydrogen in the Far Ultraviolet Spectroscopic Explorer Translucent Lines of Sight: The Full Sample}},
	volume = {180},
	year = {2009}}

@article{rachford2002,
	adsurl = {http://adsabs.harvard.edu/abs/2002ApJ...577..221R},
	author = {{Rachford}, B.~L. and {Snow}, T.~P. and {Tumlinson}, J. and {Shull}, J.~M. and {Blair}, W.~P. and {Ferlet}, R. and {Friedman}, S.~D. and {Gry}, C. and {Jenkins}, E.~B. and {Morton}, D.~C. and {Savage}, B.~D. and {Sonnentrucker}, P. and {Vidal-Madjar}, A. and {Welty}, D.~E. and {York}, D.~G.},
	doi = {10.1086/342146},
	eprint = {arXiv:astro-ph/0205415},
	journal = {\apj},
	month = sep,
	pages = {221-244},
	title = {{A Far Ultraviolet Spectroscopic Explorer Survey of Interstellar Molecular Hydrogen in Translucent Clouds}},
	volume = {577},
	year = {2002}}

@article{reeves1970,
	adsurl = {http://adsabs.harvard.edu/abs/1970Natur.226..727R},
	author = {{Reeves}, H. and {Fowler}, W.~A. and {Hoyle}, F.},
	doi = {10.1038/226727a0},
	journal = {\nat},
	month = may,
	pages = {727-729},
	title = {{Galactic Cosmic Ray Origin of Li, Be and B in Stars}},
	volume = {226},
	year = {1970}}

@article{savage1977,
	adsurl = {http://adsabs.harvard.edu/abs/1977ApJ...216..291S},
	author = {{Savage}, B.~D. and {Bohlin}, R.~C. and {Drake}, J.~F. and {Budich}, W.},
	doi = {10.1086/155471},
	journal = {\apj},
	month = aug,
	pages = {291-307},
	title = {{A survey of interstellar molecular hydrogen. I}},
	volume = {216},
	year = {1977}}

@article{sheffer2008,
	adsurl = {http://adsabs.harvard.edu/abs/2008ApJ...687.1075S},
	archiveprefix = {arXiv},
	author = {{Sheffer}, Y. and {Rogers}, M. and {Federman}, S.~R. and {Abel}, N.~P. and {Gredel}, R. and {Lambert}, D.~L. and {Shaw}, G.},
	doi = {10.1086/591484},
	eprint = {0807.0940},
	journal = {\apj},
	month = nov,
	pages = {1075-1106},
	title = {{Ultraviolet Survey of CO and H_{2} in Diffuse Molecular Clouds: The Reflection of Two Photochemistry Regimes in Abundance Relationships}},
	volume = {687},
	year = {2008}}

@article{sofia2004,
	adsurl = {http://adsabs.harvard.edu/abs/2004ApJ...605..272S},
	author = {{Sofia}, U.~J. and {Lauroesch}, J.~T. and {Meyer}, D.~M. and {Cartledge}, S.~I.~B.},
	doi = {10.1086/382592},
	eprint = {arXiv:astro-ph/0401510},
	journal = {\apj},
	month = apr,
	pages = {272-277},
	title = {{Interstellar Carbon in Translucent Sight Lines}},
	volume = {605},
	year = {2004}}

@article{sonnentrucker2007,
	adsurl = {http://adsabs.harvard.edu/abs/2007ApJS..168...58S},
	author = {{Sonnentrucker}, P. and {Welty}, D.~E. and {Thorburn}, J.~A. and {York}, D.~G.},
	doi = {10.1086/508687},
	eprint = {arXiv:astro-ph/0608557},
	journal = {\apjs},
	month = jan,
	pages = {58-99},
	title = {{Abundances and Behavior of ^{12}CO, ^{13}CO, and C_{2} in Translucent Sight Lines}},
	volume = {168},
	year = {2007}}

@article{watson1973,
	author = {Watson, William D.},
	journal = {\apj},
	pages = {L17},
	title = {The Rate of Formation of Interstellar Molecules by Ion-Molecule Reactions},
	url = {http://adsabs.harvard.edu/abs/1973ApJ...183L..17W},
	volume = {183},
	year = {1973}}

@article{welty2001,
	adsurl = {http://adsabs.harvard.edu/abs/2001ApJS..133..345W},
	author = {{Welty}, D.~E. and {Hobbs}, L.~M.},
	doi = {10.1086/320354},
	journal = {\apjs},
	month = apr,
	pages = {345-393},
	title = {{A High-Resolution Survey of Interstellar K I Absorption}},
	volume = {133},
	year = {2001}}

@ARTICLE{tielens2013,
       author = {{Tielens}, A.~G.~G.~M.},
        title = "{The molecular universe}",
      journal = {Reviews of Modern Physics},
         year = 2013,
        month = jul,
       volume = {85},
       number = {3},
        pages = {1021-1081},
          doi = {10.1103/RevModPhys.85.1021},
       adsurl = {https://ui.adsabs.harvard.edu/abs/2013RvMP...85.1021T}
}

@ARTICLE{redaelli2025,
       author = {{Redaelli}, E. and {Bovino}, S. and {Sabatini}, G. and {Arzoumanian}, D. and {Padovani}, M. and {Caselli}, P. and {Wyrowski}, F. and {Pineda}, J.~E. and {Latrille}, G.},
        title = "{Cosmic-ray ionisation rate in low-mass cores: The role of the environment}",
      journal = {\aap},
         year = 2025,
        month = oct,
       volume = {702},
          eid = {A210},
        pages = {A210},
          doi = {10.1051/0004-6361/202453198},
archivePrefix = {arXiv},
       eprint = {2508.18848},
 primaryClass = {astro-ph.GA},
       adsurl = {https://ui.adsabs.harvard.edu/abs/2025A&A...702A.210R}
}


\clearpage
\startlongtable
\begin{deluxetable}{cccc}
\tabletypesize{\tiny}
\tablecaption{Log of Observations \label{tbl_obslog}}
\tablehead{\colhead{Telescope/Instrument} & \colhead{UT Date} & \colhead{Target} & \colhead{Exposure Time} \\
\colhead{} & \colhead{} & \colhead{} & \colhead{(min)}
}
\startdata
VLT/CRIRES & 2011 Oct 29 & HD 37903 & 36 \\
VLT/CRIRES & 2011 Nov 07 & HD 37903 & 72 \\
VLT/CRIRES & 2011 Nov 30 & HD 37903 & 36 \\
VLT/CRIRES & 2011 Nov 30 & HD 47129 & 20 \\
VLT/CRIRES & 2011 Nov 30 & HD 48099 & 36 \\
VLT/CRIRES & 2011 Dec 01 & HD 42087 & 20 \\
VLT/CRIRES & 2012 Mar 13 & HD 102065 & 34 \\
IRTF/iSHELL & 2021 Oct 24 & HD 207198 & 56 \\
IRTF/iSHELL & 2021 Oct 24 & HD 216532 & 100 \\
IRTF/iSHELL & 2021 Oct 25 & HD 206165 & 40 \\
IRTF/iSHELL & 2021 Oct 25 & HD 216898 & 100 \\
IRTF/iSHELL & 2021 Oct 25 & HD 224151 & 24 \\
IRTF/iSHELL & 2021 Oct 26 & HD 206267 & 60 \\
IRTF/iSHELL & 2021 Oct 26 & HD 216532 & 96 \\
IRTF/iSHELL & 2021 Oct 26 & HD 224151 & 28 \\
IRTF/iSHELL & 2021 Oct 28 & HD 203374 & 54 \\
IRTF/iSHELL & 2021 Oct 28 & HD 216898 & 94 \\
IRTF/iSHELL & 2021 Oct 28 & HD 224151 & 30 \\
IRTF/iSHELL & 2021 Oct 29 & HD 23180 & 38 \\
IRTF/iSHELL & 2021 Oct 30 & HD 217312 & 90 \\
IRTF/iSHELL & 2021 Oct 30 & HD 224151 & 30 \\
IRTF/iSHELL & 2022 Aug 22 & HD 217312 & 90 \\
IRTF/iSHELL & 2022 Aug 22 & HD 22951 & 84 \\
IRTF/iSHELL & 2022 Aug 31 & HD 217035 & 60 \\
IRTF/iSHELL & 2022 Aug 31 & HD 281159 & 90 \\
IRTF/iSHELL & 2022 Sep 09 & HD 179406 & 70 \\
IRTF/iSHELL & 2022 Sep 09 & HD 207198 & 84 \\
IRTF/iSHELL & 2022 Sep 09 & HD 199579 & 40 \\
IRTF/iSHELL & 2022 Sep 09 & HD 281159 & 60 \\
IRTF/iSHELL & 2022 Sep 09 & HD 23180 & 30 \\
IRTF/iSHELL & 2022 Sep 09 & HD 37367 & 46 \\
IRTF/iSHELL & 2022 Sep 11 & HD 199579 & 88 \\
IRTF/iSHELL & 2022 Sep 11 & HD 281159 & 100 \\
IRTF/iSHELL & 2022 Sep 11 & HD 37367 & 60 \\
IRTF/iSHELL & 2022 Sep 13 & HD 184915 & 86 \\
IRTF/iSHELL & 2023 May 06 & HD 147933 & 50 \\
IRTF/iSHELL & 2023 Jun 03 & HD 149757 & 20 \\
IRTF/iSHELL & 2023 Jun 03 & HD 170740 & 60 \\
IRTF/iSHELL & 2023 Jun 03 & HD 179406 & 50 \\
IRTF/iSHELL & 2023 Jun 30 & HD 192639 & 40 \\
IRTF/iSHELL & 2023 Jul 27 & HD 145502 & 40 \\
IRTF/iSHELL & 2023 Jul 27 & HD 167971 & 50 \\
IRTF/iSHELL & 2023 Jul 27 & HD 170740 & 50 \\
\enddata
\end{deluxetable}
\normalsize

\clearpage
\startlongtable
\begin{deluxetable*}{cccccccc}
\tabletypesize{\scriptsize}
\tablecaption{H$_3^+$ Absorption Line Parameters \label{tbl_measurements}}
\tablehead{\colhead{Target} & \colhead{Transition} & \colhead{$v_{\rm LSR}$} & \colhead{FWHM} & \colhead{$W_\lambda$} & \colhead{$\sigma(W_\lambda)$} & \colhead{$N(J,K)$} & \colhead{$\sigma(N(J,K))$} \\
 &  & \colhead{(km s$^{-1}$)} &  \colhead{(km s$^{-1}$)} & \colhead{($10^{-6} \mu$m)} & \colhead{($10^{-6} \mu$m)} & 
 \colhead{(10$^{13}$ cm$^{-2}$)} & \colhead{(10$^{13}$ cm$^{-2}$)}
}
\startdata
HD 23180  & $R(1,1)^u$ &  7.5 &  2.6 & 0.30 & 0.11 & 1.25 & 0.47 \\ 
HD 23180  & $R(1,0)$ &  7.2 &  3.1 & 0.42 & 0.09 & 1.07 & 0.23 \\ 
HD 23180  & $R(1,1)^l$ &  6.7 &  2.8 & 0.46 & 0.11 & 2.13 & 0.50 \\ 
HD 281159 & $R(1,1)^u$ &  7.3 &  4.1 & 0.68 & 0.42 & 2.81 & 1.72 \\ 
HD 281159 & $R(1,0)$ &  4.1 &  6.0 & 0.93 & 0.45 & 2.36 & 1.13 \\ 
HD 167971 & $R(1,1)^u$ & 27.3 &  4.0 & 0.74 & 0.14 & 3.08 & 0.60 \\ 
HD 167971 & $R(1,0)$ & 25.3 &  6.6 & 1.15 & 0.22 & 2.91 & 0.55 \\ 
HD 167971 & $R(1,1)^l$ & 26.0 &  3.1 & 0.85 & 0.14 & 3.89 & 0.65 \\ 
HD 170740 & $R(1,1)^u$ &  4.4 &  4.6 & 0.46 & 0.12 & 1.89 & 0.52 \\ 
HD 170740 & $R(1,0)$ &  4.0 &  2.6 & 0.58 & 0.07 & 1.46 & 0.17 \\ 
HD 179406 & $R(1,1)^u$ &  2.8 &  4.1 & 0.53 & 0.21 & 2.21 & 0.88 \\ 
HD 179406 & $R(1,0)$ &  3.0 &  7.1 & 0.93 & 0.22 & 2.34 & 0.55 \\ 
HD 179406 & $R(1,1)^l$ &  2.2 &  6.1 & 0.66 & 0.28 & 3.00 & 1.27 \\ 
HD 203374 & $R(1,1)^u$ &  2.5 &  6.0 & 1.12 & 0.24 & 4.64 & 1.01 \\ 
HD 203374 & $R(1,0)$ &  2.6 &  4.6 & 1.09 & 0.22 & 2.76 & 0.55 \\ 
HD 203374 & $R(1,1)^l$ &  2.3 &  4.8 & 0.70 & 0.26 & 3.20 & 1.21 \\ 
HD 206165 & $R(1,1)^u$ &  1.5 &  4.3 & 0.81 & 0.18 & 3.36 & 0.75 \\ 
HD 206165 & $R(1,0)$ &  1.1 &  3.1 & 0.70 & 0.17 & 1.77 & 0.44 \\ 
HD 206267 & $R(1,1)^u$ & -0.5 &  4.5 & 0.83 & 0.24 & 3.43 & 0.98 \\ 
HD 206267 & $R(1,0)$ &  1.2 &  5.1 & 1.05 & 0.27 & 2.65 & 0.69 \\ 
HD 206267 & $R(1,1)^l$ & -0.1 &  7.7 & 1.23 & 0.27 & 5.64 & 1.25 \\ 
HD 207198 & $R(1,1)^u$ & -1.4 &  6.5 & 0.45 & 0.24 & 1.85 & 0.99 \\ 
HD 207198 & $R(1,0)$ & -1.9 &  6.0 & 0.97 & 0.18 & 2.46 & 0.45 \\ 
HD 207198 & $R(1,1)^l$ & -1.2 &  9.0 & 1.07 & 0.16 & 4.91 & 0.72 \\ 
HD 216532 & $R(1,1)^u$ &  5.3 &  6.5 & 1.92 & 0.31 & 7.96 & 1.28 \\ 
HD 216532 & $R(1,0)$ &  6.6 &  4.9 & 1.13 & 0.26 & 2.86 & 0.66 \\ 
HD 216532 & $R(1,1)^l$ &  4.1 &  5.3 & 0.91 & 0.25 & 4.16 & 1.16 \\ 
HD 216898 & $R(1,1)^u$ &  4.5 &  9.7 & 1.19 & 0.34 & 4.93 & 1.43 \\ 
HD 216898 & $R(1,0)$ &  3.5 &  5.8 & 1.02 & 0.19 & 2.58 & 0.47 \\ 
HD 224151 & $R(1,1)^u$ & -3.5 &  4.3 & 1.09 & 0.18 & 4.52 & 0.76 \\ 
HD 224151 & $R(1,0)$ & -2.9 &  4.2 & 1.41 & 0.14 & 3.57 & 0.34 \\ 
HD 224151 & $R(1,1)^l$ & -2.9 &  6.5 & 1.40 & 0.21 & 6.42 & 0.98 \\  
\enddata
\tablecomments{Parameters were determined from fitting absorption lines with gaussian functions. For sight lines where the $R(1,1)^l$ transition was not detected there is no table entry.} 
\end{deluxetable*}
\normalsize

\clearpage
\begin{deluxetable}{cccc}
\tabletypesize{\scriptsize}
\tablecaption{Analysis of H$_3^+$ Non-Detections  \label{tbl_upperlimits}}
\tablehead{\colhead{Target} & \colhead{$W_\lambda+\sigma(W_\lambda)$} & \colhead{$W_\lambda+\sigma(W_\lambda)$} & \colhead{$W_\lambda+\sigma(W_\lambda)$} \\
 & \colhead{$R(1,1)^u$} & \colhead{$R(1,0)$} & \colhead{$R(1,1)^l$} \\
 & \colhead{($10^{-6} \mu$m)} & \colhead{($10^{-6} \mu$m)} & \colhead{($10^{-6} \mu$m)}
}
\startdata
HD 22951 & $<$0.16 & $<$0.85 & $<$0.23 \\
HD 37367 & $<$0.76 & $<$0.55 & $<$0.84 \\
HD 37903 & $<$0.24 & $<$0.19 & $<$0.18 \\
HD 42087 & $<$0.30 & $<$0.48 & $<$0.21 \\
HD 47129 & $<$0.97 & $<$0.63 & $<$0.72 \\
HD 48099 & $<$1.64 & $<$0.92 & $<$0.53 \\
HD 102065 & $<$1.00 & $<$0.22 & $<$0.64 \\
HD 145502 & $<$0.53 & $<$0.42 & $<$0.22 \\
HD 147933 & $<$0.24 & $<$0.28 & $<$0.18 \\
HD 149757 & $<$0.36 & $<$0.48 & $<$0.57 \\
HD 184915 & $<$0.36 & $<$0.75 & $<$0.26 \\
HD 192639 & $<$0.94 & $<$2.12 & $<$2.89 \\
HD 199579 & $<$0.25 & $<$0.33 & $<$0.90 \\
HD 217035 & $<$10.05 & $<$3.12 & $<$2.85 \\
HD 217312 & $<$0.74 & $<$1.27 & $<$1.32 \\
\enddata
\tablecomments{The upper limits on the equivalent widths shown here are calculated as described in Section 3. }
\end{deluxetable}
\normalsize

\clearpage
\begin{deluxetable}{ccccccccccc}
\tabletypesize{\scriptsize}
\tablecaption{Line of Sight Properties \label{tbl_LoS}}
\tablehead{
\colhead{Star} & \colhead{$N({\rm H}_3^+)$} & \colhead{$\log(N({\rm H}_2))$} & \colhead{H$_2$ Ref.} & \colhead{$\log(N({\rm H}))$} & \colhead{H Ref.} & \colhead{Obs. $N_{\rm H}$} & \colhead{Map $N_{\rm H}$} & \colhead{Map/Obs.} & \colhead{$d^*$} & \colhead{$d^*$ Ref.} \\
\colhead{} & \colhead{(10$^{13}$~cm$^{-2}$)} & \colhead{} & \colhead{} & \colhead{} & \colhead{} & \colhead{(10$^{21}$~cm$^{-2}$)} & \colhead{(10$^{21}$~cm$^{-2}$)} & \colhead{} & \colhead{(pc)} & \colhead{}
}
\startdata
HD 23180 & 2.73$\pm$0.66 & 20.61$\pm$0.08 & 1 & 20.90$\pm$0.11 & 1 & 1.61 & 1.45 & 0.90 &  301 & 9 \\
HD 281159 & 5.16$\pm$2.06 & 21.09$\pm$0.15 & 2 & 21.38$\pm$0.30\tablenotemark{a} & 2 & 4.86 & 4.53 & 0.93 &  347 & 10 \\
HD 167971 & 5.98$\pm$0.82 & 20.85$\pm$0.11 & 2 & 21.60$\pm$0.30 & 2 & 5.40 & 2.02 & 0.37 & 1680 & 11 \\
HD 170740 & 3.35$\pm$0.54 & 20.86$\pm$0.07 & 2 & 21.15$\pm$0.15 & 2 & 2.86 & 2.83 & 0.99 &  250 & 9 \\
HD 179406 & 4.55$\pm$1.04 & 20.73$\pm$0.07 & 4 & 21.23$\pm$0.15 & 4 & 2.77 & 1.45 & 0.52 &  287 & 10 \\
HD 203374 & 6.81$\pm$1.16 & 20.67$\pm$0.06 & 3 & 21.20$\pm$0.05 & 3 & 2.52 & 2.10 & 0.83 &  2385 & 10 \\
HD 206165 & 5.13$\pm$0.87 & 20.77$\pm$0.03 & 5 & 21.19$\pm$0.30\tablenotemark{a} & 7\tablenotemark{b} & 2.73 & 2.20 & 0.81 &  996 & 10 \\
HD 206267 & 6.92$\pm$1.71 & 20.86$\pm$0.04 & 2 & 21.30$\pm$0.15 & 2 & 3.44 & 1.78 & 0.52 &  781 & 10 \\
HD 207198 & 6.31$\pm$2.21 & 20.83$\pm$0.09 & 3 & 21.28$\pm$0.08 & 3 & 3.26 & 2.10 & 0.64 &  980 & 11 \\
HD 216532 & 8.74$\pm$2.76 & 21.10$\pm$0.12 & 3 & 21.38$\pm$0.30\tablenotemark{a} & 3 & 4.92 & 6.10 & 1.24 &  920 & 11 \\
HD 216898 & 7.51$\pm$1.51 & 21.03$\pm$0.06 & 3 & 21.66$\pm$0.25 & 8 & 6.71 & 5.86 & 0.87 &  870 & 11 \\
HD 224151 & 8.80$\pm$1.39 & 20.53$\pm$0.05 & 3 & 21.35$\pm$0.07 & 3 & 2.92 & 1.79 & 0.61 &  1880 & 10 \\
HD 22951 & $<$2.82 & 20.46$\pm$0.08 & 1 & 21.04$\pm$0.11 & 1 & 1.67 & 1.54 & 0.92 &  369 & 10 \\
HD 37367 & $<$4.53 & 20.53$\pm$0.09 & 6 & 21.17$\pm$0.15 & 6 & 2.16 & 2.69 & 1.25 & 1274 & 10 \\
HD 37903 & $<$1.30 & 20.92$\pm$0.05 & 4 & 21.17$\pm$0.10 & 4 & 3.14 & 3.19 & 1.02 &  471 & 12 \\
HD 42087 & $<$2.18 & 20.52$\pm$0.10 & 4 & 21.39$\pm$0.11 & 4 & 3.12 & 1.28 & 0.41 & 2470 & 10 \\
HD 47129 & $<$4.90 & 20.55$\pm$0.07 & 1 & 21.08$\pm$0.18 & 1 & 1.91 & 0.98 & 0.51 & 1271 & 10 \\
HD 48099 & $<$4.76 & 20.29$\pm$0.04 & 1 & 21.15$\pm$0.15 & 1 & 1.80 & 0.77 & 0.42 & 1289 & 10 \\
HD 102065 & $<$3.51 & 20.50$\pm$0.05 & 2 & 20.54$\pm$0.30\tablenotemark{a} & 2 & 0.98 & 1.10 & 1.12 &  200 & 9 \\
HD 145502 & $<$2.07 & 19.89$\pm$0.09 & 1 & 21.15$\pm$0.15 & 1 & 1.57 & 1.08 & 0.69 &  170 & 9 \\
HD 147933 & $<$1.53 & 20.57$\pm$0.09 & 1 & 21.81$\pm$0.08 & 1 & 7.20 & 4.90 & 0.68 &  170 & 9 \\
HD 149757 & $<$2.70 & 20.65$\pm$0.04 & 1 & 20.72$\pm$0.02 & 1 & 1.42 & 1.21 & 0.85 &  135 & 10 \\
HD 184915 & $<$3.10 & 20.31$\pm$0.06 & 1 & 20.90$\pm$0.13 & 1 & 1.20 & 1.16 & 0.96 &  498 & 10 \\
HD 192639 & $<$9.25 & 20.69$\pm$0.03 & 2 & 21.32$\pm$0.12 & 2 & 3.07 & 2.26 & 0.74 & 2130 & 11 \\
HD 199579 & $<$1.84 & 20.51$\pm$0.07 & 3 & 21.04$\pm$0.11 & 3 & 1.74 & 3.37 & 1.93 &  940 & 11 \\
HD 217035 & $<$21.00 & 20.93$\pm$0.07 & 3 & 21.46$\pm$0.12 & 3 & 4.59 & 2.07 & 0.45 &  720 & 11 \\
HD 217312 & $<$6.28 & 20.79$\pm$0.06 & 3 & 21.48$\pm$0.09 & 3 & 4.25 & 1.62 & 0.38 &  600 & 11 \\
\enddata
\tablecomments{H$_3^+$ column densities are the sum of the values in the $(J,K)$=(1,0) and (1,1) states presented in Table \ref{tbl_measurements}. In sight lines where H$_3^+$ absorption is not detected, upper limits are displayed. References for H$_2$ column densities---all directly measured from H$_2$ absorption in the UV---are as follows: (1) \citet{savage1977}; (2) \citet{rachford2002}; (3) \citet{shull2021};  (4) \citet{rachford2009}; (5) \citet{pan2005}; (6) \citet{jenkins2009}. For most sight lines atomic H is directly measured from Ly-$\alpha$ absorption, but in some cases it must be estimated. References for H column densities use the same numbering as H$_2$, and further include: (7) \citet{pan2004}; (8) \citet{vandeputte2023}. The observed total hydrogen column, Obs. $N_{\rm H}$, the integrated total hydrogen column out to the adopted distance of the background star ($d^*$) from the Gaia extinction maps, Map $N_{\rm H}$, and their ratio are also given. For several sight lines, integrating the total hydrogen column in the \citet{edenhofer2024} maps to the distance of the background star reported by \citet{bailer-jones2021} results in a value of Map $N_{\rm H}$ that is significantly different from Obs. $N_{\rm H}$. In some cases, this disagreement can be improved by ``moving'' the star to the other side of a nearby interstellar cloud. Sight lines with an entry of (9) in the $d^*$ Ref. column are cases where we moved the star from the location reported by \citet{bailer-jones2021} to the other side of a density peak in the \citet{edenhofer2024} map to bring Map/Obs. closer to unity. All other adopted distances were taken directly from the following works: (10) \citet{bailer-jones2021}; (11) \citet{shull2019}; (12) \citet{vanleeuwen2007}.}
\tablenotetext{a}{The atomic hydrogen column density was not directly measured, but estimated using $N({\rm H}_2)$, $E(B-V)$, and the relationship $(N({\rm H})+2N({\rm H_2}))/E(B-V)=5.8\times10^{21}$~cm$^{-2}$~mag$^{-1}$.}
\tablenotetext{b}{$N({\rm H})$ is not reported in this paper, but $E(B-V)$ used for the calculation above is given.}
\end{deluxetable}
\normalsize


\clearpage
\begin{deluxetable}{ccc|cc|ccc}
\tabletypesize{\scriptsize}
\tablecaption{Inferred Cloud Properties \label{tbl_zeta}}
\tablehead{
 & & \colhead{3D-PDR Model} & \multicolumn{2}{c}{Analytical, $dE$} & \multicolumn{3}{c}{Analytical, C$_2$} \\
\colhead{Target Star} & \colhead{Cloud Distance} & \colhead{$\zeta({\rm H_2})$} & \colhead{$n_{\rm H}$} & \colhead{$\zeta({\rm H_2})$} & \colhead{C${_2}$ Ref.} & \colhead{$n_{\rm H}$} & \colhead{$\zeta({\rm H_2})$} \\
 & \colhead{(pc)} & \colhead{(10$^{-17}$~s$^{-1}$)} & \colhead{(cm$^{-3}$)} & \colhead{(10$^{-17}$~s$^{-1}$)} & & \colhead{(cm$^{-3}$)} & \colhead{(10$^{-17}$~s$^{-1}$)}
}
\startdata
HD 23180 & 148 & 3.4$^{+0.9}_{-0.9}$ & 32 & 6.0$\pm$2.2 & 1 & 41 & 7.6$\pm$2.4 \\
HD 281159 & 308 & 5.0$^{+4.5}_{-4.3}$ & 102 & 9.5$\pm$5.4 & 2 & 72 & 6.7$\pm$4.1 \\
HD 27778 & 153 & 7.5$^{+0.8}_{-0.8}$ & 44 & 11.8$\pm$3.3 & 2 & 64 & 17.1$\pm$3.4 \\
HD 170740 & 230 & 3.2$^{+0.6}_{-0.5}$ & 51 & 5.3$\pm$1.6 & 2 & 70 & 7.3$\pm$1.9 \\
HD 179406 & 226 & 4.4$^{+1.2}_{-1.0}$ & 36 & 7.6$\pm$2.6 & 2 & 162 & 34.1$\pm$14.2 \\
HD 203374 & 427 & 4.7$^{+1.1}_{-0.8}$ & 16 & 4.6$\pm$1.4 & ... & ... & ... \\
HD 206165 & 427 & 2.7$^{+0.6}_{-0.6}$ & 10 & 2.0$\pm$0.5 & ... & ... & ... \\
HD 206267 & 462 & 2.5$^{+0.7}_{-0.7}$ & 12 & 2.7$\pm$0.9 & 2 & 132 & 30.2$\pm$8.1 \\
HD 207198 & 425 & 3.7$^{+1.4}_{-1.4}$ & 17 & 3.8$\pm$1.7 & 2 & 56 & 12.4$\pm$5.0 \\
HD 224151 & 366 & 11.5$^{+3.8}_{-2.8}$ & 16 & 8.0$\pm$2.3 & ... & ... & ... \\
\hline
HD 24398 & 147 & 7.6$^{+0.7}_{-0.7}$ & 38 & 12.6$\pm$3.7 & 1 & 28 & 9.2$\pm$2.0 \\
HD 24534 & 146 & 7.0$^{+1.1}_{-1.0}$ & 67 & 14.9$\pm$3.9 & 2 & 70 & 15.5$\pm$2.8 \\
HD 73882 & 897 & 2.7$^{+0.3}_{-0.1}$ & 24 & 4.5$\pm$1.3 & 2 & 132 & 24.7$\pm$5.0 \\
HD 110432 & 196 & 9.0$^{+0.5}_{-0.5}$ & 38 & 10.5$\pm$2.2 & 3 & 32 & 8.9$\pm$2.3 \\
HD 154368 & 201 & 6.0$^{+1.1}_{-1.0}$ & 119 & 20.7$\pm$7.6 & 2 & 80 & 14.0$\pm$4.4 \\
HD 210839 & 448 & 3.8$^{+1.0}_{-0.8}$ & 15 & 3.7$\pm$1.0 & 3 & 27 & 6.6$\pm$2.6 \\
\enddata
\tablecomments{Results in the top portion of the table are from new observations presented in this paper, while results in the bottom portion of the table are from \citet{obolentseva2024}, and are included for completeness. Column 3 presents cosmic-ray ionization rates inferred from the 3D-PDR modeling analysis, while columns 5 and 8 present ionization rates inferred from equation (\ref{eq_crir}) using gas densities derived from the differential extinction maps and C$_2$ rotation analysis, respectively, which are themselves given in columns 4 and 7. References for C$_2$ observations used to determine gas density using the methods of \citet{neufeld2024} are as follows: (1) \citet{fan2024}; (2) \citet{neufeld2024}, (3) \citet{sonnentrucker2007}. All values of the cosmic-ray ionization rate discussed throughout this paper refer to the 3D-PDR model results shown in column 3.}
\end{deluxetable}

\clearpage
\begin{deluxetable}{cccc}
\tabletypesize{\scriptsize}
\tablecaption{Calculated Cloud Properties \label{tbl_zeta_n}}
\tablehead{
\colhead{Target Star} & \colhead{Cloud Distance} & \colhead{$N({\rm H_2)}$ est.} &  \colhead{$\zeta({\rm H_2})/n_{\rm H}$} \\
 & \colhead{(pc)} & \colhead{($10^{20}$ cm$^{-2}$)} & \colhead{(10$^{-18}$~cm~$^3$~s$^{-1}$)} 
}
\startdata
HD 41117 & 721, 986, 1093 & ... & 2.7$^{+1.0}_{-1.0}$ \\
HD 43384 & 565, 1089 & 5.2$^{+2.1}_{-3.1}$ & 1.7$^{+1.1}_{-0.7}$ \\
HD 167971 & 111, 208, 1370 & 2.9$^{+1.2}_{-1.8}$ & 4.9$^{+3.0}_{-2.1}$ \\
HD 216532 & 801, 862 & 3.7$^{+1.5}_{-2.3}$ & 5.4$^{+3.7}_{-2.8}$ \\
HD 216898 & 801, 859 & 3.0$^{+1.2}_{-1.8}$ & 5.0$^{+3.2}_{-2.3}$ \\
\hline
HD 22951 & 148, 306 & ... & $<2.3$ \\
HD 37367 & 221 & ... & $<3.0$ \\
HD 37903 & 417 & ... & $<0.4$ \\
HD 42087 & 1104 & ... & $<1.6$ \\
HD 47129 & 829, 1018 & ... & $<3.2$ \\
HD 48099 & 989 & ... & $<4.6$ \\
HD 102065 & 196 & ... & $<2.5$ \\
HD 145502 & 107, 150 & ... & $<5.4$ \\
HD 147933 & 150 & ... & $<1.1$ \\
HD 149757 & 106 & ... & $<1.6$ \\
HD 184915 & 137 & ... & $<3.5$ \\
HD 192639 & 809 & ... & $<3.6$ \\
HD 199579 & 821, 911 & ... & $<1.2$ \\
HD 217035 & 274, 351 & ... & $<5.4$ \\
HD 217312 & 275, 347 & ... & $<2.1$ \\
\enddata
\tablecomments{The top portion of the table contains sight lines where H$_3^+$ was detected but the sight line was not suitable for modeling. HD~41117 from \citet{albertsson2014} is reanalyzed here using this method for consistency, as \citet{obolentseva2024} found that the H$_3^+$ absorption cannot be localized to a specific gas density peak along the line of sight. For sight lines where the H$_3^+$ absorption is well-matched to CH absorption in velocity space, the estimated H$_2$ column density in the absorbing cloud is also presented. Because the value of $\zeta({\rm H_2})/n_{\rm H}$ cannot be attributed to a specific cloud, we list the distances to all density peaks where the H$_3^+$ absorption may be located. The bottom portion of the table contains the same information, but for sight lines where H$_3^+$ is not detected.}
\end{deluxetable}


\clearpage
\begin{figure}
\epsscale{1.0}
\plotone{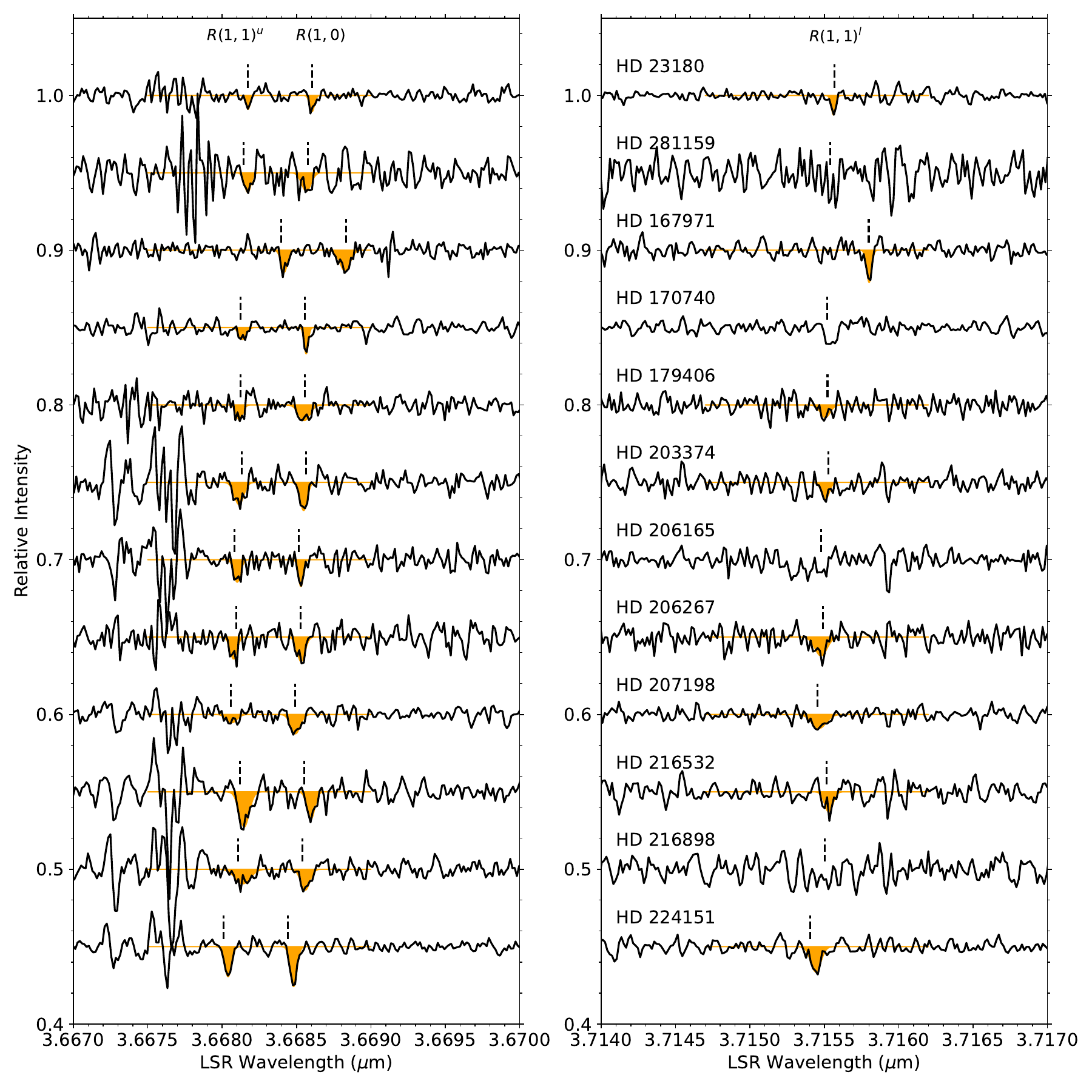}
\caption{Spectra in the left panel are focused on the $R(1,1)^u$ and $R(1,0)$ transitions of H$_3^+$, while spectra in the right panel are focused on the $R(1,1)^l$ transition. All spectra are normalized, and have been shifted vertically for clarity. Dashed vertical lines above spectra mark the expected location of H$_3^+$ absorption, based on previous observations of absorption from different molecular and/or atomic species along these sight lines. All wavelengths have been shifted into the LSR frame. The large noise feature near 3.6676~$\mu$m is due to a strong atmospheric CH$_4$ absorption line. Orange shaded regions show the gaussian fits to the H$_3^+$ absorption lines. Spectra in the right panel lacking orange regions were not fit due to either high noise levels or known contamination of the $R(1,1)^l$ transition.}
\label{fig_h3p_spectra}
\end{figure}


\clearpage
\begin{figure}
\epsscale{1.0}
\plotone{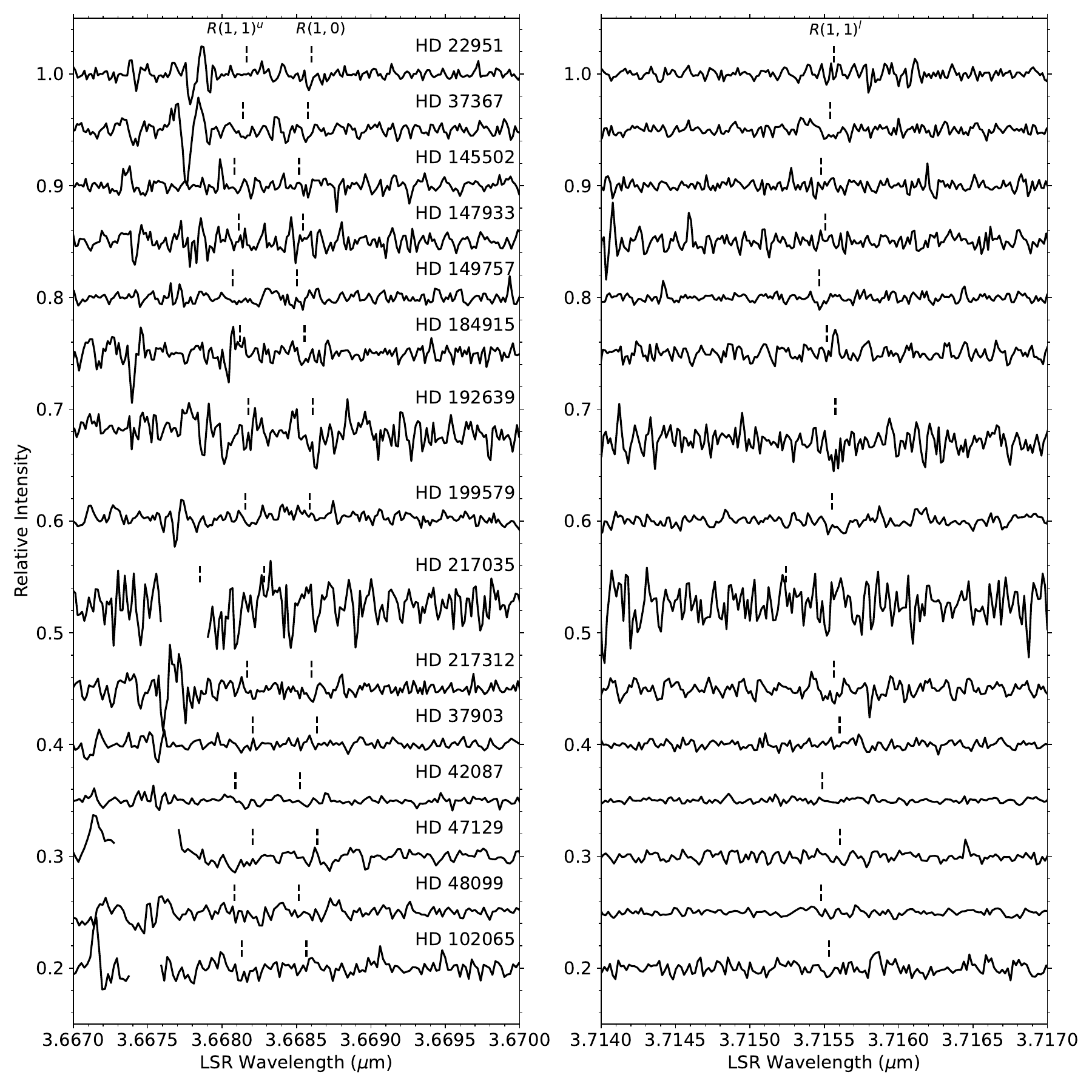}
\caption{Same as Figure \ref{fig_h3p_spectra}, but showing sight lines where H$_3^+$ absorption is not detected. Gaps in spectra indicate regions where high noise levels caused by strong atmospheric absorption have been masked out. The bottom five spectra are from VLT/CRIRES observations made in 2011-2012.}
\label{fig_h3p_spectra_no}
\end{figure}

\clearpage
\begin{figure}
\epsscale{0.5}
\plotone{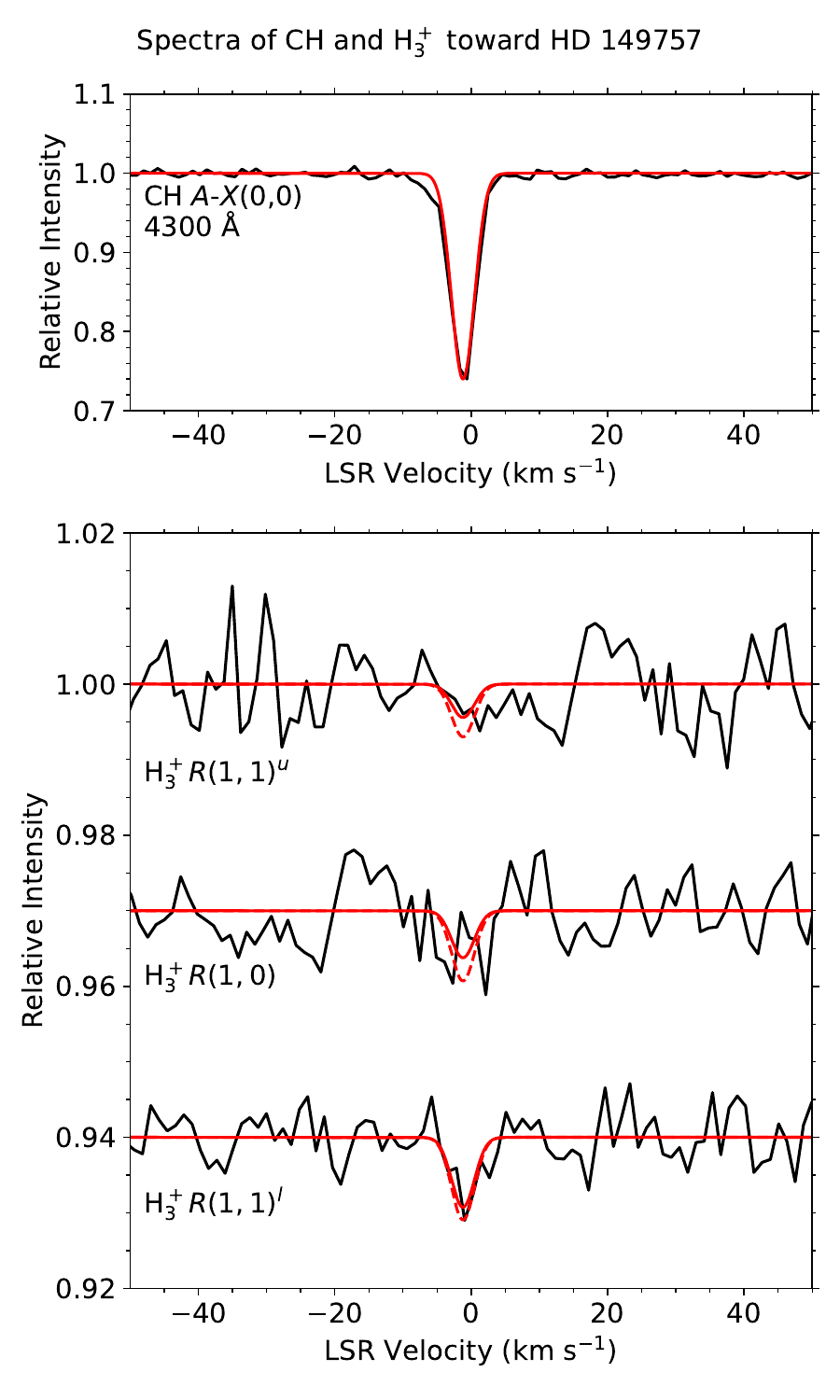}
\caption{The top panel shows a normalized spectrum of HD 149757 near the $A$-$X$(0,0) transition of CH at 4300.308~\AA\ in black, observed using UVES at the VLT. The red curve is a fit to the CH absorption line. The bottom panel shows normalized spectra of HD 149757 near the $R(1,1)^u$, $R(1,0)$, and $R(1,1)^l$ transitions of H$_3^+$ in black, with spectra shifted in the vertical direction for clarity. Solid red curves show the gaussian fits with line center and line width fixed to the results of the CH fit, as described in Section \ref{sect_spectral_analysis}. Dashed red curves show the gaussian fits plus 1$\sigma$ uncertainties, and the integrated areas defined by the dashed curves correspond to the upper limits on equivalent widths given in Table \ref{tbl_upperlimits}.}
\label{fig_example_CHfit}
\end{figure}

\clearpage
\begin{figure}
\epsscale{1.15}
\plotone{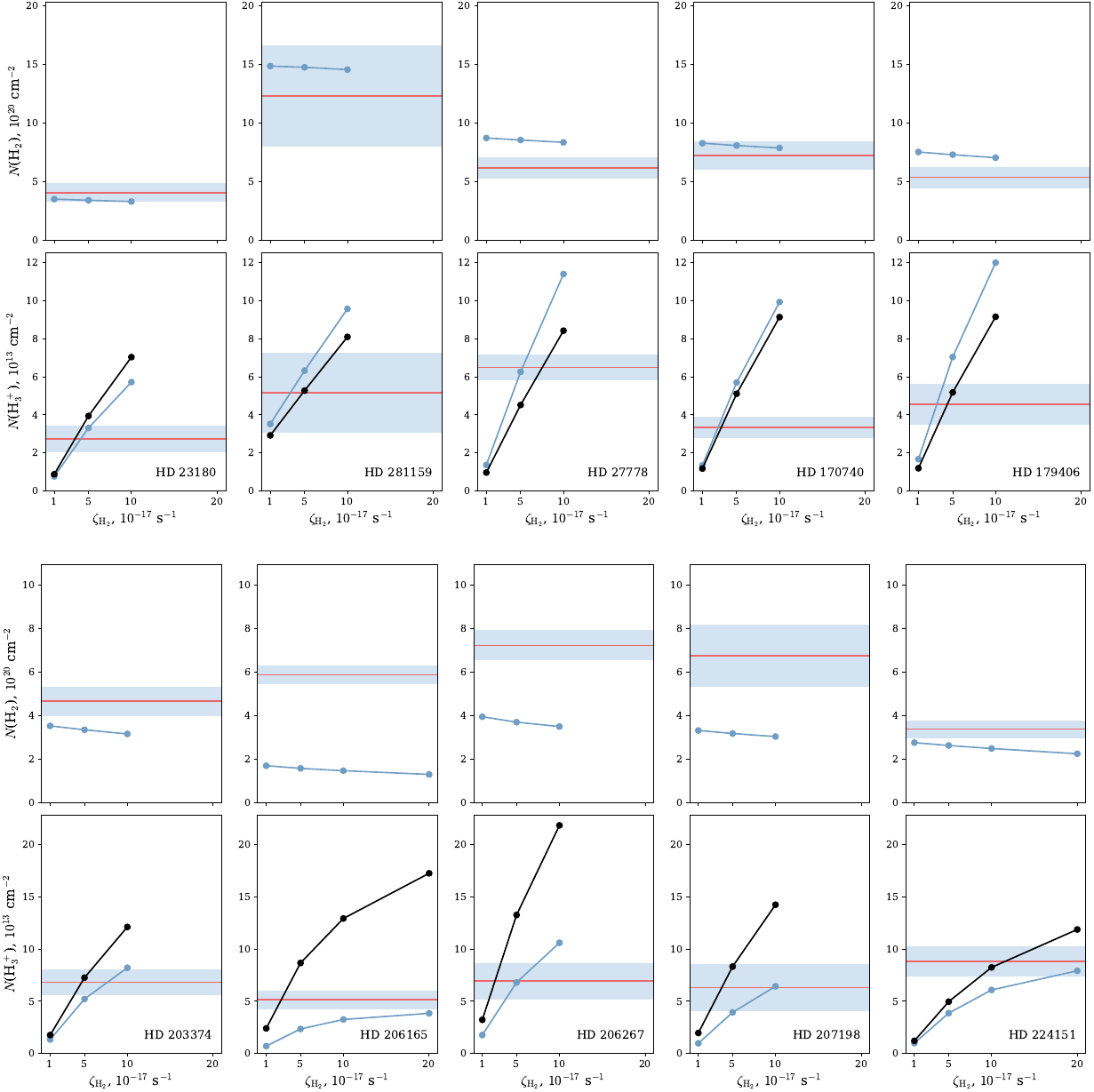}
\caption{These panels compare the observed H$_2$ and H$_3^+$ column densities in a given sight line to column densities predicted by models of that sight line with different cosmic-ray ionization rates. There are two panels for each sight line, with the top panel showing H$_2$ and the bottom panel showing H$_3^+$. Horizontal red lines and blue shaded regions mark the observed column densities and 1$\sigma$ uncertainties. Blue circles connected by lines show the column densities returned by the model as a function of $\zeta({\rm H_2})$. In the bottom panel for each sight line, black circles mark the ``corrected'' H$_3^+$ column density, accounting for the difference between observed and predicted H$_2$ column density as described in Section \ref{sec_sim3DPDR}. The intersection of the black curve and red line marks the inferred cosmic-ray ionization rate for each cloud, and is the value reported in Table \ref{tbl_zeta}. Note that for HD~281159 (top row, second panel) the simulation is for the cloud at $\sim300$~pc, and it assumes a contribution to the total H$_3^+$ column density from the cloud at $\sim150$~pc equal to that measured in the nearby HD~23180 sight line (see Section \ref{sec_sim3DPDR}). This is why the modeled $N({\rm H_3^+})$ for HD~281159 approaches the observed HD~23180 $N({\rm H_3^+})$ as the cosmic-ray ionization rate (in the far cloud) approaches zero.}
\label{fig_modelresults}
\end{figure}

\clearpage
\begin{figure}
\epsscale{1.15}
\plotone{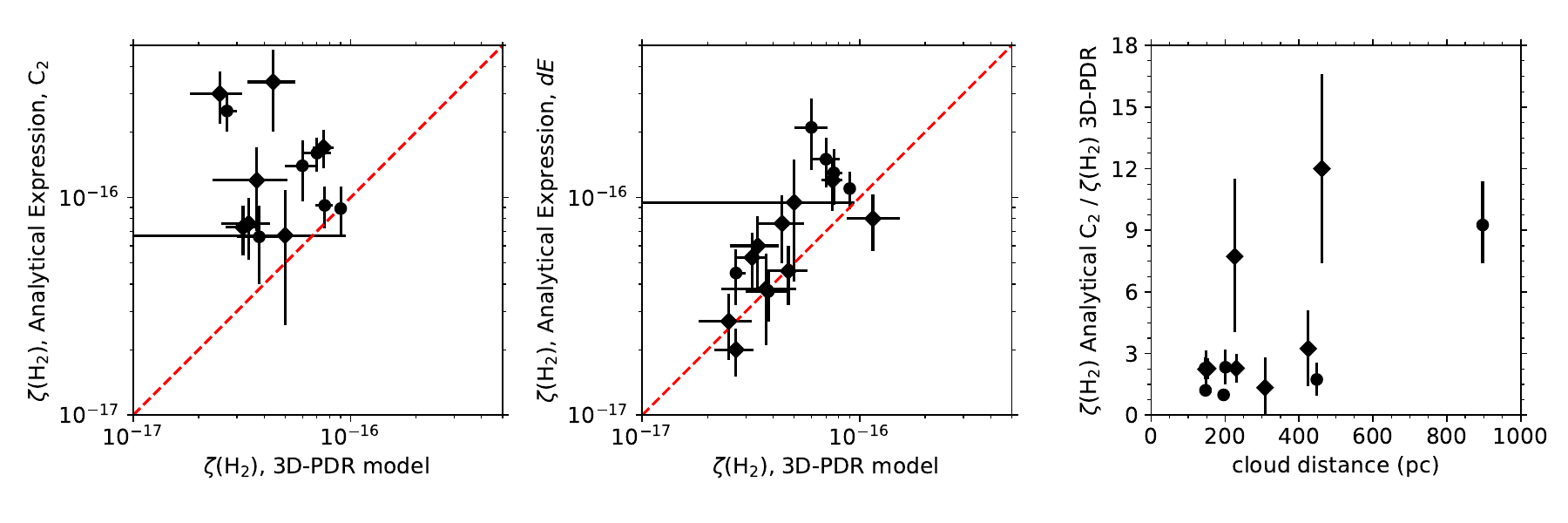}
\caption{The left panel compares cosmic-ray ionization rates inferred from equation (\ref{eq_crir}) using gas densities derived from C$_2$ to those inferred using the 3D-PDR model. The center panel compares cosmic-ray ionization rates inferred from equation (\ref{eq_crir}) using peak gas densities derived from the \citet{edenhofer2024} differential extinction map to those inferred using the 3D-PDR model. The right panel is the ratio of the ionization rates in the left panel as a function of distance to the gas cloud. Filled circles are from \citet{obolentseva2024} while filled diamonds are from this work. The dashed red diagonal lines indicate a one-to-one correlation.}
\label{fig_method_compare}
\end{figure}

\clearpage
\begin{figure}
\epsscale{1.0}
\plotone{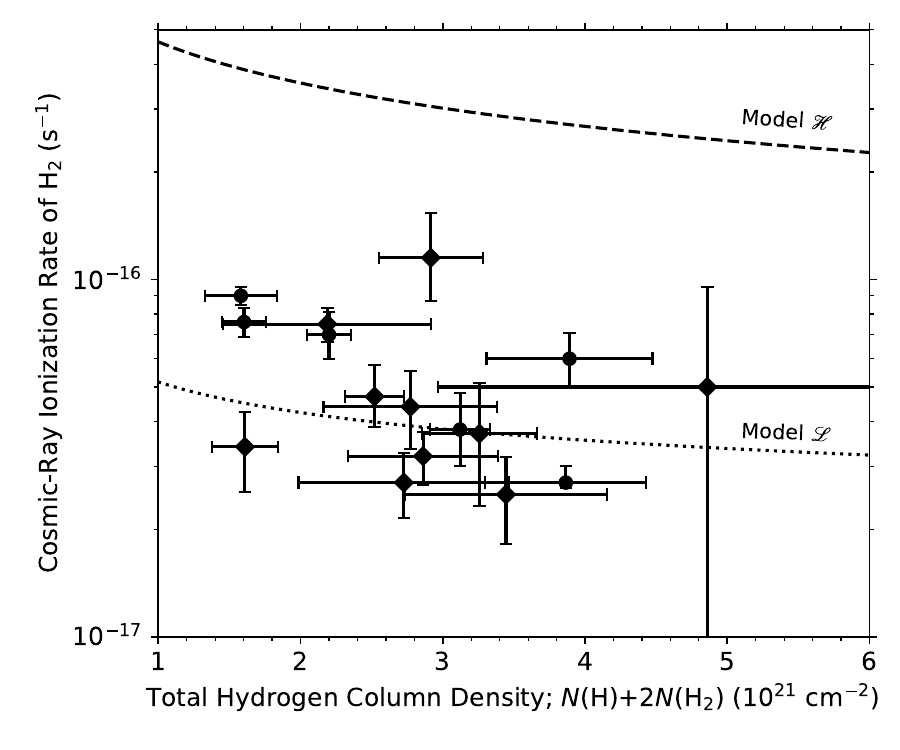}
\caption{For sight lines where the cosmic-ray ionization rate has been determined via 3D-PDR modeling, $\zeta({\rm H_2})$ is plotted against the total hydrogen column density. The dashed and dotted black curves show the predicted relationships for the Model $\mathscr{H}$ and Model $\mathscr{L}$ cosmic-ray spectra of \citet{padovani2018}, respectively. Filled circles are from \citet{obolentseva2024} while filled diamonds are from this work. Note that the points from \citet{obolentseva2024} are in slightly different locations than in their Figure 17 for three reasons: 1) we plot the {\it corrected} optimum value of $\zeta({\rm H_2})$ (blue points in their figure); 2) we simply take the measured sum of $N({\rm H})+2N({\rm H_2})$, not the optimum value described in their Section 6.2; 3) we have excluded the HD~41117 point for reasons already discussed. Uncertainties on $N_{\rm H}$ are based on observational uncertainties in the measurements of $N({\rm H})$ and $N({\rm H_2})$. Error bars for $\zeta({\rm H_2})$ are based only on uncertainties in $N({\rm H}_3^+)$, and more details about other potential sources of uncertainty are provided in Section \ref{sec_discussion}.}
\label{fig_zeta_vs_NH}
\end{figure}

\clearpage
\begin{figure}
\epsscale{0.95}
\plotone{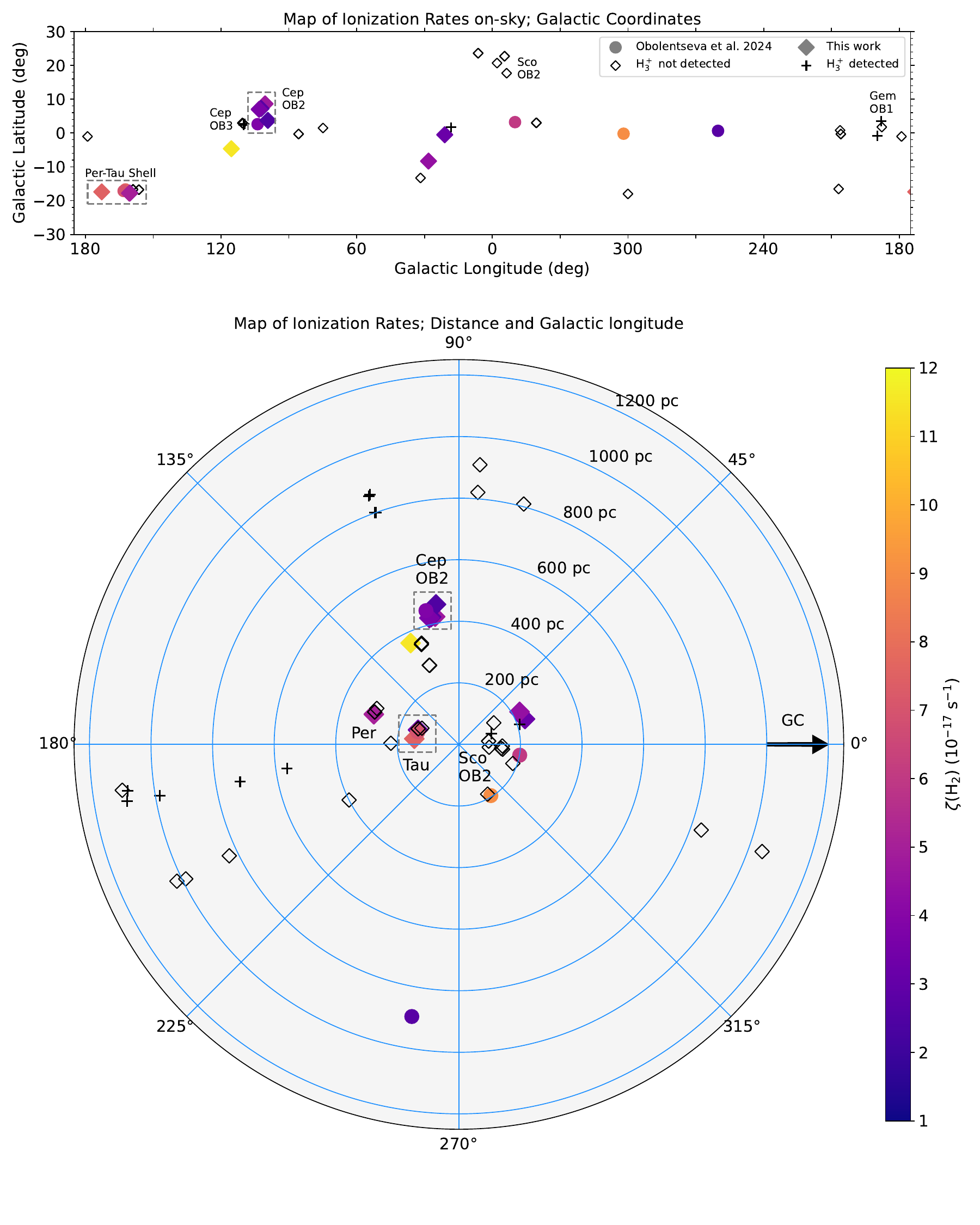}
\caption{The top panel shows our target sight lines in Galactic longitude and latitude. Empty symbols denote non-detections of H$_3^+$, while filled symbols have the value of the inferred cosmic-ray ionization rate indicated by color, using the color bar in the bottom panel. Crosses indicate sight lines where H$_3^+$ is detected, but for which the 3D-PDR modeling is not performed. The bottom panel shows a view of the Galactic Plane from above, with the Sun at the center and the direction toward the Galactic Center at right. Galactic longitude increases in the counterclockwise direction from right, and the map extends to 1.25~kpc from the Sun, the extent of the differential extinction maps from \citet{edenhofer2024}. Individual clouds where we have inferred cosmic-ray ionization rates are marked by symbols, with ionization rates indicated by color. Two regions---Per-Tau and Cep OB2---are crowded in these maps, and Figure \ref{fig_crirmapzoom} contains zoomed-in views of these regions, both in Galactic coordinates on-sky, and in heliocentric cartesian coordinates. The dashed grey boxes in the top and bottom panels here mark the regions appearing in Figure \ref{fig_crirmapzoom}.}
\label{fig_crirmap}
\end{figure}

\clearpage
\begin{figure}
\epsscale{1.0}
\plotone{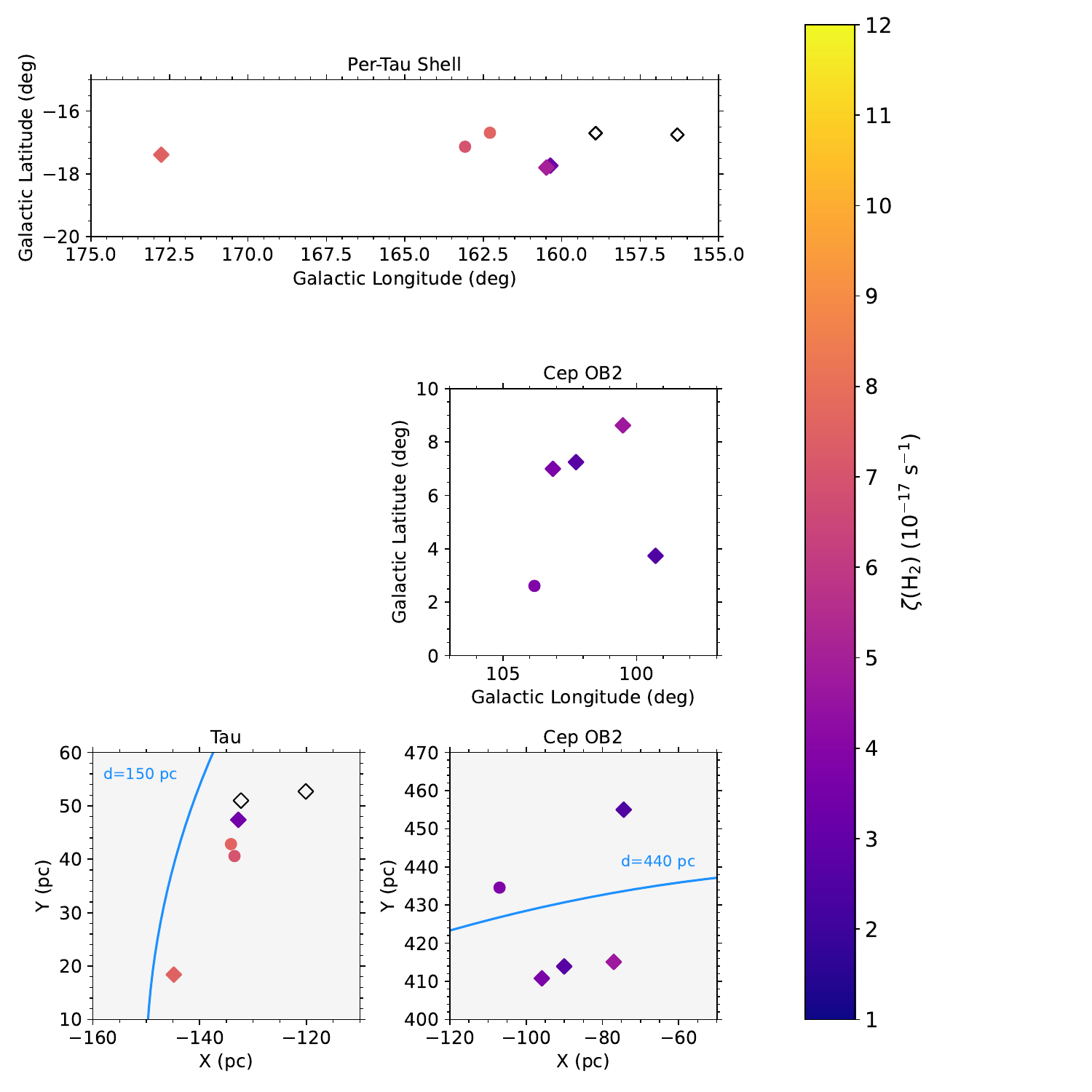}
\caption{These are zoomed-in views of the regions marked by dashed gray boxes in Figure \ref{fig_crirmap}. The top panel shows Galactic coordinates (on-sky) for the Per-Tau shell, while the middle panel shows the same for the Cep OB2 region. The bottom row shows heliocentric cartesian coordinates (distance from Sun in $X$ and $Y$ directions), with the Tau and Cep OB2 regions on the left and right sides, respectively. Blue arcs show constant distance from the Sun within the $X,Y$ plane, and are similar to the concentric circles in the bottom panel of Figure \ref{fig_crirmap}. Note that the bottom left panel only shows the near (Tau) side of the Per-Tau shell at about 150~pc, and that one of the data points from the top panel is ``missing'' because it is in the far (Per) side of the shell at about 300~pc.}
\label{fig_crirmapzoom}
\end{figure}


\appendix

\restartappendixnumbering

\section{Gas Distributions in and Around Target Sight Lines}



\subsection{Sight lines with 3D-PDR modeling applied.}

\clearpage
\begin{figure}
\epsscale{1.0}
\plotone{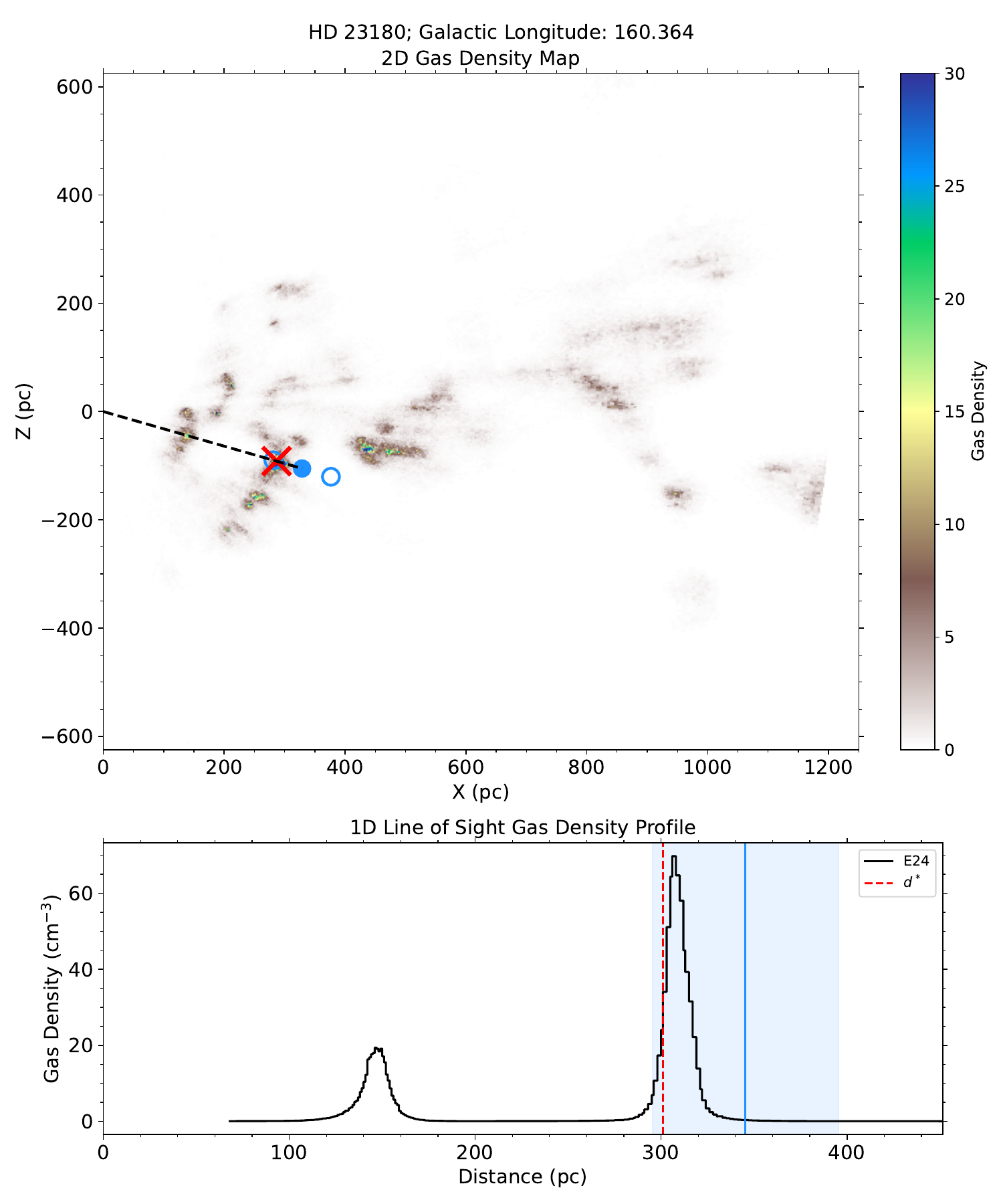}
\caption{The top panel shows the 2D distribution of gas from one realization of the differential extinction maps of \citet{edenhofer2024} in a plane of constant Galactic longitude that includes both the Sun and the background star. The black dashed line shows the line of sight to the star, with our adopted distance marked by a red $\times$. The filled blue circle marks the median distance to the star derived from Gaia EDR3 \citep{bailer-jones2021}, while the open blue circles mark the 16\% and 84\% distances. In the bottom panel, the black curve shows the mean distribution of gas densities along the line of sight from all 12 differential extinction realizations of \citet{edenhofer2024}. The vertical blue line and shaded region again mark the Gaia EDR3 median and 16--84\% distance ranges, respectively, while the vertical dashed red line marks our adopted distance (see Table \ref{tbl_LoS}).}
\label{fig_gaia_hd23180}
\end{figure}

\clearpage
\begin{figure}
\epsscale{1.0}
\plotone{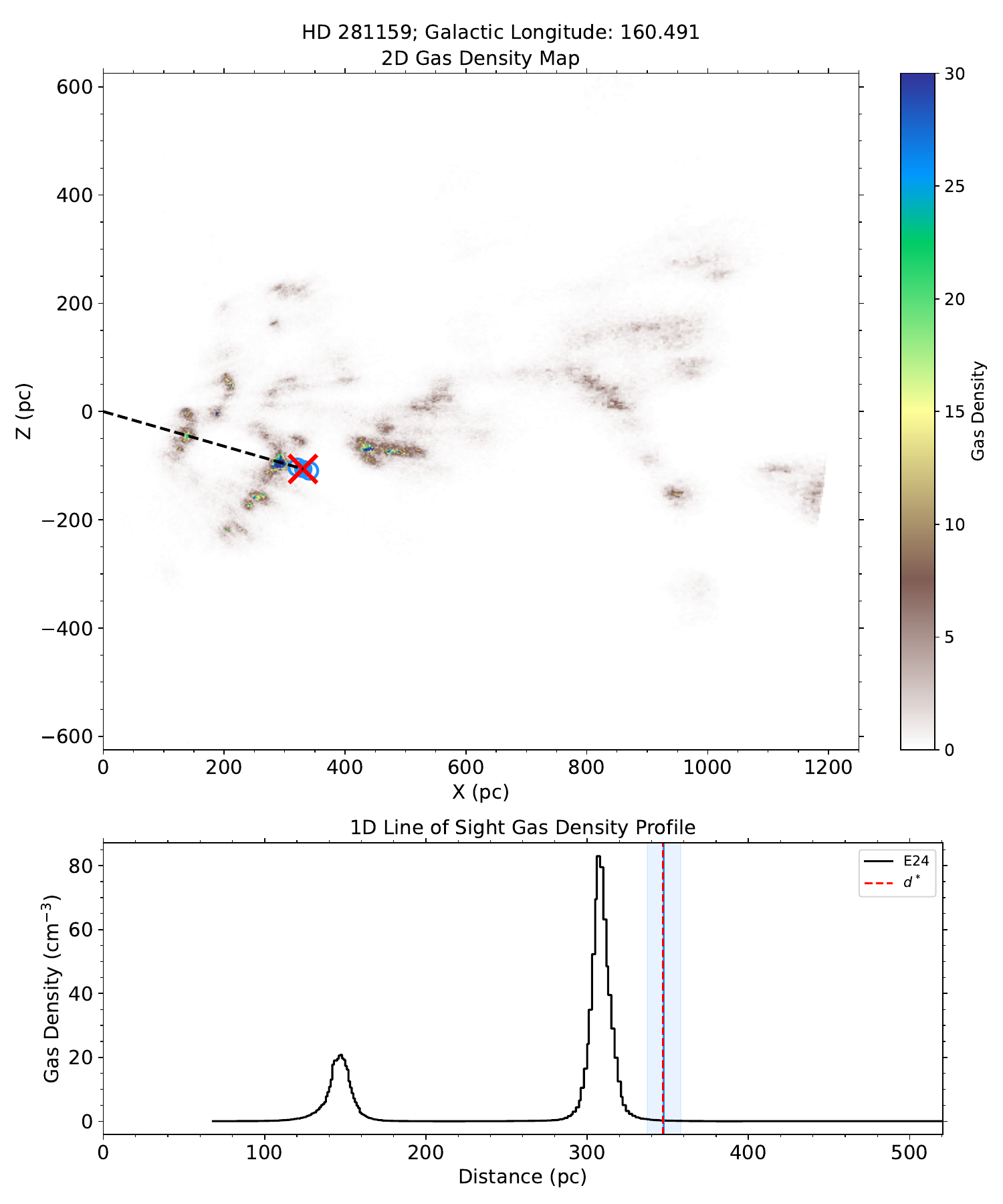}
\caption{Same as Figure \ref{fig_gaia_hd23180}, but for the sight line toward HD~281159.}
\label{fig_gaia_hd281159}
\end{figure}

\clearpage
\begin{figure}
\epsscale{1.0}
\plotone{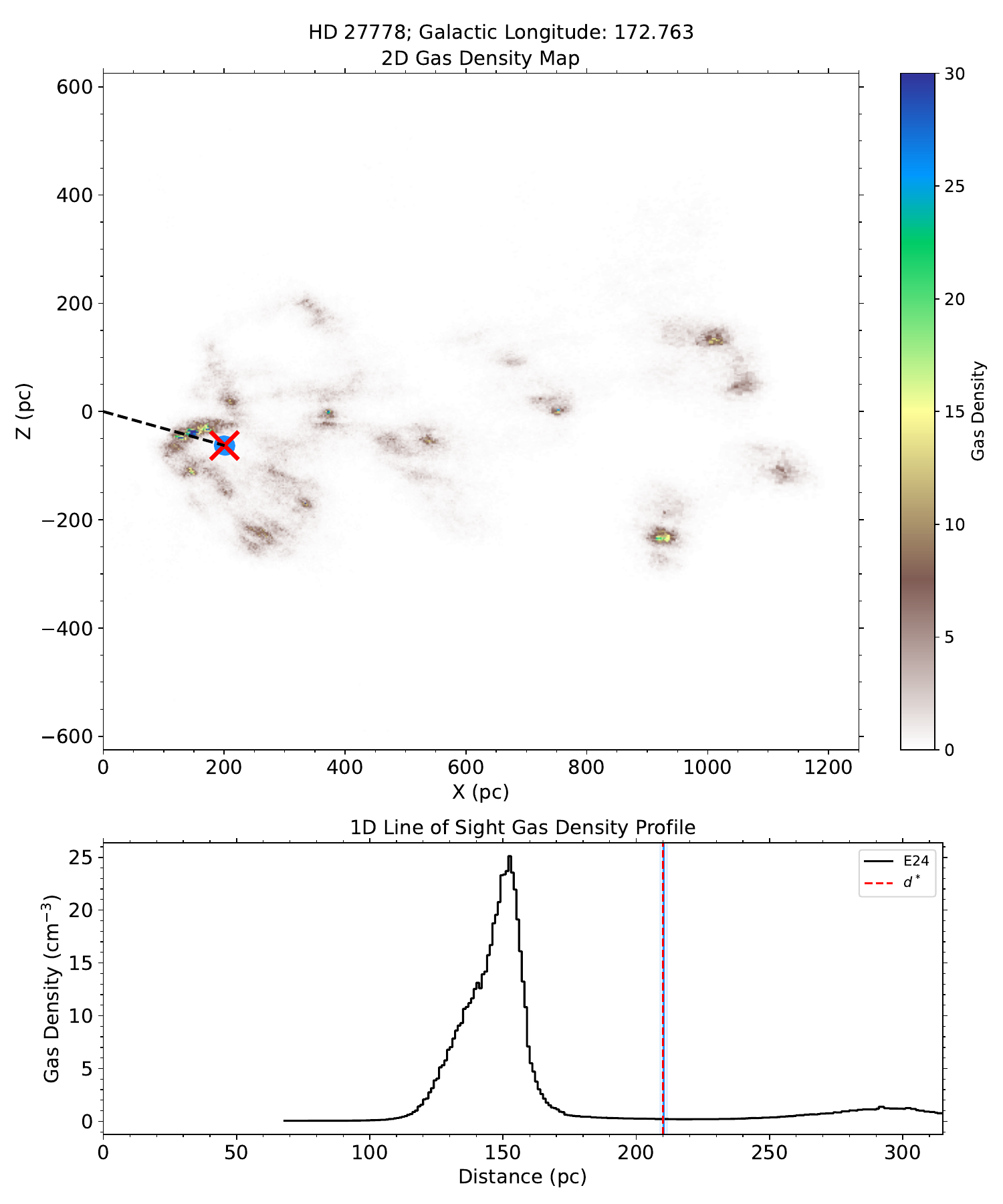}
\caption{Same as Figure \ref{fig_gaia_hd23180}, but for the sight line toward HD~27778.}
\label{fig_gaia_hd27778}
\end{figure}

\clearpage
\begin{figure}
\epsscale{1.0}
\plotone{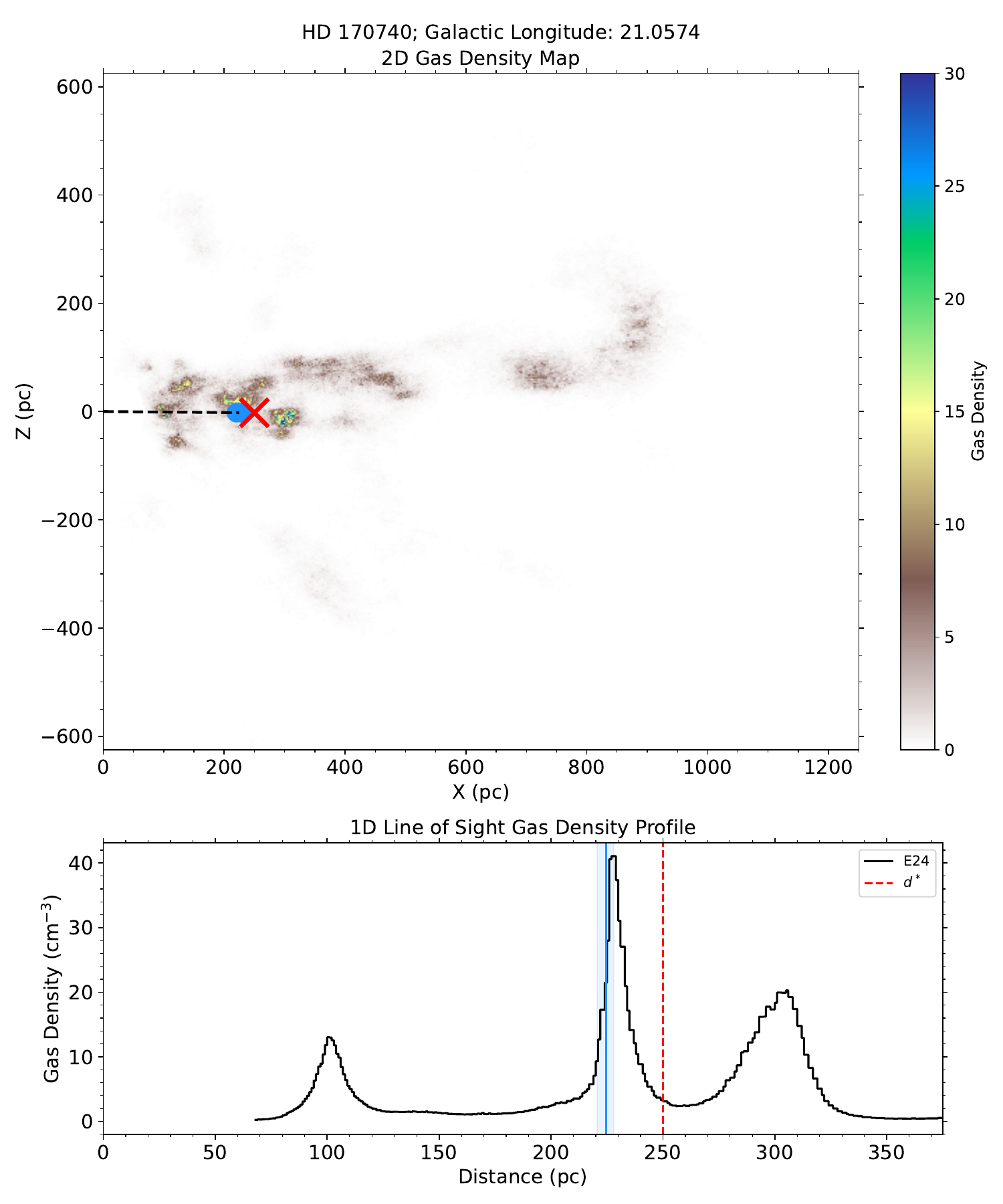}
\caption{Same as Figure \ref{fig_gaia_hd23180}, but for the sight line toward HD~170740.}
\label{fig_gaia_hd170740}
\end{figure}

\clearpage
\begin{figure}
\epsscale{1.0}
\plotone{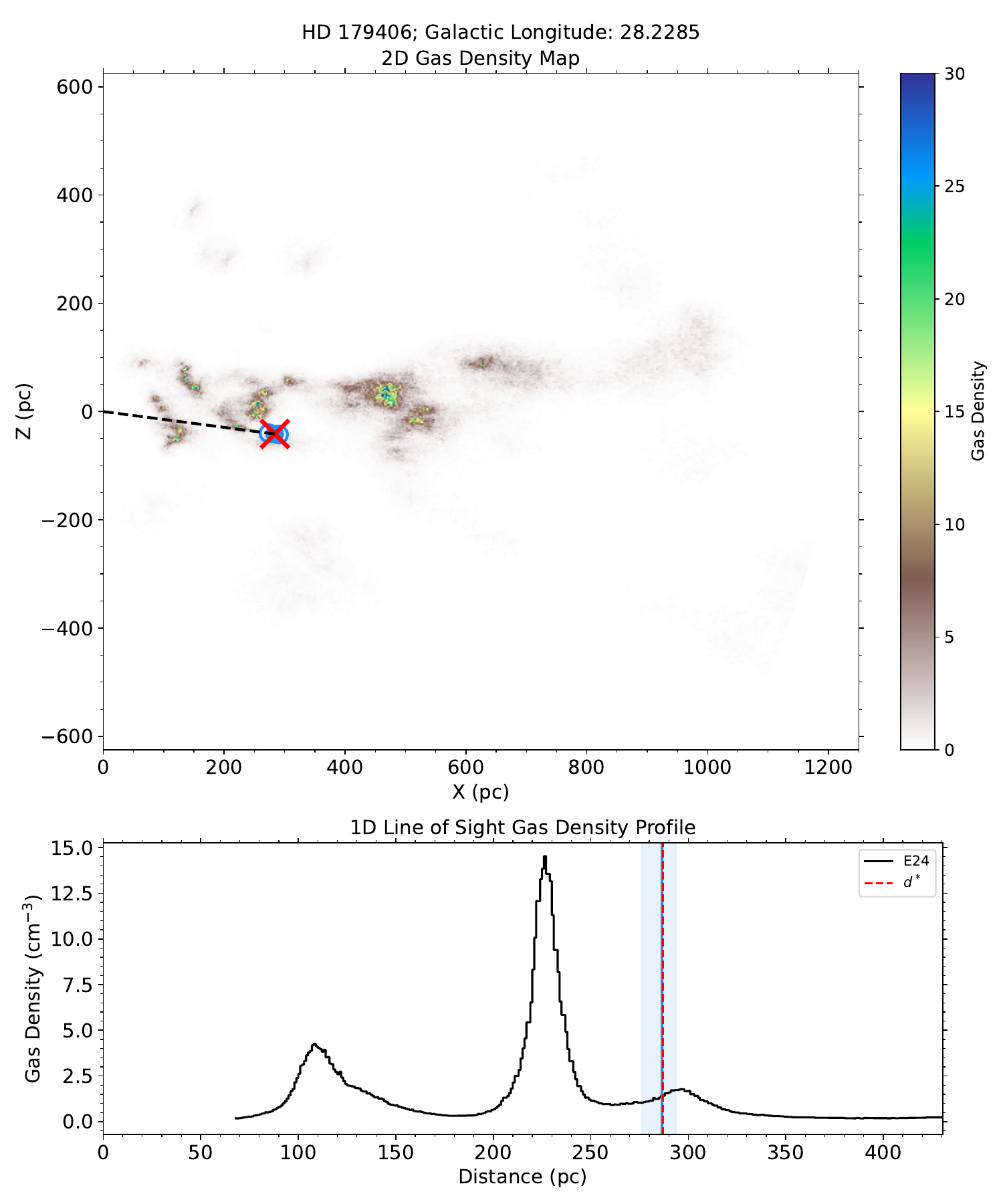}
\caption{Same as Figure \ref{fig_gaia_hd23180}, but for the sight line toward HD~179406.}
\label{fig_gaia_hd179406}
\end{figure}

\clearpage
\begin{figure}
\epsscale{1.0}
\plotone{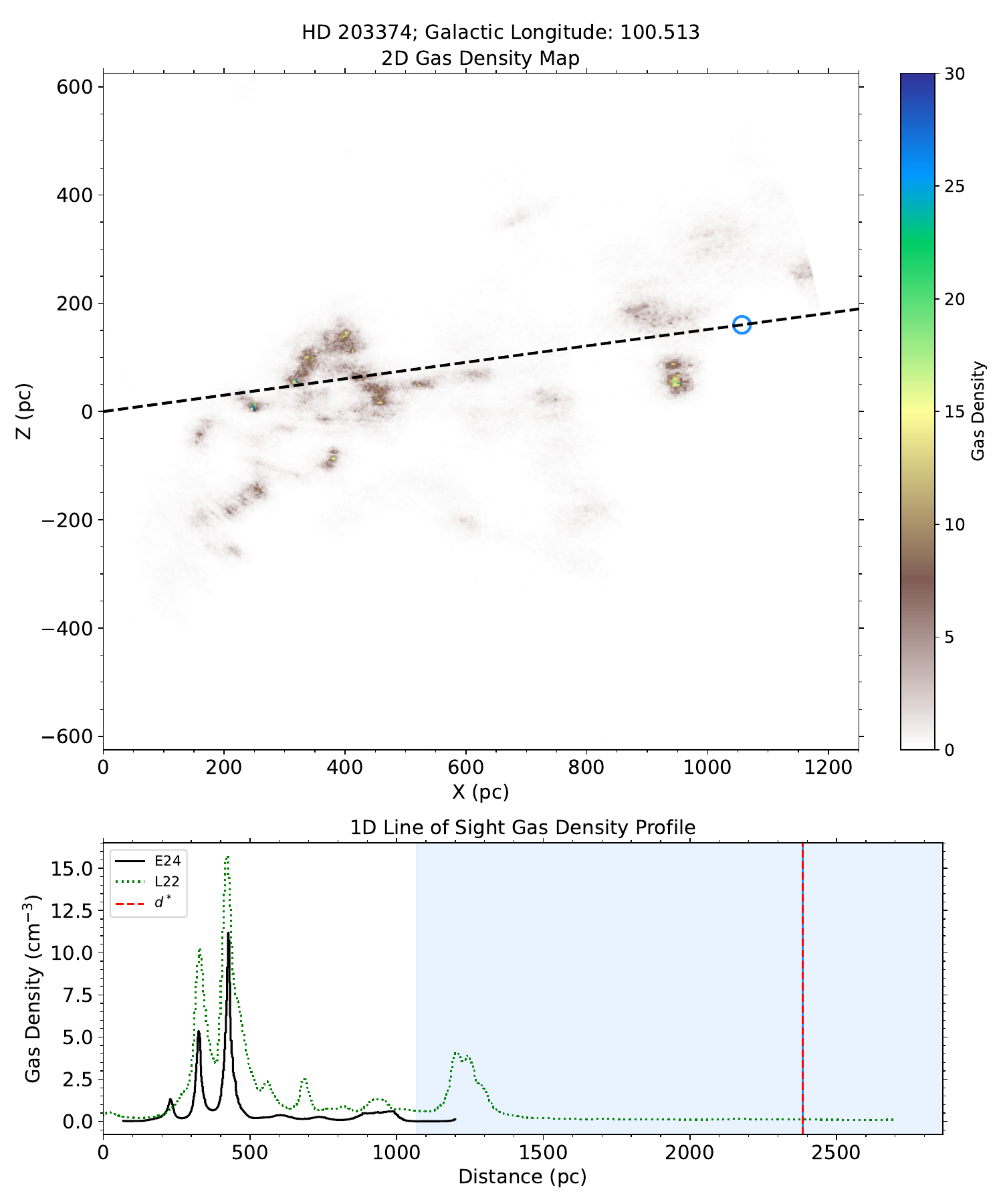}
\caption{Same as Figure \ref{fig_gaia_hd23180}, but for the sight line toward HD~203374. The Gaia EDR3 median and 84\% distances are beyond the extent of the figure.}
\label{fig_gaia_hd203374}
\end{figure}

\clearpage
\begin{figure}
\epsscale{1.0}
\plotone{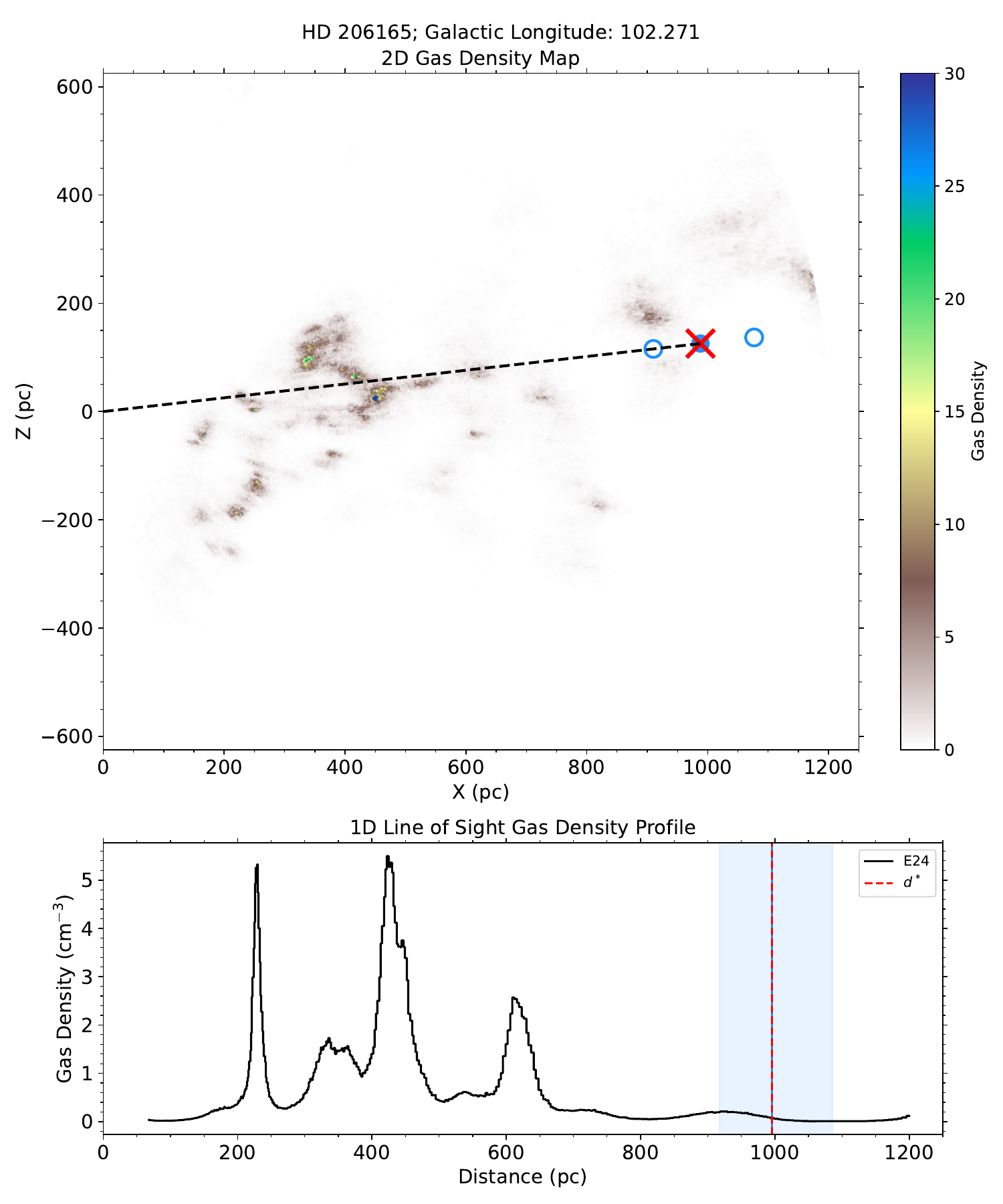}
\caption{Same as Figure \ref{fig_gaia_hd23180}, but for the sight line toward HD~206165.}
\label{fig_gaia_hd206165}
\end{figure}

\clearpage
\begin{figure}
\epsscale{1.0}
\plotone{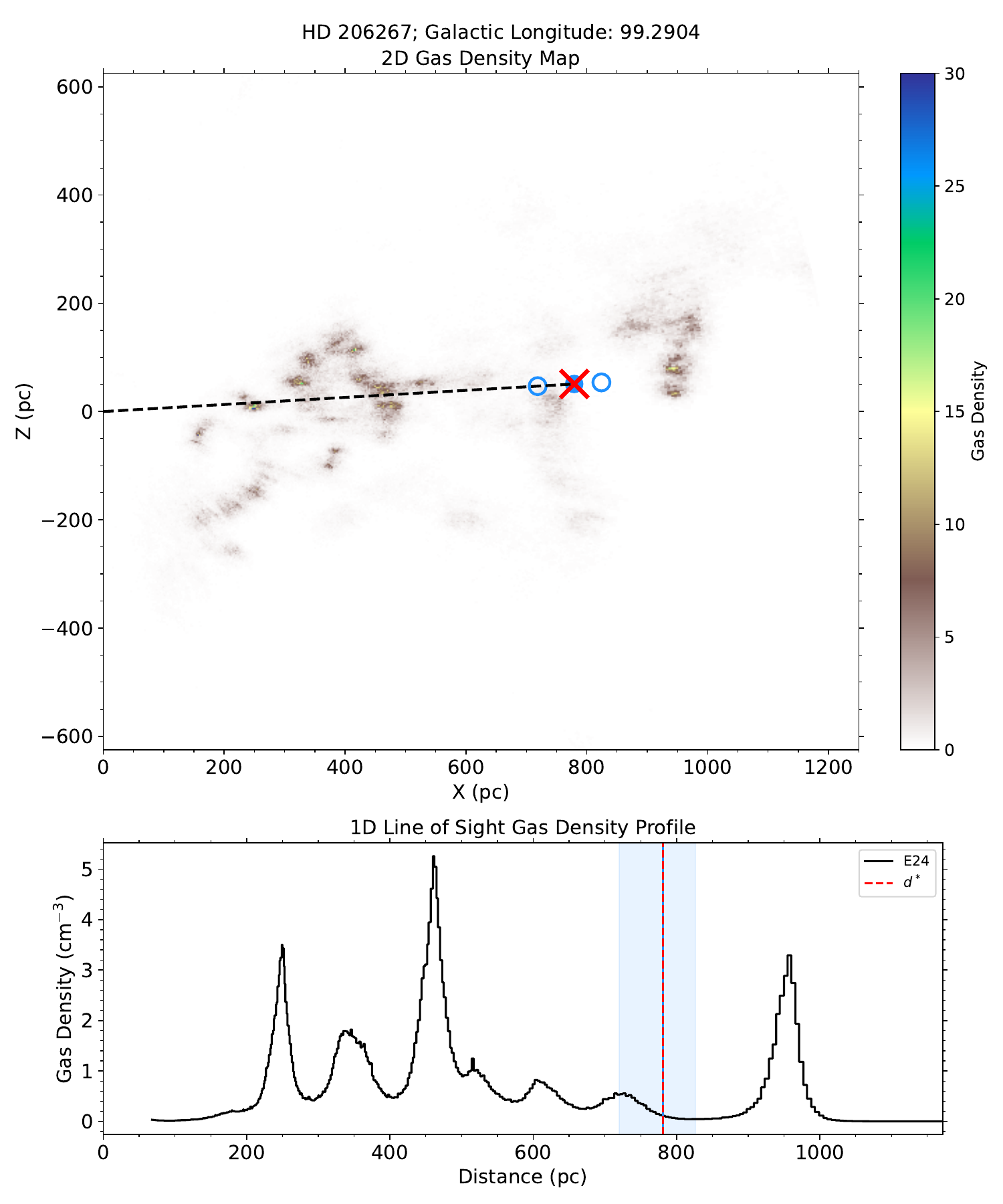}
\caption{Same as Figure \ref{fig_gaia_hd23180}, but for the sight line toward HD~206267.}
\label{fig_gaia_hd206267}
\end{figure}

\clearpage
\begin{figure}
\epsscale{1.0}
\plotone{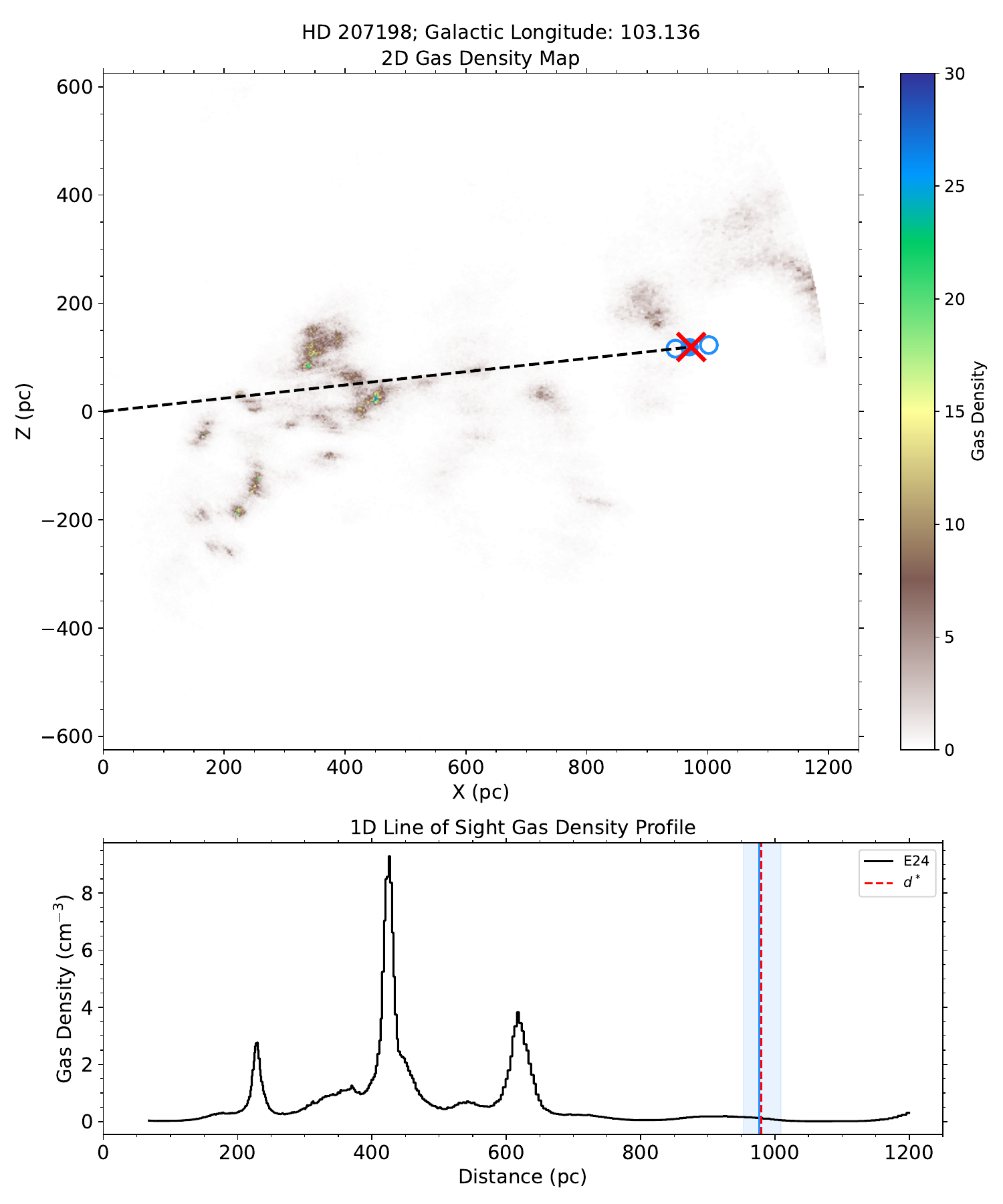}
\caption{Same as Figure \ref{fig_gaia_hd23180}, but for the sight line toward HD~207198.}
\label{fig_gaia_hd207198}
\end{figure}

\clearpage
\begin{figure}
\epsscale{1.0}
\plotone{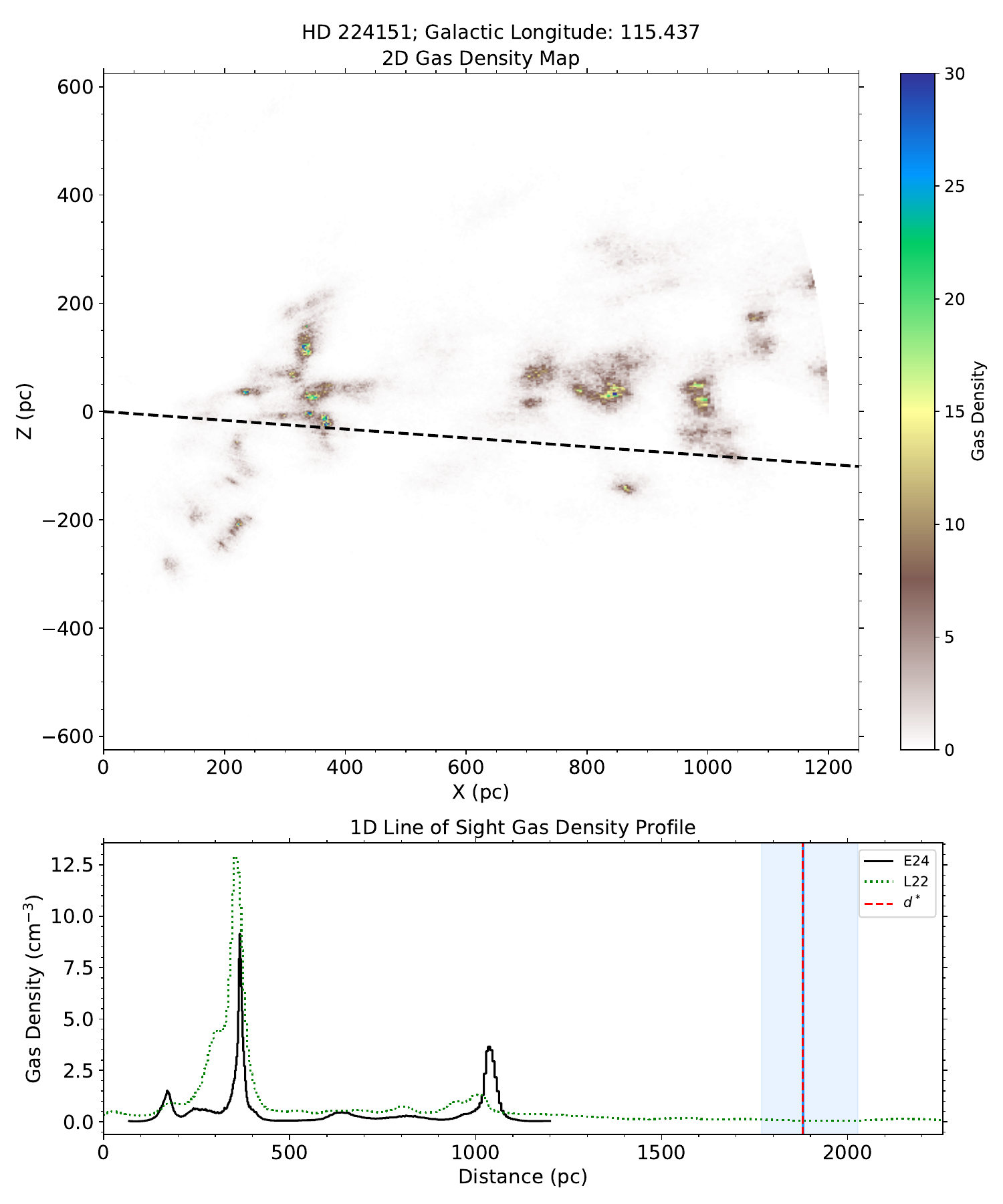}
\caption{Same as Figure \ref{fig_gaia_hd23180}, but for the sight line toward HD~224151. All Gaia EDR3 distance estimates are beyond the extent of the figure.}
\label{fig_gaia_hd224151}
\end{figure}

\subsection{Sight Lines with Trihydrogen Cation Detections that Use Analytical Method}

\clearpage
\begin{figure}
\epsscale{1.0}
\plotone{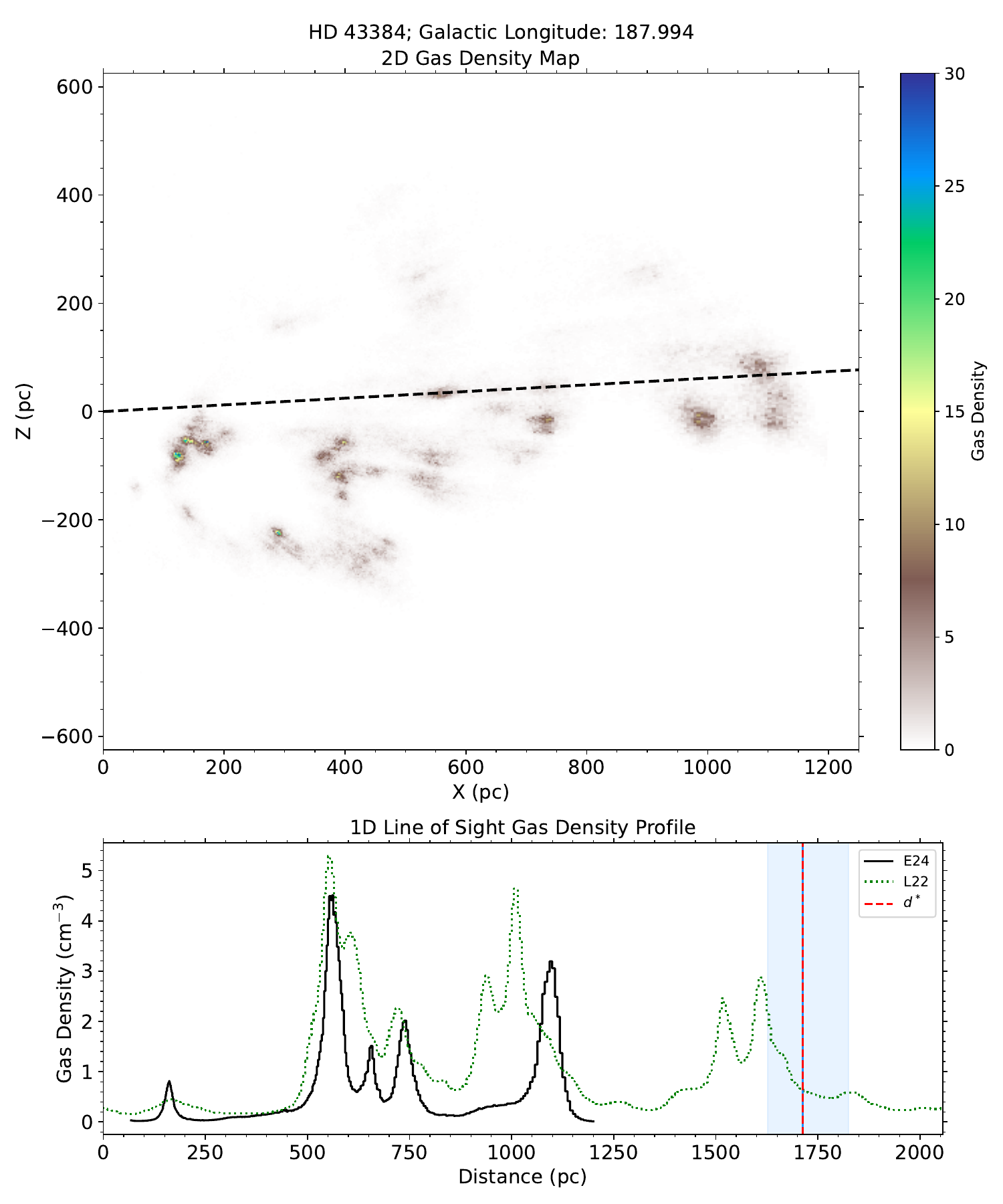}
\caption{Similar to Figure \ref{fig_gaia_hd23180}, but for the sight line toward HD~43384. At $d=1713$~pc the star itself is beyond the extent of the \citet{edenhofer2024} extinction map, and so does not appear in the 2D map. In the bottom panel, the green dotted curve is the density distribution from the \citet{lallement2022} map, which extends beyond 1250~pc but at sparser resolution. The vertical blue line and shaded region again mark the Gaia EDR3 median and 16--84\% distance ranges, respectively, while the vertical dashed red line marks our adopted distance.}
\label{fig_gaia_hd43384}
\end{figure}

\clearpage
\begin{figure}
\epsscale{1.0}
\plotone{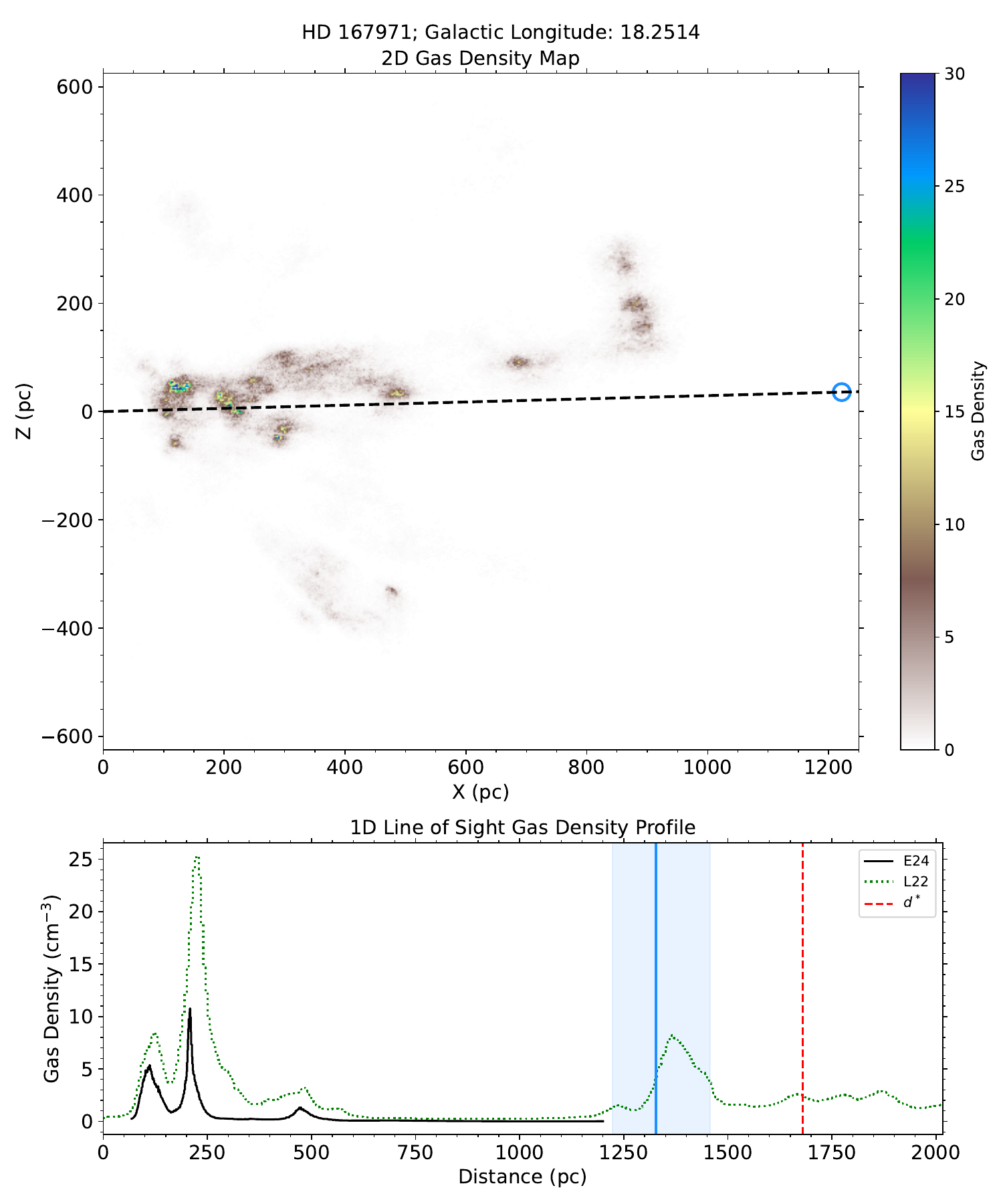}
\caption{Same as Figure \ref{fig_gaia_hd43384}, but for the sight line toward HD~167971.}
\label{fig_gaia_hd167971}
\end{figure}

\clearpage
\begin{figure}
\epsscale{1.0}
\plotone{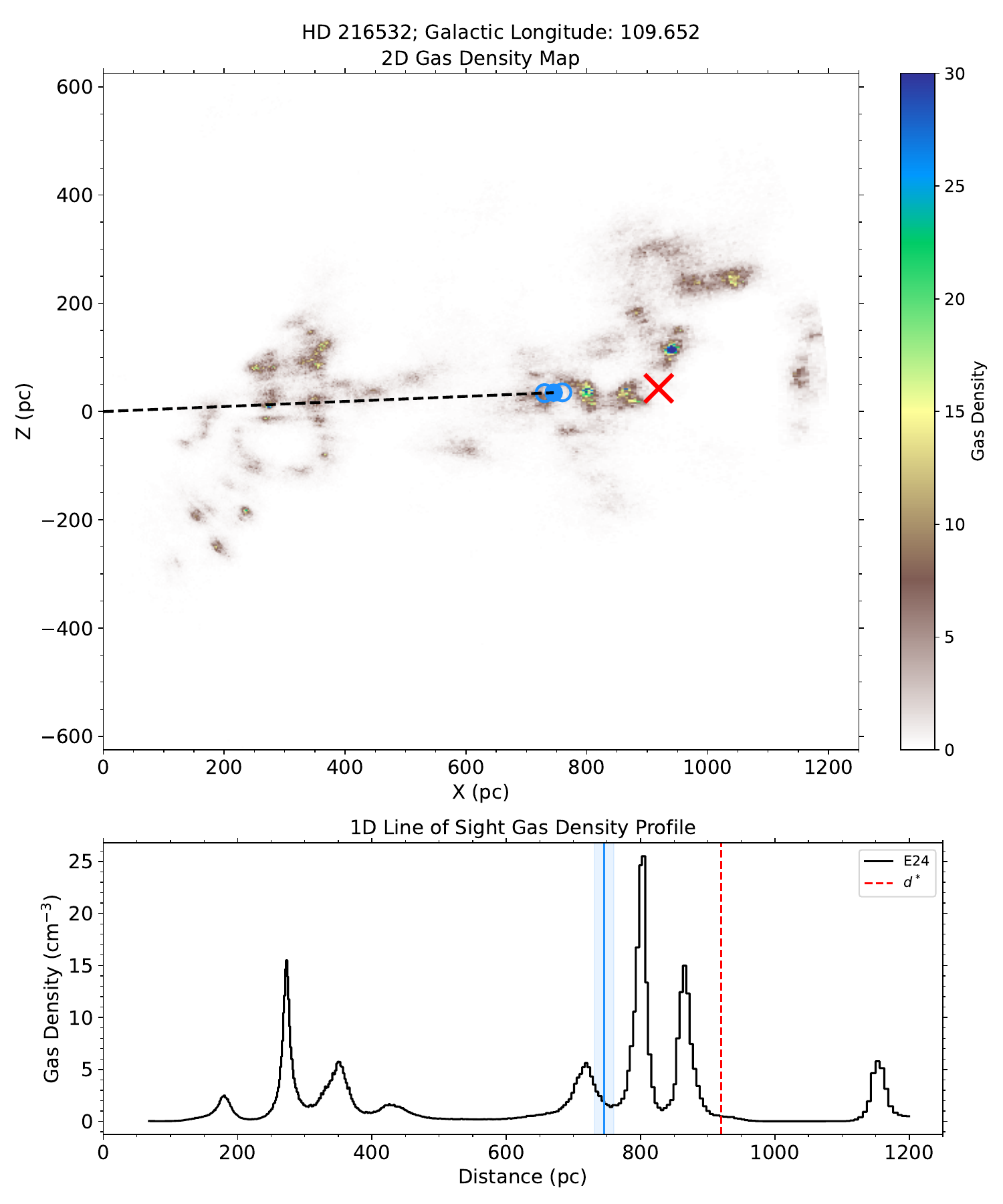}
\caption{Same as Figure \ref{fig_gaia_hd23180}, but for the sight line toward HD~216532.}
\label{fig_gaia_hd216532}
\end{figure}

\clearpage
\begin{figure}
\epsscale{1.0}
\plotone{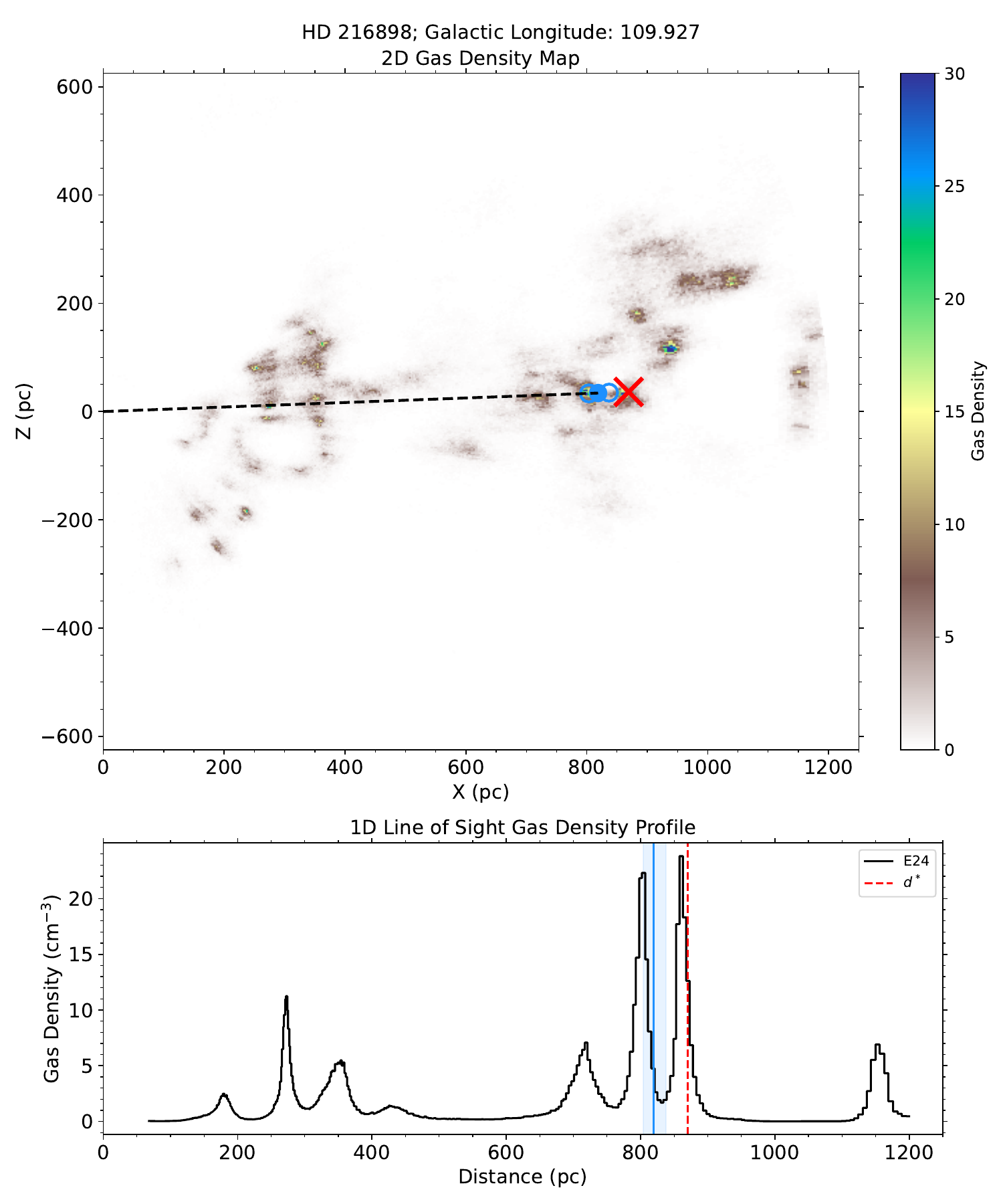}
\caption{Same as Figure \ref{fig_gaia_hd23180}, but for the sight line toward HD~216898.}
\label{fig_gaia_hd216898}
\end{figure}

\subsection{Sight Lines with Trihydrogen Cation Non-Detections}

\clearpage
\begin{figure}
\epsscale{1.0}
\plotone{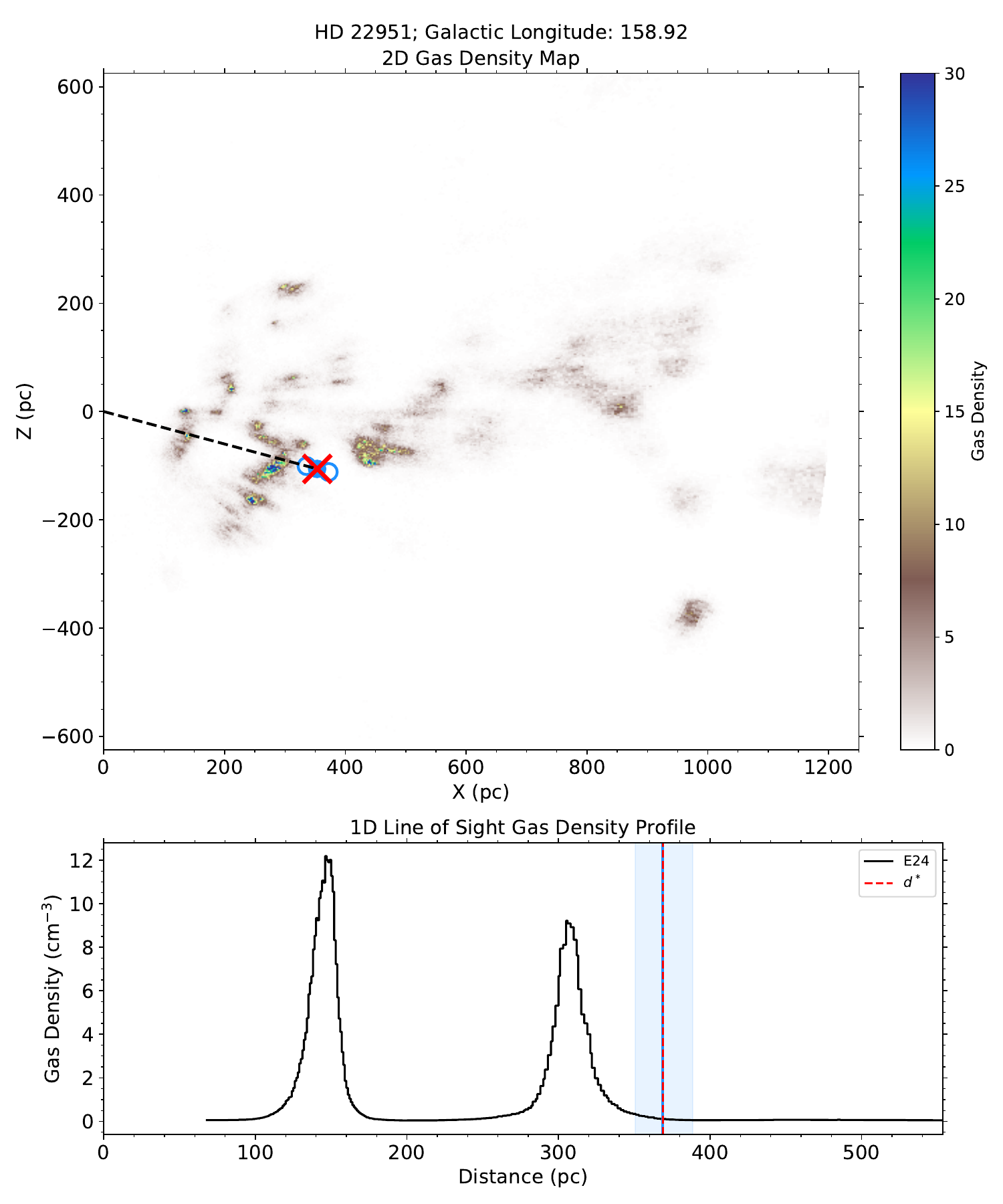}
\caption{Same as Figure \ref{fig_gaia_hd23180}, but for the sight line toward HD~22951.}
\label{fig_gaia_hd22951}
\end{figure}

\clearpage
\begin{figure}
\epsscale{1.0}
\plotone{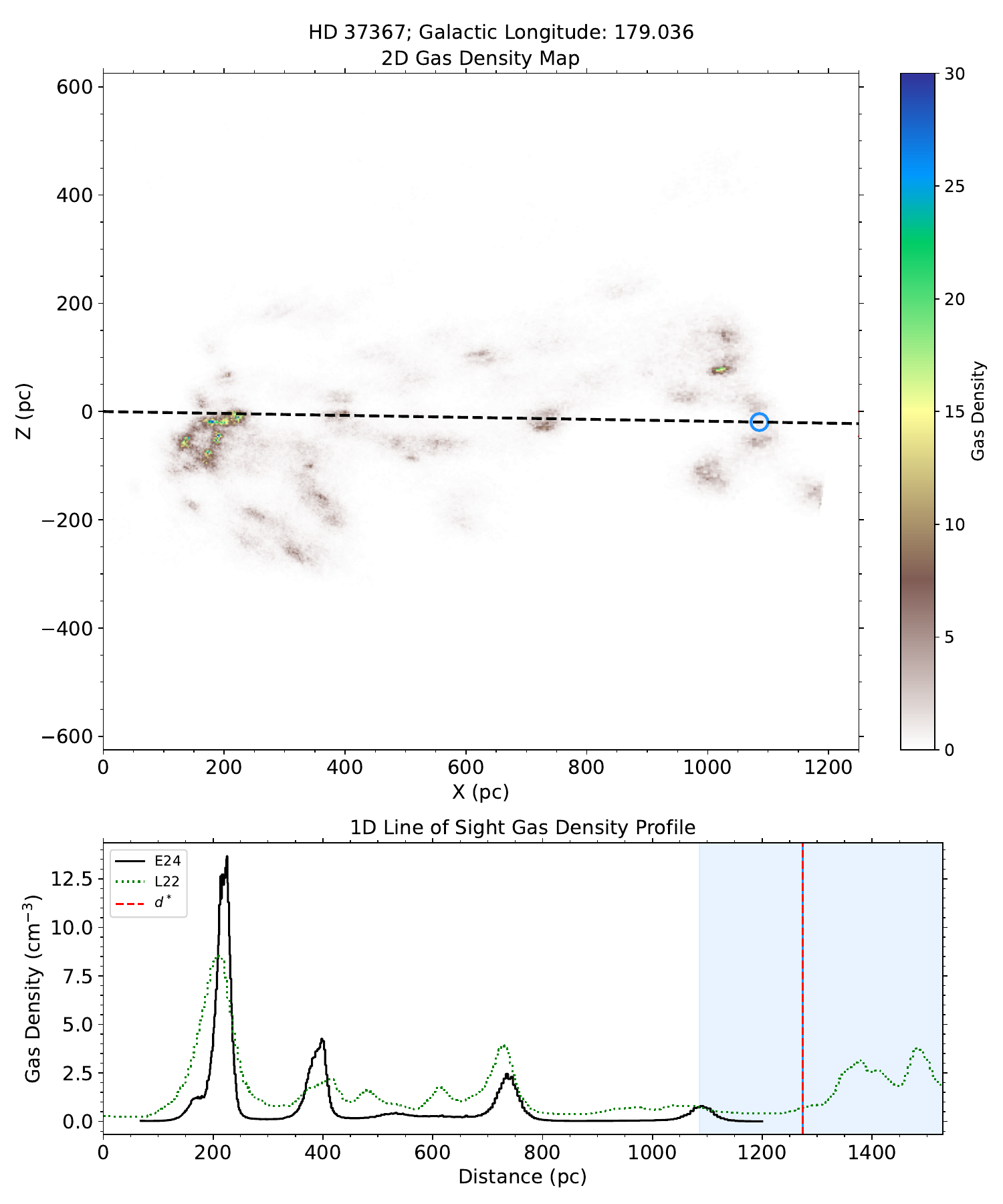}
\caption{Same as Figure \ref{fig_gaia_hd43384}, but for the sight line toward HD~37367. At $d=1274$~pc the star is beyond the limits of the 2D map presented in the top panel.}
\label{fig_gaia_hd37367}
\end{figure}

\clearpage
\begin{figure}
\epsscale{1.0}
\plotone{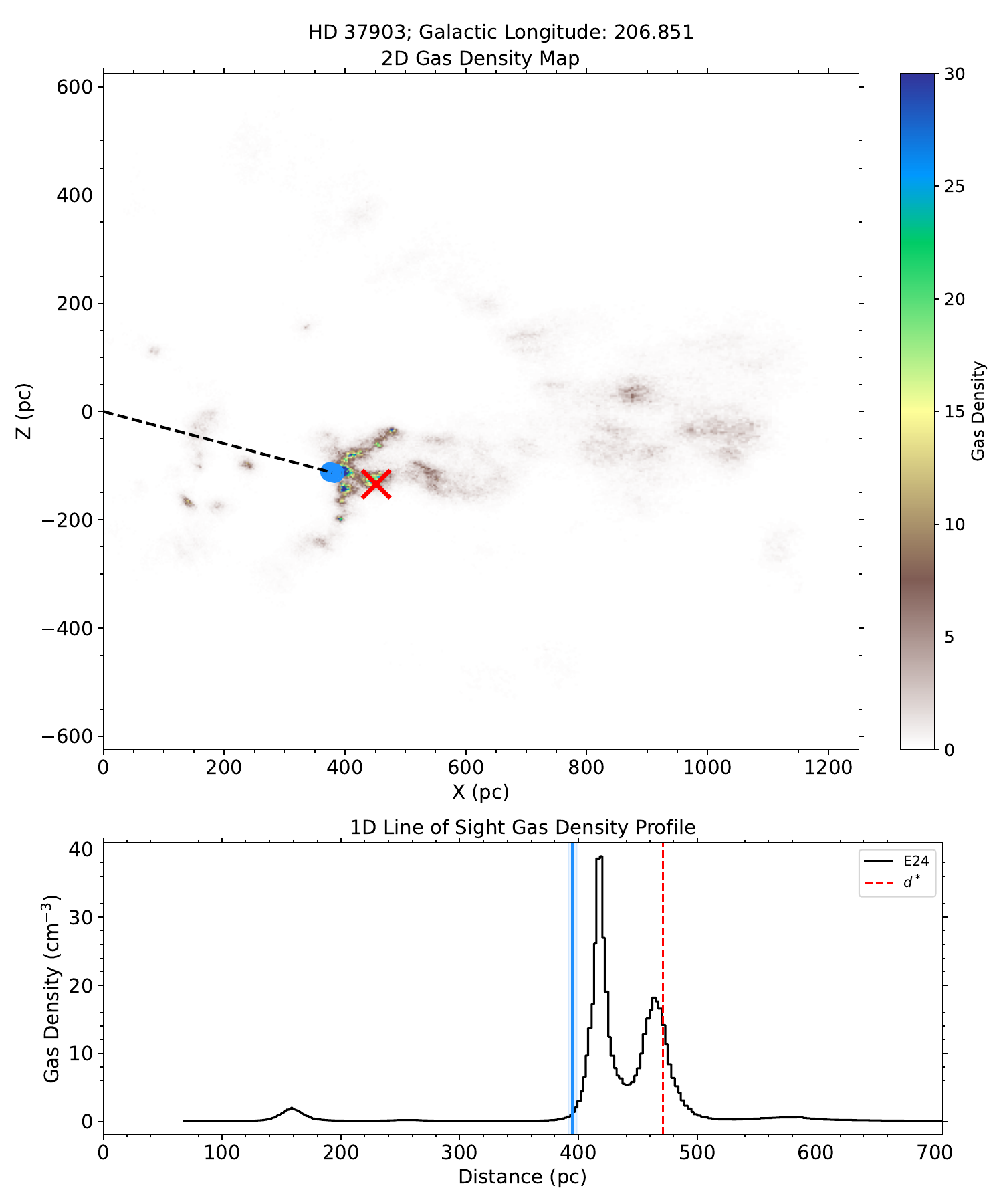}
\caption{Same as Figure \ref{fig_gaia_hd23180}, but for the sight line toward HD~37903.}
\label{fig_gaia_hd37903}
\end{figure}

\clearpage
\begin{figure}
\epsscale{1.0}
\plotone{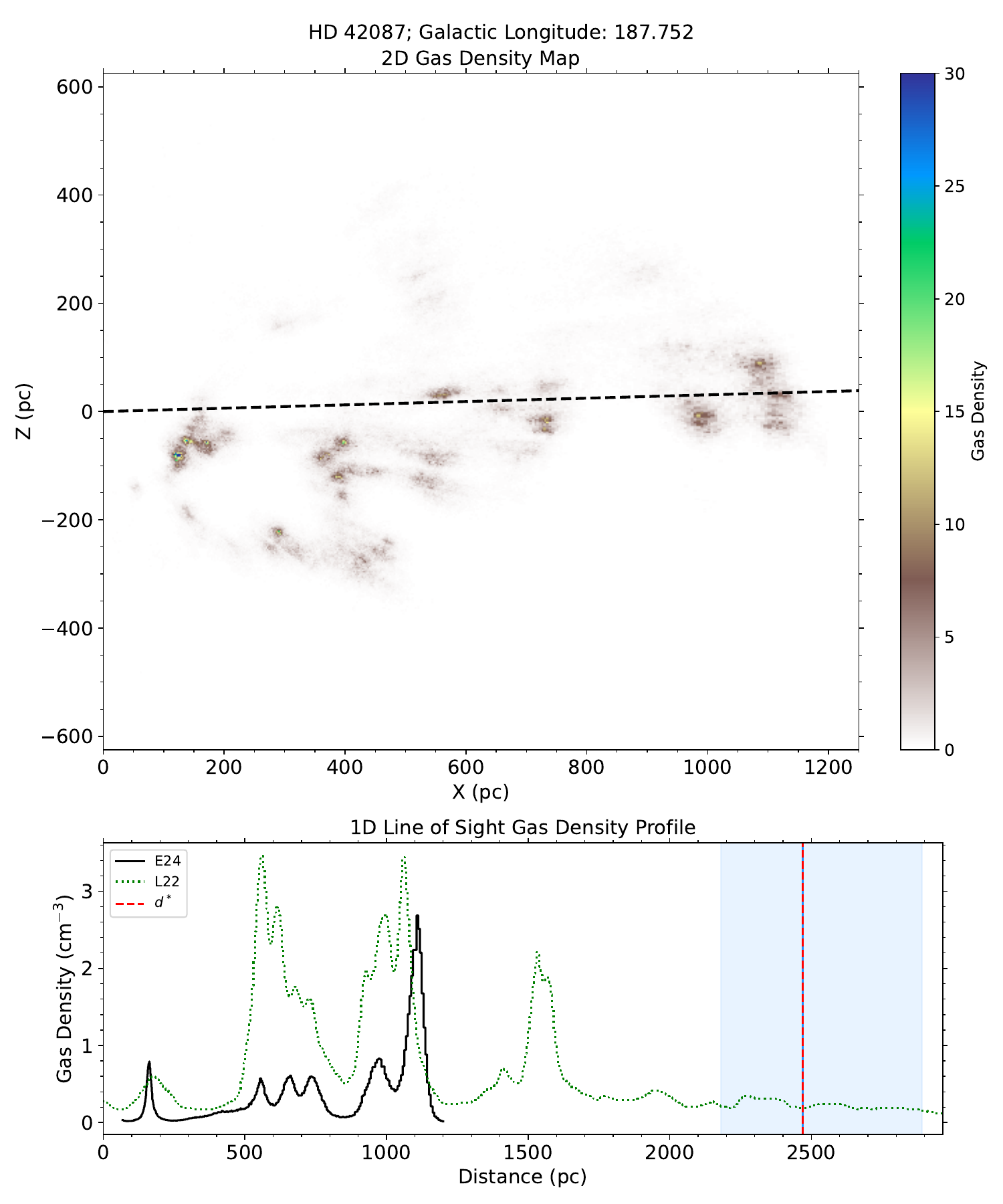}
\caption{Same as Figure \ref{fig_gaia_hd43384}, but for the sight line toward HD~42087. At $d=2470$~pc the star is beyond the limits of the 2D map presented in the top panel.}
\label{fig_gaia_hd42087}
\end{figure}

\clearpage
\begin{figure}
\epsscale{1.0}
\plotone{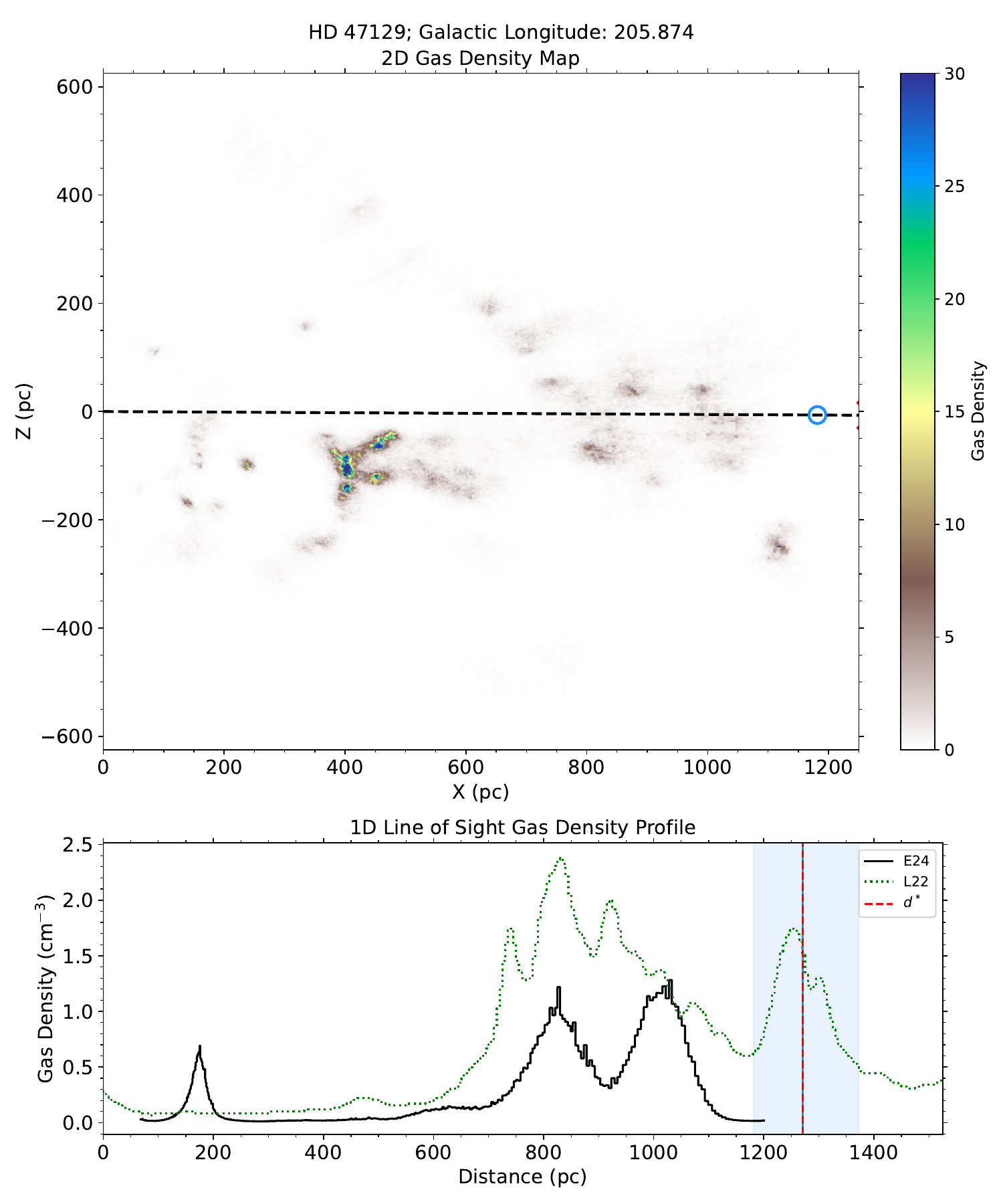}
\caption{Same as Figure \ref{fig_gaia_hd43384}, but for the sight line toward HD~47129. At $d=1271$~pc the star is beyond the limits of the 2D map presented in the top panel.}
\label{fig_gaia_hd47129}
\end{figure}

\clearpage
\begin{figure}
\epsscale{1.0}
\plotone{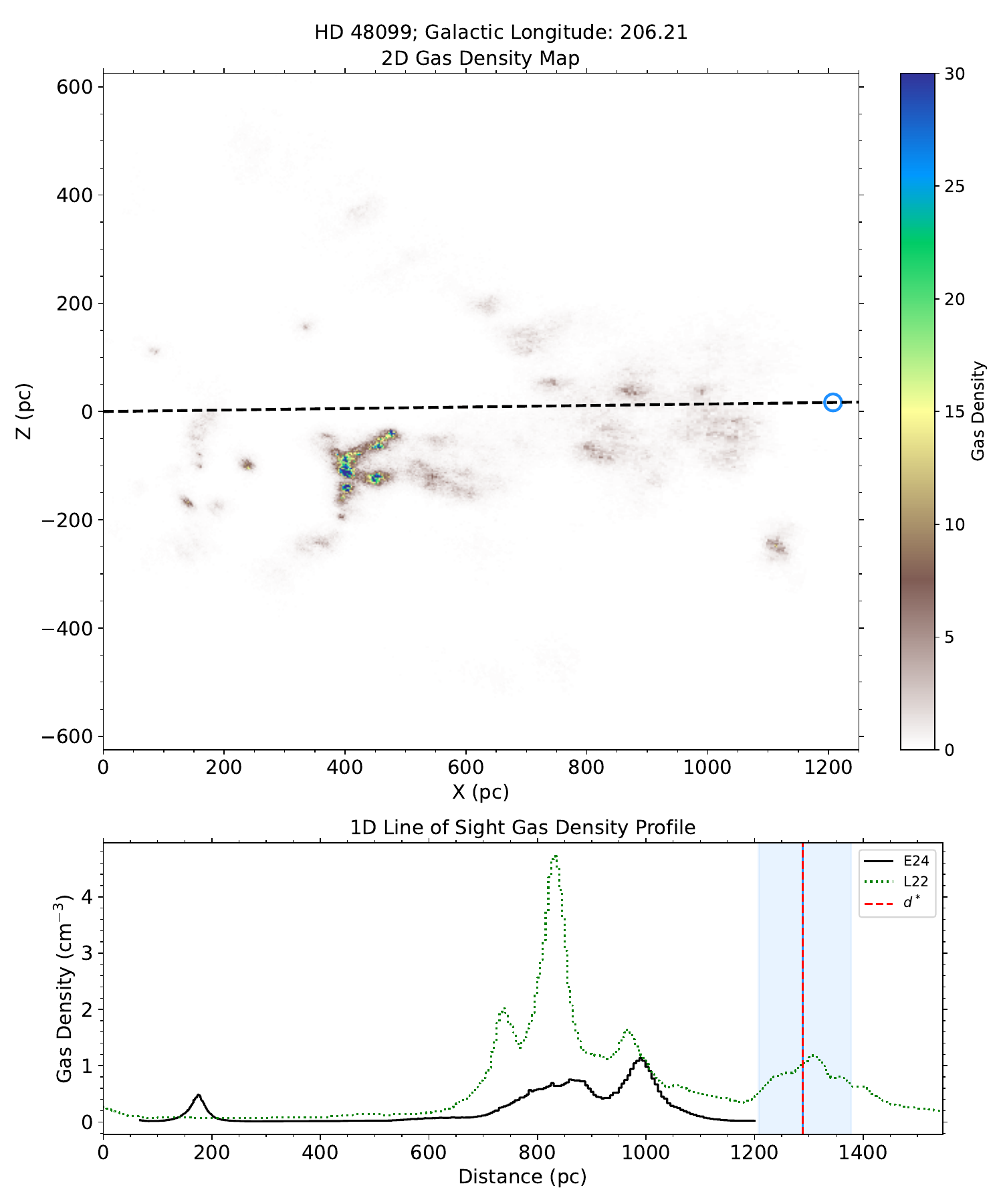}
\caption{Same as Figure \ref{fig_gaia_hd43384}, but for the sight line toward HD~48099. At $d=1289$~pc the star is beyond the limits of the 2D map presented in the top panel.}
\label{fig_gaia_hd48099}
\end{figure}

\clearpage
\begin{figure}
\epsscale{1.0}
\plotone{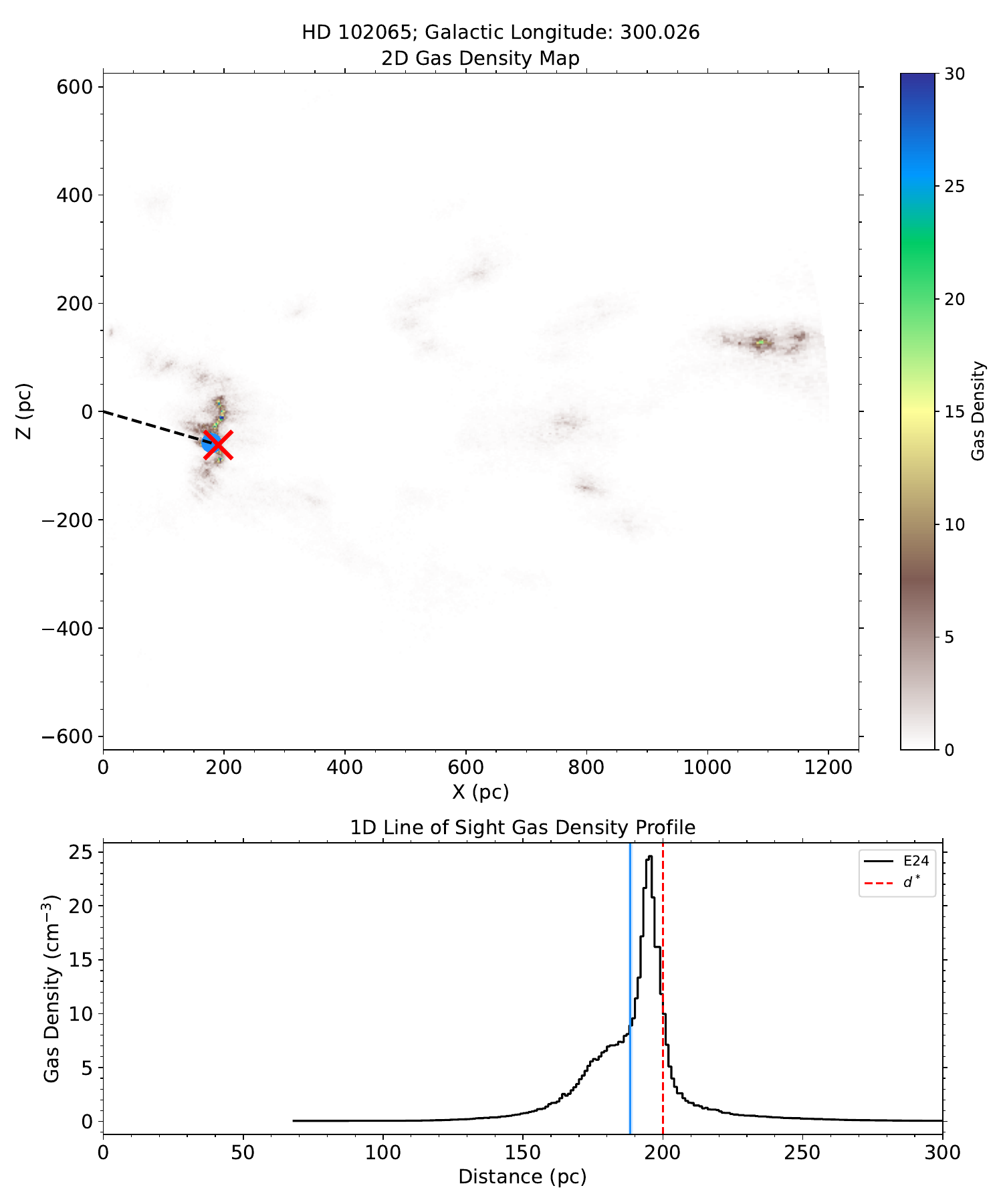}
\caption{Same as Figure \ref{fig_gaia_hd23180}, but for the sight line toward HD~102065.}
\label{fig_gaia_hd102065}
\end{figure}

\clearpage
\begin{figure}
\epsscale{1.0}
\plotone{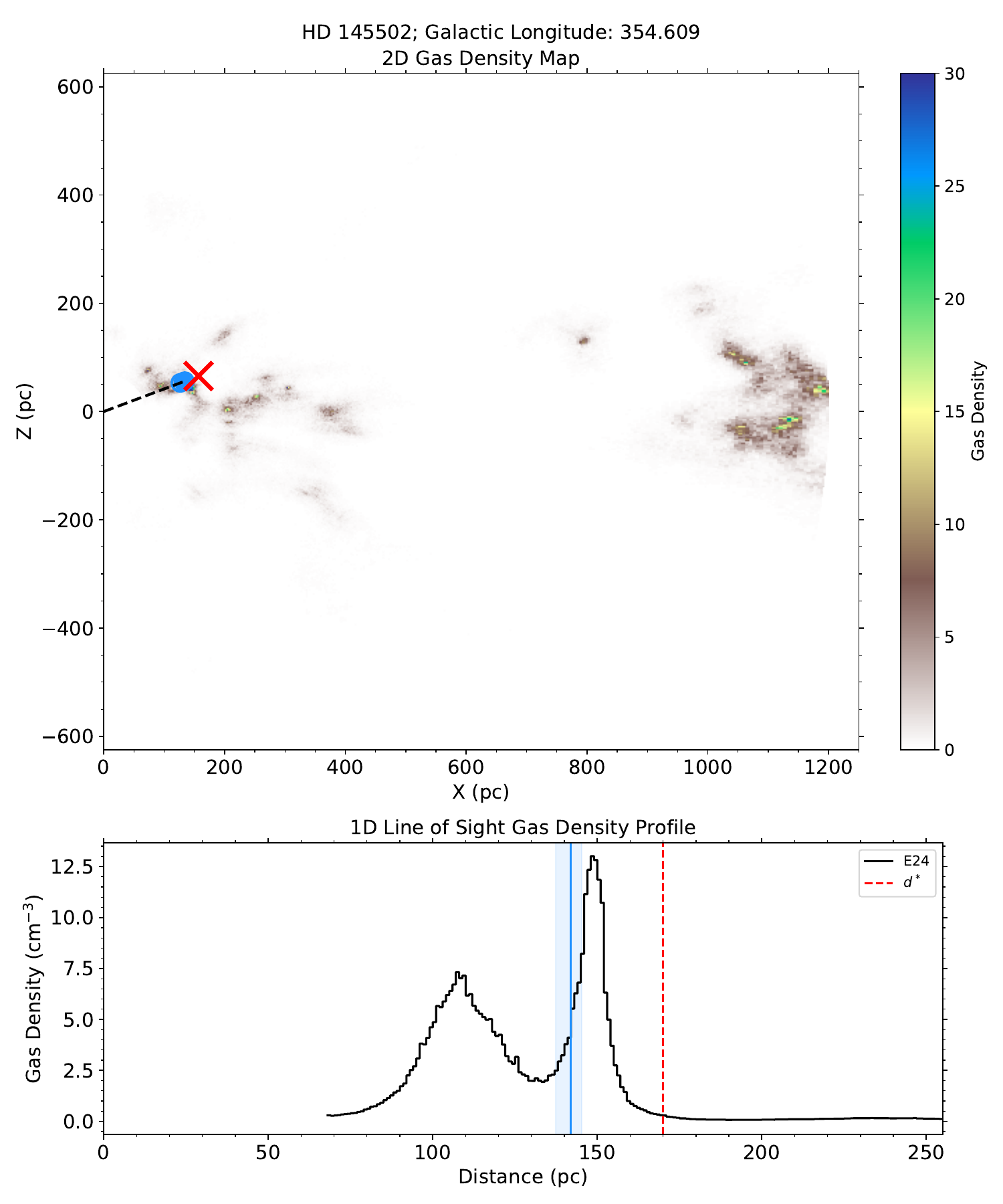}
\caption{Same as Figure \ref{fig_gaia_hd23180}, but for the sight line toward HD~145502}
\label{fig_gaia_hd145502.}
\end{figure}

\clearpage
\begin{figure}
\epsscale{1.0}
\plotone{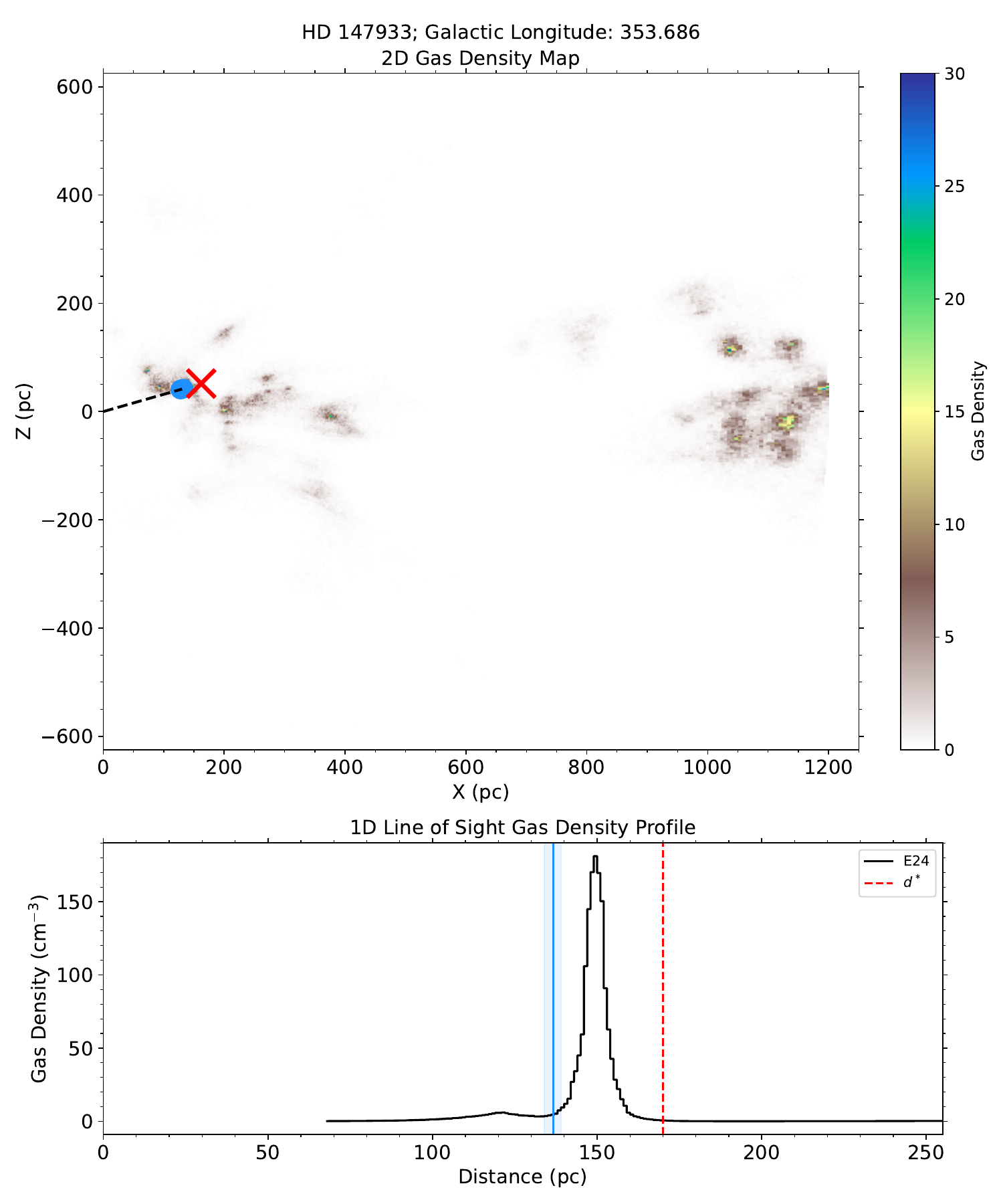}
\caption{Same as Figure \ref{fig_gaia_hd23180}, but for the sight line toward HD~147933.}
\label{fig_gaia_hd147933}
\end{figure}

\clearpage
\begin{figure}
\epsscale{1.0}
\plotone{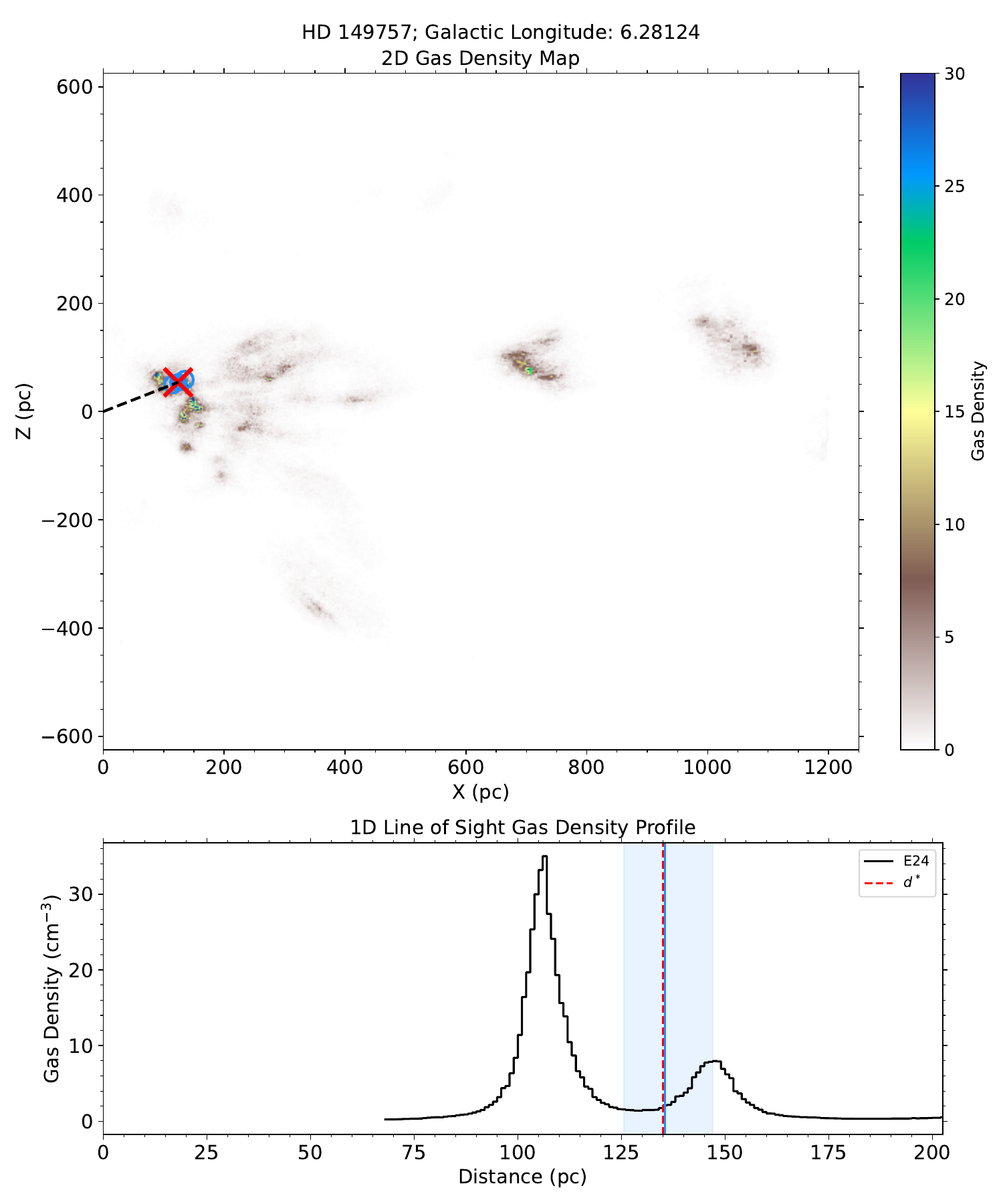}
\caption{Same as Figure \ref{fig_gaia_hd23180}, but for the sight line toward HD~149757.}
\label{fig_gaia_hd149757}
\end{figure}

\clearpage
\begin{figure}
\epsscale{1.0}
\plotone{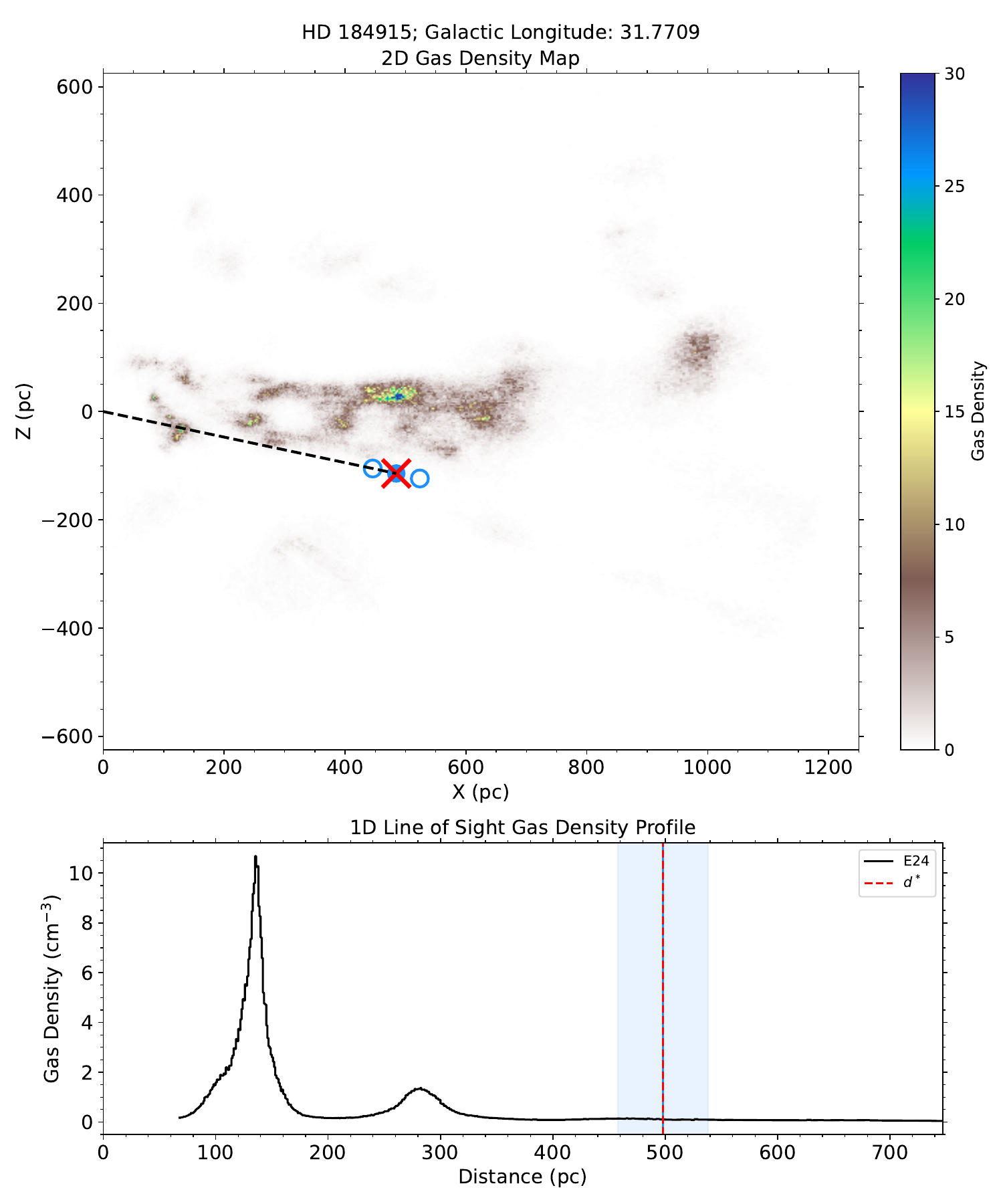}
\caption{Same as Figure \ref{fig_gaia_hd23180}, but for the sight line toward HD~184915.}
\label{fig_gaia_hd184915}
\end{figure}

\clearpage
\begin{figure}
\epsscale{1.0}
\plotone{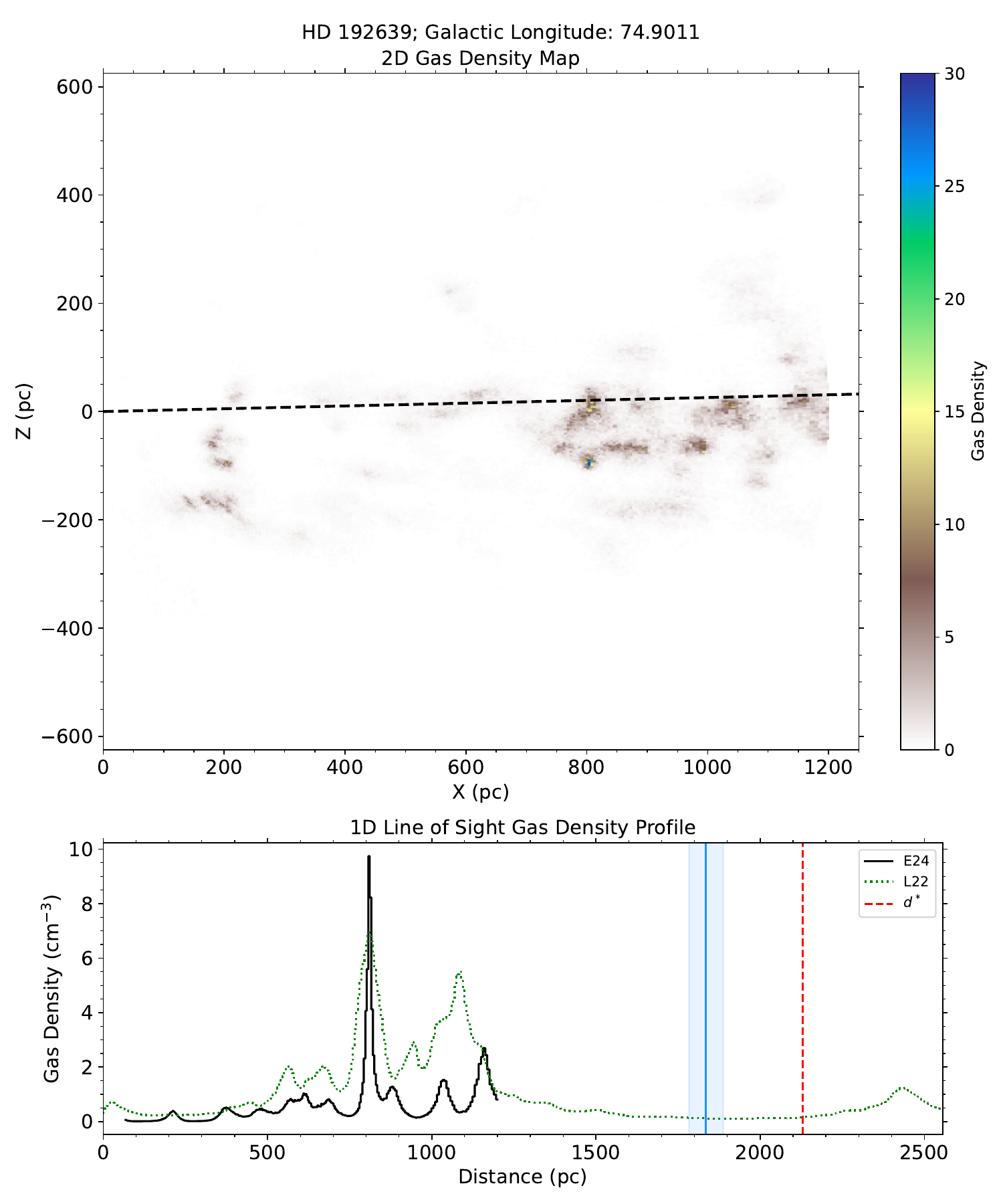}
\caption{Same as Figure \ref{fig_gaia_hd43384}, but for the sight line toward HD~192639. At $d=2130$~pc the star is beyond the limits of the 2D map presented in the top panel.}
\label{fig_gaia_hd192639}
\end{figure}

\clearpage
\begin{figure}
\epsscale{1.0}
\plotone{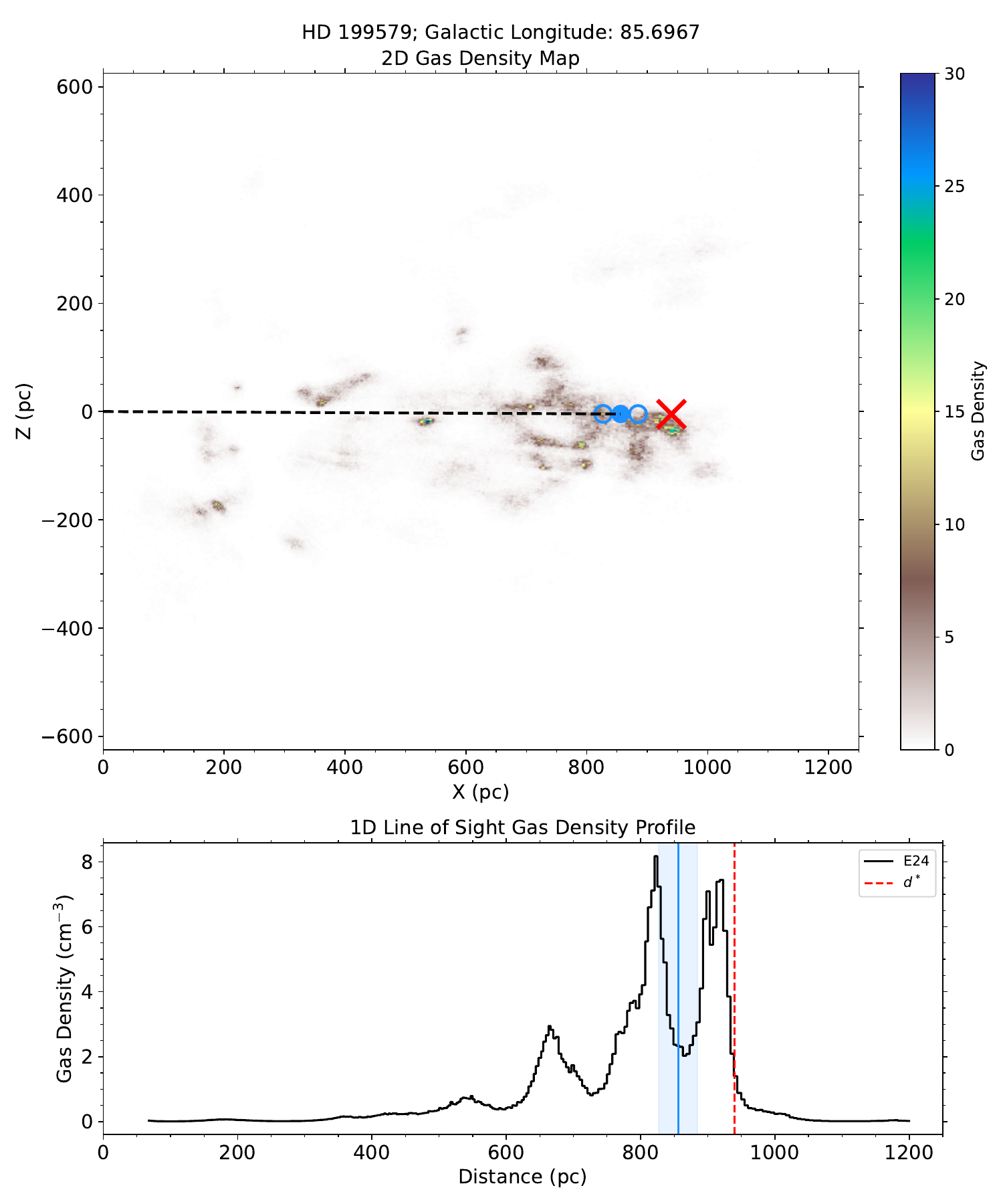}
\caption{Same as Figure \ref{fig_gaia_hd23180}, but for the sight line toward HD~199579.}
\label{fig_gaia_hd199579}
\end{figure}

\clearpage
\begin{figure}
\epsscale{1.0}
\plotone{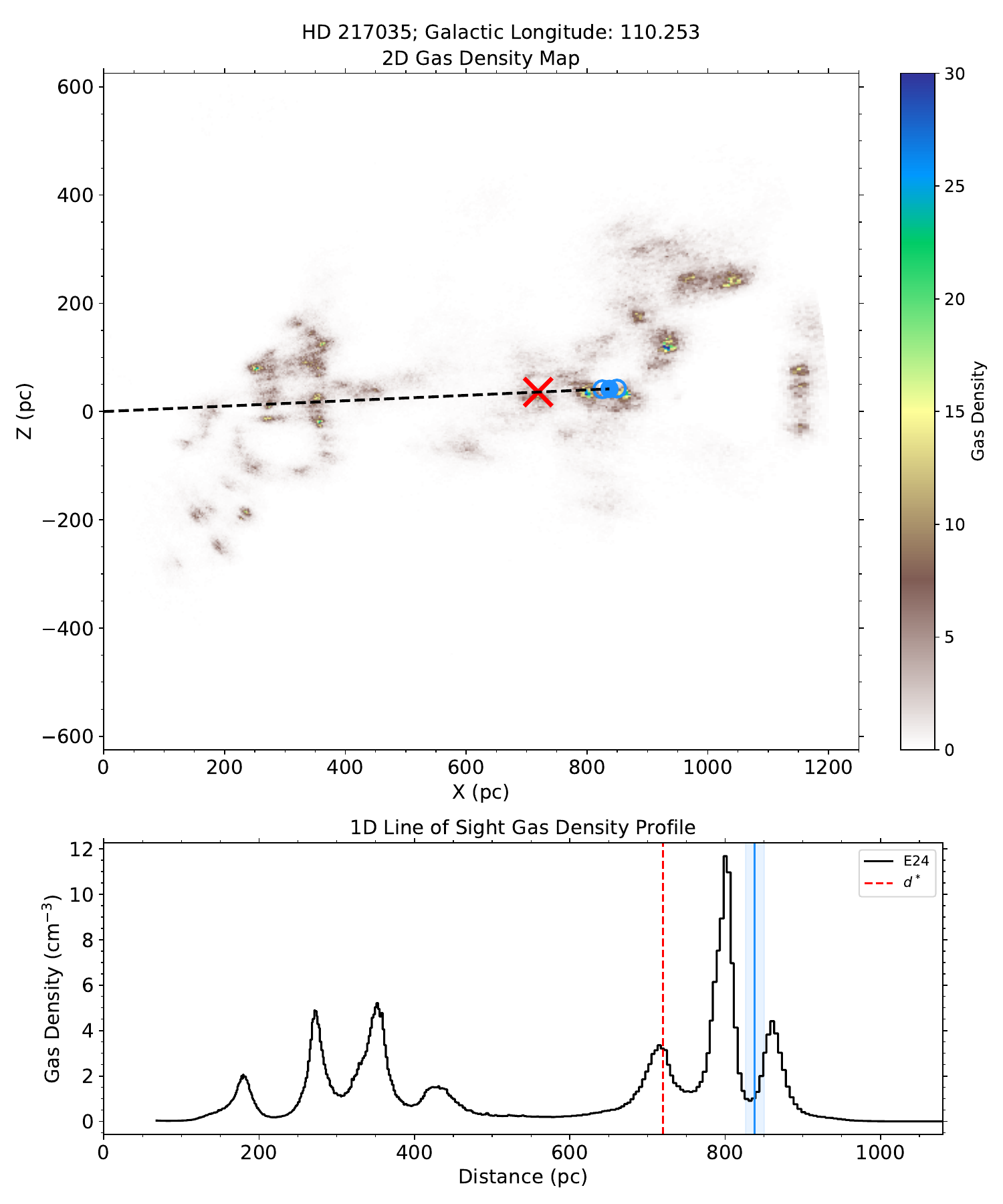}
\caption{Same as Figure \ref{fig_gaia_hd23180}, but for the sight line toward HD~217035.}
\label{fig_gaia_hd217035}
\end{figure}

\clearpage
\begin{figure}
\epsscale{1.0}
\plotone{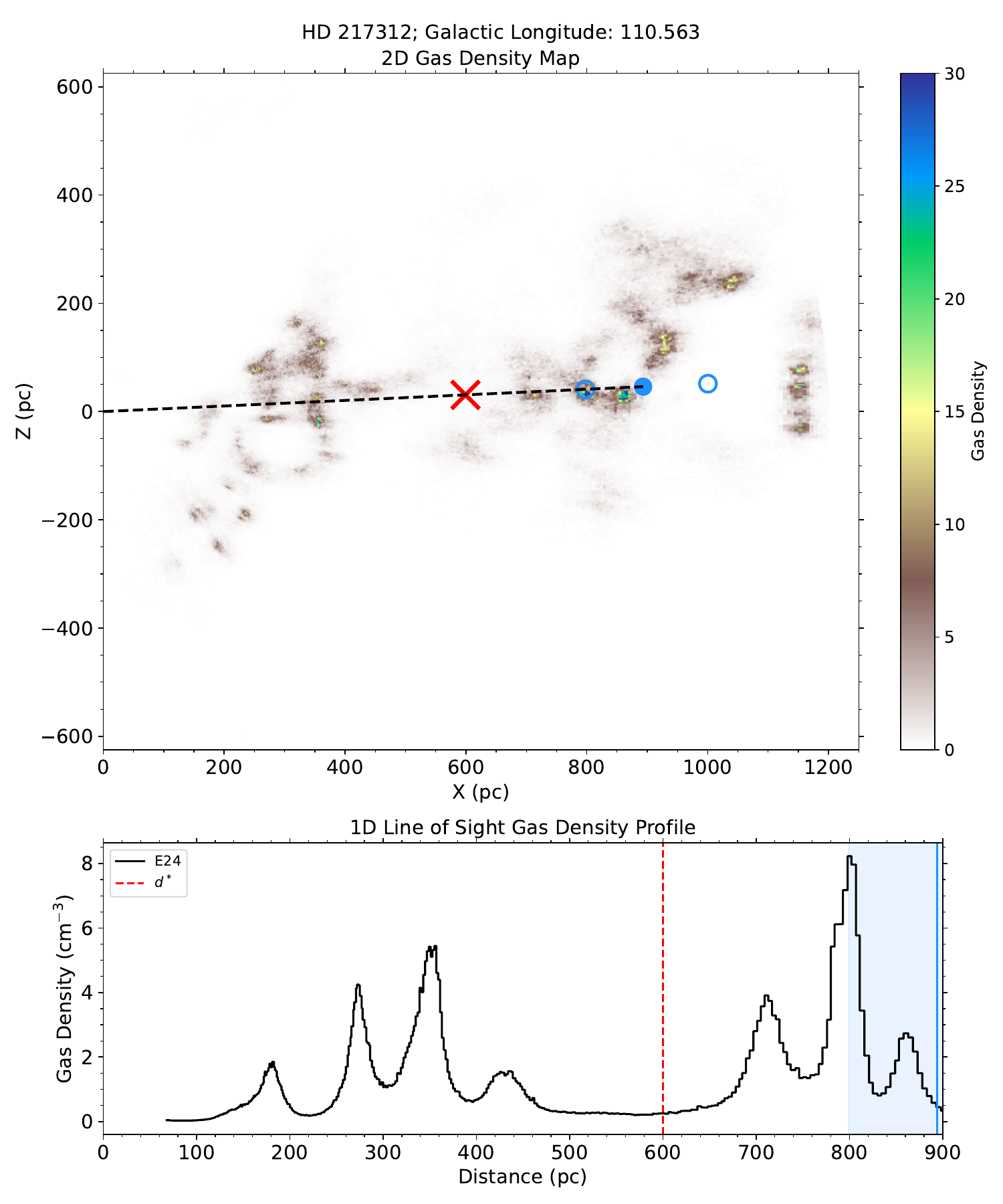}
\caption{Same as Figure \ref{fig_gaia_hd23180}, but for the sight line toward HD~217312.}
\label{fig_gaia_hd217312}
\end{figure}

\end{document}